\documentclass[a4paper]{amsart}
\usepackage[T1]{fontenc}

\usepackage{amsmath,amssymb,epsf}
\usepackage{amsfonts,amsbsy,amscd,stmaryrd}
\usepackage{latexsym,amsopn}
\usepackage{amsthm}
\usepackage{esint}
\usepackage{mathrsfs}
\usepackage{float}
\usepackage{graphicx}
\usepackage{xcolor}
\definecolor{Green}{rgb}{0.,1.,0.}
\definecolor{Blue}{rgb}{0.,0.,1.}
\definecolor{Red}{rgb}{1.,0.,0.}
\newcounter{smallarabics}
\newenvironment{arabicenumerate}
{\begin{list}{{\normalfont\textrm{(\arabic{smallarabics})}}}
  {\usecounter{smallarabics}\setlength{\itemindent}{0cm}
   \setlength{\leftmargin}{5ex}\setlength{\labelwidth}{4ex}
   \setlength{\topsep}{0.75\parsep}\setlength{\partopsep}{0ex}
   \setlength{\itemsep}{0ex}}}
{\end{list}}

\newcounter{smallroman}

\newcommand{\ben}{\begin{arabicenumerate}}  
\newcommand{\een}{\end{arabicenumerate}}

\def\init{\setcounter{equation}{0}}

\newtheorem{theoreme}{Theorem }[section]
\newtheorem{proposition}[theoreme]{Proposition}
\newtheorem{lemma}[theoreme]{Lemma}
\newtheorem{definition}[theoreme]{Definition}

\newtheorem{remark}[theoreme]{Remark}
\newtheorem{example}[theoreme]{Example}
\newcommand{\beq}{\begin{equation}}
\newcommand{\eeq}{\end{equation}}

\newcommand{\bex}{\begin{example}}
\newcommand{\eex}{\end{example}}
\def\bel{\begin{lemma}}
\def\eel{\end{lemma}}
\def\bet{\begin{theoreme}}
\def\eet{\end{theoreme}}
\def\bed{\begin{definition}}
\def\eed{\end{definition}}
\def\ber{\begin{remark}}
\def\eer{\end{remark}}

\def\rr{{\mathbb R}}

\def\cc{{\mathbb C}}
\def\nn{{\mathbb N}}

\def\Im{{\rm Im}\,}
\def\Re{{\rm Re}\,}

\def\cl{{\rm cl}}
\def\bar{\overline}

\def\cinf{C^\infty}
\def\c0inf{C_0^\infty}

\def\proof{
\noindent{\bf Proof.}\ \ }
\def\ch{{\mathfrak h}}

\def\cZ{{\mathcal Z}}
\def\cY{{\mathcal Y}}

\def\cS{{\mathcal S}}

\def\cD{{\mathcal D}}

\def\cM{{\mathcal M}}

\def\CCR{{\rm CCR}}

\def\i{{\rm i}}

\def\Dom{{\rm Dom}}

\def\qed{$\Box$\medskip}

\def \p{ \partial}

\def\12{\frac{1}{2}}
\def\14{\frac{1}{4}}

\def\supp{{\rm supp}\,}

\def\e{{\rm e}}

\def\bbbone{{\mathchoice {\rm 1\mskip-4mu l} {\rm 1\mskip-4mu l}
{\rm 1\mskip-4.5mu l} {\rm 1\mskip-5mu l}}}
\def\one{\bbbone}
\def\cH{{\mathcal H}}

\def\ii{{\rm j}}

\def\w{{\rm w}}

\def\sgn{{\rm sgn}}

\def\Ker{{\rm Ker}}
\def\coinf{C_{\rm c}^{\infty}}

\def\cF{{\mathcal F}}

\def\cX{{\mathcal X}}

\def\cK{{\mathcal K}}

\def\rw{{\rm rw}}

\def \p{ \partial}

\def\12{\frac{1}{2}}

\def\e{{\rm e}}

\def\cH{{\mathcal H}}

\def\bep{\begin{proposition}}
\def\eep{\end{proposition}}

\def\b{{\rm b}}

\def\w{{\rm w}}

\newcommand{\mat}[4]{\left(\begin{array}{cc}#1 &#2  \\ #3 &#4 \end{array}\right)}

\def\t{{\scriptscriptstyle\#}}
\def\otimesal{\mathop{\hbox{\raise 1.5 ex
  \hbox{$\scriptscriptstyle\rm al$}
\kern -0.92 em \hbox{$\otimes$}}}}
\def\oplusal{\mathop{\hbox{\raise 1.5 ex
  \hbox{$\scriptscriptstyle\rm al$}
\kern -0.92 em \hbox{$\oplus$}}}}
\def\Gammal{\hbox{\raise 1.68 ex 
\hbox{$\scriptscriptstyle\rm al$}\kern -0.50 em $\Gamma$}}
\def\Bal{\hbox{\raise 1.68 ex 
\hbox{$\scriptscriptstyle\rm  al$}\kern -0.50 em $B$}}
\def\CARal{{\rm C\hskip 0.25 em \hbox{\raise 1.72 ex 
\hbox{$\scriptscriptstyle\rm al$}\kern -0.57 em A}R}}
\def\t{{\scriptscriptstyle\#}}

\def\cE{{\mathcal E}}
\makeatletter
\newcommand*{\defeq}{\mathrel{\rlap{%
                     \raisebox{0.3ex}{$\m@th\cdot$}}%
                     \raisebox{-0.3ex}{$\m@th\cdot$}}%
                     =}
                     
\makeatother
\makeatletter
\newcommand*{\eqdef}{=\mathrel{\rlap{%
                     \raisebox{0.3ex}{$\m@th\cdot$}}%
                     \raisebox{-0.3ex}{$\m@th\cdot$}}%
                     }
\makeatother
\def\Sol{{\rm Sol}_{\rm sc}}
\newcommand{\tra}[1]{\mskip-6mu\upharpoonright_{#1}\mskip+4mu}
\newcommand{\col}[2]{\begin{pmatrix}#1 \\#2\end{pmatrix}}
\def\dito{\!\cdot\!}
\def\bS{\mathbb{S}}
\def\b2{\frac{\beta}{2}}

\def\dyn{{\rm dyn}}

  \def\bz{\bar{z}}
  \def\bro{\bar{\rho}}
  \def\bZ{\bar{\cZ}}
\def\rk{{\bf k}}\def\rh{{\bf h}}
\def\rg{{\bf g}}
\def\cB{\mathcal{B}}
\def\cN{\mathcal{N}}
\def\Me{M^{\rm eucl}}
\def\r{{\rm r}}
\def\l{{\rm l}}
\def\tf{\tilde{f}}

\def\tosim{\xrightarrow{\sim}}
\def\dual{\!\cdot \!}
\newcommand{\traa}[1]{\mskip-6mu\upharpoonright_{#1}}
\def\ext{{\rm ext}}
\def\cR{\mathcal{R}}
\def\newphi{\tilde{\phi}}
\def\newf{\tilde{f}}
\def\neww{\tilde{w}}
\def\newh{\tilde{h}}
\def\newH{\tilde{H}}
\def\newE{\tilde{E}}
\def\newcE{\tilde{\cE}}
\def\newK{\tilde{K}}
\def\newk{\tilde{k}}
\def\newrho{\tilde{\varrho}}
\def\newQ{\tilde{Q}}
\def\newq{\tilde{q}}
\def\newP{\tilde{P}}
\def\newcD{\tilde{\cD}}
\def\newcY{\tilde{\cY}}
\def\newomega{\tilde{\omega}}
\def\newu{\tilde{u}}
\def\newgamma{\tilde{\gamma}}
\def\newg{\tilde{g}}
\def\newS{\tilde{S}}
\def\newcH{\tilde{\cH}}
\def\newlambda{\tilde{\lambda}}

\def\newc{\tilde{c}}
\def\thE{{\hat \cE}}

\def\tf{\tilde{f}}

\def\bxi{\overline{\xi}}
\def\Calderon{Calder\'{o}n }
\def\rw{{\bf w}}
\def\zero{{\mskip-4mu{\rm\textit{o}}}}

\let\origmaketitle\maketitle
\def\maketitle{
  \begingroup
  \def\uppercasenonmath##1{} 
  \let\MakeUppercase\relax 
	\origmaketitle
  \endgroup
	}
\begin{document}
\pagestyle{plain}

\title{\large The Hartle-Hawking-Israel state on  spacetimes with stationary bifurcate Killing horizons
}
\author{}
\address{Universit\'e Paris-Sud XI, D\'epartement de Math\'ematiques, 91405 Orsay Cedex, France}
\email{christian.gerard@math.u-psud.fr}
\date{April 2021}
\author{\normalsize Christian \textsc{G\'erard} }
\thanks{\emph{Acknowledgments.} We would like to thank  Micha{\l} Wrochna and Ko Sanders for useful discussions.}
\begin{abstract}
We consider  a free massive quantized Klein-Gordon field in a spacetime $(M, \rg)$ containing a  stationary bifurcate Killing horizon, i.e. a bifurcate Killing horizon whose Killing vector field is globally time-like in the right wedge $\mathcal{M}^{+}$ associated to the horizon. 

We prove the existence of the {\em Hartle-Hawking-Israel} vacuum state, which is a pure state on the whole spacetime whose restriction to $\cM^{+}$ is a thermal state $\omega_{T_{\rm H}}$ for the time-like Killing field at Hawking temperature $T_{\rm H}=\kappa(2\pi)^{-1}$, where $\kappa$ is the surface gravity of the horizon.

We show that the HHI state is a Hadamard state and is the unique Hadamard state which is equal to  the double $T_{\rm H}^{-1}$-KMS state in the double wedge $\mathcal{M}^{-}\cup \mathcal{M}^{+}$.  We construct the HHI state by Wick rotation in Killing time coordinates, using  the notion of  the \Calderon projector for elliptic boundary value problems.
\end{abstract}

\maketitle

\section{Introduction}\label{sec0}\init

In this paper we consider  a free quantized Klein-Gordon field in a spacetime $(M, \rg)$ containing a stationary  bifurcate Killing horizon.  The problem we are interested in originates from  the works by Hartle and Hawking \cite{HH} and Israel \cite{I}. 

Hartle and Hawking considered  a free Klein-Gordon field in the  Schwarzschild spacetime, and used formal path integral arguments to construct a distinguished Feynman propagator,  which essentially amounts to constructing a distinguished state. This Feynman propagator is obtained by    Wick rotation arguments,  from a corresponding inverse for an elliptic operator. They showed that the state obtained from this formal procedure is a thermal state in the exterior region (or right wedge) at Hawking temperature $T_{\rm H}= \kappa(2\pi)^{-1}$, where $\kappa$ is the surface gravity of the horizon. Israel discussed the extension of this state to the left wedge of the Kruskal extension of the Schwarzschild spacetime.
We refer the reader to \cite[Chap. 5]{W} for a more detailed discussion.

The first rigorous construction of this so-called Hartle-Hawking-Israel  (HHI) state in the double wedge region of the Kruskal spacetime is due to Kay \cite{K3}. This construction was valid for any temperature, the resulting state being an example of a {\em double KMS state}.

 Later on Kay and Wald \cite{KW} and Kay \cite{K4} made a systematic study of  states for linear scalar quantum fields in a globally hyperbolic spacetime with a bifurcate Killing horizon. They gave a rigorous definition of  the {\em Hadamard condition} and emphasized its importance in this context.
 
They constructed subalgebras $\mathcal{A}$, resp.  $\mathcal{A}^{L/R}$ of the free field algebra  which are in some sense attached to the horizon resp. to its left/right portions, and showed that  all states which are both invariant under  the Killing isometries and Hadamard near the  horizon have the same   restriction to these subalgebras. This restriction must moreover be a KMS state (with respect to the Killing vector field generating the horizon) at the Hawking temperature on the right algebra $\mathcal{A}^{R}$.

They also showed  non-existence results in suitable globally hyperbolic regions of the maximally extended  Schwarzschild-de Sitter  and Kerr spacetimes, (which both contain bifurcate Killing horizons), by proving that  there does not exist {\em any} invariant state (not necessarily KMS) which is of Hadamard form near the horizon. (In the case of the Kerr spacetime the result needs a physically reasonable, but  still unproved hypothesis of existence of some superradiant solutions of the wave equation.)

 The first global construction of the HHI state  is due to Sanders \cite{S1}, who considered spacetimes with a {\em static} bifurcate Killing horizon, i.e.  such that the Killing vector field $V$ is {\em static} in the exterior region. Sanders proved in \cite{S1} the existence of the HHI state and showed that it is a pure Hadamard state. The proof in \cite{S1} relied on  the  {\em Wick rotation} in the Killing time coordinates, which was also the basis for the heuristic arguments in \cite{HH} and which we will also use in this paper.
 
 In \cite{G} we gave another proof of the Hadamard property of the HHI state in the situation considered in  \cite{S1}, by  combining the Wick rotation  with a tool which is familiar in elliptic boundary value problems, namely the {\em \Calderon projector}, see \ref{sec0.2.2}, which was also used in \cite{GW2} to construct {\em analytic Hadamard states} on general analytic spacetimes.  
  The use of \Calderon projectors allows to construct the HHI state directly on a Cauchy surface $\Sigma$ and avoids needing to consider its behavior near the Killing horizon.
 
In the present paper we consider the more general {\em stationary case}, and  give a construction of the HHI state for spacetimes with a  stationary bifurcate Killing horizon.  At  the end of Subsect. \ref{sec0.1} we will comment on the relationships between the situations considered  in \cite{KW} or  \cite{S1} and the one considered in  this paper.

 \subsection{Results}\label{sec0.1}
 We now present in greater detail the main result of this paper.
\subsubsection{Bifurcate Killing horizons}\label{bifurco}
Let $(M, \rg)$ be a globally hyperbolic spacetime with a complete Killing vector field $V$.   We use the 'mostly $+$' convention for the metric signature.

$(M, \rg)$ admits a {\em bifurcate Killing horizon}  see \cite{B, KW}, if the 
{\em bifurcation surface} $\mathcal{B}= \{x\in M : V(x)=0\}$ is an orientable submanifold of codimension $2$, and if there exists a smooth space-like Cauchy surface $\Sigma$ containing $\mathcal{B}$.  We will always assume that $\mathcal{B}$ is {\em compact and connected}.

$M$ splits then into  four globally hyperbolic regions, the {\em right/left wedges} ${\mathcal M}^{+}$, ${\mathcal M}^{-}$ and the {\em future/past cones} $\mathcal{F}$, $\mathcal{P}$, each invariant under the flow of  $V$. 

The {\em bifurcate Killing horizon} is  then $\mathcal{H}= \p (\mathcal{F}\cup \mathcal{P})$. 
 An important object related with the bifurcate Killing horizon   is its {\em surface gravity} $\kappa$, which is a scalar, constant over all of $\mathcal{H}$.

One also assumes the existence of a {\em wedge reflection} $R: M\to M$ which is an isometry of $(\cM^{-}\cup U\cup\cM^{+}, \rg)$, where  $U$ is a neighborhood of $\cB$ in $M$, such that  $R\circ R= Id$, $R= Id$ on $\cB$, $R$ reverses the time orientation and $R^{*}V= V$.
In concrete situations, the left wedge $\cM^{-}$ is actually constructed by reflection of the right wedge $\cM^{+}$, so the existence of a wedge reflection does not seem to be such a strong hypothesis.

The bifurcate Killing horizon $\mathcal{H}$ is {\em stationary}  resp. {\em static} if  there exists a Cauchy surface satisfying the above requirements such that, in addition, $V$ is {\em time-like} on  $\Sigma\setminus \mathcal{B}$, resp. orthogonal to $\Sigma\setminus \mathcal{B}$.  For technical reasons, we require $V$ to be {\em uniformly time-like} near infinity on $\Sigma$, see Subsect. \ref{sec10.23}. This condition is imposed only far away from the bifurcation surface $\mathcal{B}$ and will hold for example if $(M, \rg)$ is asymptotically flat near spatial infinity.

We consider on $(M,\rg)$ a free quantum Klein-Gordon field associated to the Klein-Gordon equation
\[
-\Box_{\rg}\phi(x)+ m(x)\phi(x)=0,
\]
where $m\in\cinf(M, \rr)$ is invariant under $V$ and $R$. We assume that $m(x)\geq m_{0}^{2}>0$ i.e.  the Klein-Gordon field is massive. 

\subsubsection{The double $\beta$-KMS state}
Since $({\mathcal M}^{+}, \rg, V)$ is a stationary spacetime, there exists (see \cite{S2}) for any $\beta>0$ a {\em thermal state} $\omega_{\beta}$  at temperature $\beta^{-1}$ with respect to the group of Killing isometries of $({\mathcal M}^{+}, \rg)$ generated by $V$. 

The wedge reflection $R: \cM^{+}\tosim \cM^{-}$ allows to extend $\omega_{\beta}$ to the {\em double} $\beta$-{\em KMS state} $\omega_{\rm D}$ on $\cM^{+}\cup \cM^{-}$. This extension exists for any $\beta>0$ and is a pure state in $\cM^{+}\cup \cM^{-}$.

\subsubsection{Main result}
We prove in this paper the following theorem. 
\begin{theoreme}\label{thm1.1}
Let $(M, \rg, V)$ be a globally hyperbolic spacetime with a stationary bifurcate Killing horizon and a wedge reflection. Let $P= -\Box_{\rg}+V$ be a Klein-Gordon operator invariant under the Killing vector field $V$ and the wedge reflection $R$. Assume moreover that  the  Cauchy surface $\Sigma$ containing $\cB$ can be chosen so that conditions {\rm (H1)---(H4)}  in Subsects. \ref{sec10.1}, \ref{sec10.22}, \ref{sec10.23} are satisfied.

Then there exists a state $\omega_{\rm  HHI}$ for $P$ in  $(M, \rg)$ called the {\em Hartle-Hawking-Israel state} such that:
\ben
\item $\omega_{\rm HHI}$ is a pure Hadamard state in $M$,
\item  the restriction of $\omega_{\rm HHI}$ to $\cM^{+}\cup \cM^{-}$ is the double $\beta$-KMS state $\omega_{\rm D}$ at Hawking temperature $T_{\rm H}= \kappa(2\pi)^{-1}$ where $\kappa$ is the surface gravity of the horizon,
\item $\omega_{\rm HHI}$ is the unique extension of $\omega_{\rm D}$ such that its spacetime covariances $\Lambda^{\pm}$ map $\coinf(M)$ into $\cinf(M)$ continuously. In particular it is the unique Hadamard extension of $\omega_{\rm D}$.
\een 
\end{theoreme}
Thm. \ref{thm1.1} will be proved in Sects. \ref{sec14} and \ref{secoto}.

\subsubsection{Some comments}
Let  us now comment on the hypotheses {\rm (H1)---(H4)}  that we impose in this paper.  Conditions {\rm (H1i)} and {\rm H2i)} are the standard conditions needed to setup the problem, namely existence of a bifurcate Killing horizon with a wedge reflection, and invariance of the Klein-Gordon operator under the Killing field and wedge reflection. 

Condition {\rm (H1ii)} requires that the bifurcation surface $\cB$ is compact and connected. 

The fact that $\cB$  is compact allows to introduce Gaussian normal coordinates to $\cB$ in the Cauchy surface $\Sigma$ and to construct the smooth extension $(M^{\rm eucl}_{\rm ext}, \rg^{\rm eucl}_{\rm ext})$  of the Wick rotated manifold $(\bS_{\beta}\times \Sigma^{+}, \rg^{\rm eucl})$, (for $\beta$ equal to the inverse Hawking temperature), where $\rg^{\rm eucl}$ is the complex metric obtained from $\rg$ by Wick rotation in the Killing time coordinate.

This assumption excludes of course the Minkowski spacetime, when  the Killing vector field $X$ is a boost generator, for which the above results are very well-known. 

The fact that $\cB$ is connected implies that the surface gravity is  constant on $\cB$.   This excludes for example the extended Schwarzschild-de Sitter spacetime,  see \cite{KW},  where the bifurcation surface has two connected components. It is quite likely that  the results of this paper can be extended if $\mathcal{B}$ is compact with several connected components, with the same surface gravity. 

Finally let us recall that  for  the extended Kerr spacetime, 
 the Killing vector field  becomes space-like at space-like infinity.  The horizon is not stationary and hence the Kerr spacetime is also outside the scope of this paper. Note that it has been shown in \cite[Sect. 6.2]{KW}  that  if the Killing field is not everywhere time-like then  there exists  no associated KMS state in the exterior region, hence  in particular no HHI state.   
 
Condition {\rm( H2ii)} requires that the Klein-Gordon field is massive. Note that in \cite{S1}, the weaker condition $m(x)>0$ was assumed. 
 
Conditions {\rm (H3 H4)} concern the  behavior of  the metric $\rg$ and of the Killing vector field $V$ near infinity on $\Sigma$. {\rm (H4)}, i.e. completeness of $(\Sigma, \rh)$, where $\rh$ is the Riemannian metric induced by $\rg$ on $\Sigma$ is needed to prove the purity of the HHI state.

\subsection{Main ideas of the construction}\label{sec0.2}
We now  outline  the construction of the HHI state $\omega_{\rm HHI}$.  We look for $\omega_{\rm HHI}$ as an extension to $M$ of the double $\beta$-KMS state $\omega_{\rm D}$ on $\cM^{-}\cup\cM^{+}$, where $\beta^{-1}= \kappa(2\pi)^{-1}$ is the Hawking temperature.  The first step consists in understanding in sufficient detail the $\beta$-KMS state in $\cM^{+}$.

Writing the metric $\rg$ in $\cM^{+}$ using the Killing time coordinate associated to $V$ and $\Sigma$,  $\cM^{+}$ is identified with $\rr\times \Sigma^{+}$, where $\Sigma^{+}= \cM^{+}\cap \Sigma$ and the metric $\rg$ becomes
\beq\label{e0.1}
\rg= - N^{2}(y)dt^{2}+ \rh_{ij}(y)(dy^{i}+ \rw^{i}(y)dt)(dy^{j}+ \rw^{j}(y)dt),
\eeq
where $N$ is the lapse function, $\rw$ the shift vector field, $\rh$ the induced metric on $\Sigma$. The Killing field $V$ is simply $\frac{\p }{\p t}$. The fact that $V$ is time-like in $\cM^{+}$ is equivalent to the inequality $N^{2}(y)> \rw^{i}(y)\dual \rh_{ij}(y) \rw^{j}(y)$ for $y\in \Sigma^{+}$.

The Klein-Gordon operator $P$ associated to $\rg$ can be written as:
\beq\label{e0.2}
P= (\p_{t}+w^{*})N^{-2}(\p_{t}-w)+ h_{0},
\eeq
where $w= \rw^{i}\dual \p_{y^{i}}$, (considered as a first order differential operator),  and $h_{0}= \nabla^{*} \rh^{-1}\nabla + m$ is an elliptic operator on $\Sigma$. 
\subsubsection{The Wick rotation}\label{sec0.2.1}
The {\em Wick rotation} consists in replacing $t$ by $\i s$ and produces the {\em complex metric}
\beq\label{e0.3}
\rg^{\rm eucl} = N^{2}(y)ds^{2}+ \rh_{ij}(y)(dy^{i}+ \i \rw^{i}(y)ds)(dy^{j}+ \i \rw^{j}(y)ds).
\eeq
In the static case considered in \cite{S1, G} $\rw$ vanishes and $\rg^{\rm eucl}$ is Riemannian. The fact that $\rg^{\rm eucl}$ is now a complex metric causes several new difficulties. Performing the same transformation on $P$ yields the {\em Wick rotated operator}
\[
K= -(\p_{s}+ \i w^{*})N^{-2}(\p_{s}+ \i w)+ h_{0}.
\]
There are several different linear operators that can be associated to the formal expression $K$. The first one consists in working on $L^{2}(\rr\times\Sigma^{+})$, using the sesquilinear form
\[
Q_{\infty}(u, u)= \| N^{-1}\p_{s}u\|^{2}+ (u | hu)- \i (N^{-1}\p_{s}u|N^{-1} wu)- \i (N^{-1}wu| N^{-1}\p_{s}u), 
\] 
where $h= h_{0}- w^{*}N^{-2}w$, with $\Dom Q_{\infty}=\coinf(\rr\times \Sigma^{+})$.
Another possibility is to work on $L^{2}(\bS_{\beta}\times \Sigma^{+})$ where $\bS_{\beta}= [-\b2, \b2[$ is the circle of length $\beta$. The sesquilinear form $Q_{\beta}$ has the same expression as $Q_{\infty}$ but the domain is now $ \Dom Q_{\beta}=\coinf(\bS_{\beta}\times \Sigma^{+})$, which corresponds to imposing $\beta-$periodic boundary conditions on $K$.

Since we have assumed that $V$ is  {\em uniformly time-like} near infinity, see Subsect. \ref{sec10.23}, one can show that the sesquilinear forms $Q_{\infty}$, $Q_{\beta}$ are closeable and {\em sectorial} and hence generate {\em injective} linear operators $K_{\infty}$, $K_{\beta}$. Their 
inverses $K_{\infty}^{-1}$, $K_{\beta}^{-1}$ are then well defined between abstract Sobolev spaces, using the Lax-Milgram theorem.

\subsubsection{\Calderon projectors}\label{sec0.2.2}
Let $\Omega\subset \rr^{n}$ an open set with smooth {\em compact} boundary and $P = P(x, \p_{x})$ a second order  elliptic operator on $\rr^{n}$. Let us set  $\Omega^{+}=\Omega, \Omega^{-}= \rr^{n}\setminus \Omega^{\rm cl}$. If $u\in \overline{\cD'}(\Omega^{\pm})$ is an extendible distribution in $\Omega^{\pm}$ such that $Pu=0$ in $\Omega^{\pm}$ then its traces $\gamma^{\pm}u= \col{u\traa{\p \Omega}}{\pm \nu\dual d u\traa{\p \Omega}}$, where $\nu$ is a vector field transverse to $\p \Omega$,  are well defined in $\cD'(\p\Omega; \cc^{2})$. 

The spaces $Z^{\pm}= \{\gamma^{\pm}u: u\in  \overline{\cD'}(\Omega^{\pm}), \ Pu=0\hbox{ in }\Omega^{\pm}\}$ are two complementary  spaces in $\cD'(\Omega; \cc^{2})$ and the associated projections $c^{\pm}$ are called {\em \Calderon projectors} for $P$ and $\p \Omega$. Assuming $P$ to be invertible, they can be easily expressed in terms of $P^{-1}$, see eg . Def. \ref{def13.1} below. 

The expressions giving $c^{\pm}$ still make sense if $\p \Omega$ is not compact, for example as operators $c^{\pm}: \cE'(\p\Omega; \cc^{2})\to \cD'(\p \Omega; \cc^{2})$ but they are not projections on $\cE'(\p\Omega; \cc^{2})$ since they do not preserve this space. Nevertheless we will still call $c^{\pm}$ \Calderon projectors.

We will use \Calderon projectors denoted by $c_{\beta}^{\pm}$, $\beta\in ]0, +\infty]$ for the open sets 
$\Omega_{\infty}= ]0, +\infty[\times \Sigma^{+}$, $\Omega_{\beta}= ]0, \b2[\times \Sigma^{+}$ and elliptic operators $K_{\infty}$, $K_{\beta}$,  $\nu$ being the exterior unit normal  for $\rg^{\rm eucl}$ to $\p \Omega_{\beta}$, $\beta\in ]0, +\infty]$. Note that $\nu$ is a {\em complex} vector field, but its imaginary part is tangent to $\p \Omega_{\beta}$.  
The expression of $c_{\beta}^{\pm}$ in terms of the inverse $K_{\beta}^{-1}$ is given in  Subsect. \ref{sec12.6}.

\subsubsection{Vacuum and double $\beta$-KMS states}\label{sec0.2.3}
If $\beta= \infty$, the boundary $\p \Omega_{\infty}$ equals $\Sigma^{+}$, and one can try to construct a state in $\cM^{+}$ by defining its covariances on $\Sigma^{+}$ as 
\[
\lambda_{\infty}^{\pm}= \pm q\circ c_{\infty}^{\pm}, 
\]
where $q= \mat{0}{1}{1}{0}$ is the charge defining the symplectic structure on the space $\coinf(\Sigma^{+}; \cc^{2})$ of Cauchy data on $\Sigma^{+}$. It turns out that $\lambda_{\infty}^{\pm}$ are actually the covariances of the {\em vacuum state} $\omega_{\rm vac}$ in $\cM^{+}$.

Of course the study of   the vacuum state  $\omega_{\rm vac}$, corresponding to $\beta= \infty$,  is not necessary for the construction of the HHI state, but gives a nice introduction to the more complicated case $\beta<\infty$.

If $\beta<\infty$, the boundary $\p \Omega_{\beta}$ has  two components, both isomorphic to $\Sigma^{+}$. The state $\omega_{\rm D}$ obtained similarly from the \Calderon projectors $c_{\beta}^{\pm}$ is now the {\em double} $\beta$-{\em KMS state} $\omega_{\rm D}$ in $\cM^{-}\cup \cM^{+}$, modulo the identification of $\Sigma^{+}$ with $\Sigma^{-}$ by the wedge reflection.

The proof of these  facts takes up a large part of the paper.  First of all we reduce ourselves to the situation $N(y)= 1$ by considering $\tilde{P}= NPN$ and $\tilde{K}_{\beta}= NK_{\beta}N$, the last identity taking a rather transparent form if we use the framework of sesquilinear forms, see 
Subsect. \ref{sec12.5}.
The covariances of $\omega_{\rm vac}$, $\omega_{\rm D}$ for the Klein-Gordon operator $P$ can similarly be deduced from those of the analogous states $\tilde{\omega}_{\rm vac}$, $\tilde{\omega}_{\rm D}$ for $\tilde{P}$. 

The operator $\tilde{P}$ can be written as $(\p_{t}+\tilde{w}^{*})(\p_{t}-\tilde{w})+ \tilde{h}_{0}$, and the computations of $\tilde{\omega}_{\rm vac}$, $\tilde{\omega}_{\rm D}$ can be done by reducing the Klein-Gordon equation $\tilde{P}\tilde{\phi}=0$ to a first order system $\p_{t}f- \i H f=0$, see  Sects. \ref{sec8}, \ref{sec9}. This system  is an example of a {\em stable symplectic dynamics}, which is studied in Sects. \ref{sec1}, \ref{sec3}.

\subsubsection{The surface gravity and the extended Euclidean metric}\label{sec0.2.4}
All the constructions up to now are valid for any value of the inverse temperature $\beta$.   
If $\beta= (2\pi)\kappa^{-1}$, i.e.  if $\beta^{-1}$ equals the {\em Hawking temperature} $\kappa(2\pi)^{-1}$, where $\kappa$ is the {\em surface gravity} of the horizon, one can show that  $(\bS_{\beta}\times \Sigma^{+},\rg^{\rm eucl})$  has a unique extension $(M_{\rm ext}^{\rm eucl},\rg_{\rm ext}^{\rm eucl})$, which corresponds exactly to passing from polar to cartesian coordinates in the plane.

\subsubsection{The Hartle-Hawking-Israel state}\label{sec0.2.5}
The open set $]0, \b2[\times \Sigma^{+}$ extends as  an  open set $\Omega_{\rm ext}$ with boundary isomorphic to the {\em full Cauchy surface} $\Sigma$.  The Wick rotated operator $K_{\beta}$ extends as an elliptic operator $K_{\rm ext}$ acting on $M^{\rm eucl}_{\rm ext}$, and one can consider the \Calderon projectors $c^{\pm}_{\rm ext}$ associated to $K_{\rm ext}$ and $\Omega_{\rm ext}$. 

One defines the covariances on $\Sigma$
\[
\lambda_{\rm HHI}^{\pm}= \pm q\circ c^{\pm}_{\rm ext},
\]
and one can rather easily show that $\lambda_{\rm HHI}^{\pm}$ are the covariances of a quasi-free state $\omega_{\rm HHI}$ defined on the whole of $M$. One uses that the restriction of $\lambda^{\pm}_{\rm HHI}$ to $\coinf(\Sigma\setminus\cB)$ are precisely the covariances of the double $\beta$-KMS state $\omega_{\rm D}$, and some continuity properties of \Calderon projectors and density results in Sobolev spaces, see Subsect. \ref{sec14.2}. 

One can also prove that the HHI state $\omega_{\rm HHI}$ is a Hadamard state, by an argument already used in \cite{G} in the static case, relying on the fact that the covariances of any Hadamard state on $\Sigma$ are matrices of pseudodifferential operators. 

The proof of the purity of $\omega_{\rm HHI}$ is more complicated and relies on arguments from \cite{GW2}.

\subsection{Notations}\label{sec0.10}
We now collect some notation.

  We set $\langle \lambda\rangle= (1+ \lambda^{2})^{\12}$ for $\lambda\in \rr$.
 
  We write $A\Subset B$ if $A$ is relatively compact in $B$.
 
  If $X,Y$ are sets and $f:X\to Y$ we write  $f: X \xrightarrow{\sim}Y$ if $f$ is
bijective.  If $X, Y$ are equiped with topologies, {we write $f:X\to Y$ if the map is continuous, and $f: X \xrightarrow{\sim}Y$ if it is a homeomorphism.}

\subsubsection{Duals and antiduals}\label{sec1.1.1}
Let $\cX$ be a real vector space. Its dual will be denoted by  $\cX^{\t}$.
Let $\cY$ be a complex vector space. We denote by $\cY_{\rr}$ its {\em real form}, i.e.  $\cY$ as a  vector space over $\rr$. 
We denote by $\cY^{\t}$ its dual, i.e.  the space of $\cc-$linear forms on $\cY$  and by $\cY^{*}$ its anti-dual, i.e.  the space of $\cc-$antilinear forms on $\cY$. 

We denote by $\bar{\cY}$ the {\em conjugate vector space} to $\cY$, i.e.  $\bar{\cY}= \cY_{\rr}$ as a $\rr-$vector space, equiped with the complex structure $-\i$, if $\i\in L(\cY_{\rr})$ is the complex structure  of $\cY$. The (anti-linear) identity map $Id: \cY\to \bar{\cY}$ will be denoted by $y\mapsto \bar{y}$, i.e.  $\bar{y}$ equals $y$ but considered as an element of $\bar{\cY}$.

If $\cY$ is a Hilbert space, then $\bar{\cY}$ inherits also a Hilbert space structure by
\[
(\bar{y}_{1}| \bar{y}_{2})_{\bar{\cY}}\defeq \overline{(y_{1}| y_{2})_{\cY}}.
\]
By definition we have $\cY^{*}= \bar{\cY}^{\t}$.
Note that we have a $\cc-$linear identification  $\overline{\cY^{\t}}\sim \bar{\cY}^{\t}$ defined as follows: if $y\in \cY$ and $w\in \cY^{\t}$ then
\[
\bar{w}\dito \bar{y}\defeq \overline{w\dito y}
\]
This identifies $\bar{w}\in \overline{\cY^{\t}}$ with an element of $\bar{\cY}^{\t}$. Similarly we have a $\cc-$linear identification $\bar{\cY}^{*}\sim \overline{\cY^{*}}$.
\subsubsection{Linear operators}\label{sec1.1.2}
If $\cX_{i}$, $i= 1, 2$ are real or complex vector spaces and $a\in L(X_{1}, X_{2})$ we denote by $a^{\t}\in L(\cX_{2}^{\t}, \cX_{1}^{\t})$ its transpose.  If $\cY_{i}$, $i=1,2$ are complex vector spaces  we denote by $a^{*}\in L(\cY_{2}^{*}, \cY_{1}^{*})$ its adjoint, and by $\bar{a}\in L(\bar{\cY}_{1}, \bar{\cY}_{2})$ its {\em conjugate}, defined by $\bar{a}\,\bar{y}_{1}= \overline{ a y_{1}}$. With the above identifications we have $a^{*}= \bar{a}^{\t}= \overline{a^{\t}}$.
\subsubsection{Bilinear and sesquilinear forms}\label{sec1.1.3}
If $\cX$ is a real or complex vector space, a bilinear form on $\cX$ is given by  $a\in L(\cX, \cX^{\t})$, its action on a couple $(x_{1}, x_{2})$ is denoted by $x_{1}\dito a x_{2}$. We denote by $L_{\rm s/a}(\cX, \cX^{\t})$ the symmetric/antisymmetric forms on $\cX$.  $a$ is {\em non-degenerate} if $\Ker\, a= \{0\}$.
An antisymmetric, non-degenerate form $\sigma$ is called a {\em symplectic form} on $\cX$.

Similarly if $\cY$ is a complex vector space, a sesquilinear form on $\cY$ is given by  $a\in L(\cY, \cY^{*})$, its action on  a couple $(y_{1}, y_{2})$ is denoted by  $\bar{y}_{1}\dito a y_{2}$, the last notation being a reminder that $\cY^{*}= \bar{\cY}^{\t}$. We denote by $L_{\rm h/a}(\cY, \cY^{*})$ the Hermitian/anti-Hermitian forms on $\cY$. Non-degenerate forms are defined as in the real case. An anti-Hermitian, non-degenerate  form $\sigma$ is called a (complex) {\em symplectic form} on $\cY$.

If $a\in L(\cY, \cY^{*})$ then $\bar{a}\in L(\bar{\cY}, \bar{\cY^{*}})$ and with the above identifications we have
$y_{1}\dual \bar{a}\bar{y}_{2}= \overline{\bar{y}_{1}\dual a y_{2}}$ for $y_{1}, y_{2}\in \cY$.

\subsubsection{Linear operators   on Hilbert spaces}\label{scaleof}

  The domain of a closed, densely defined operator  $a$  on a Hilbert space $\cH$ will be denoted by $\Dom a$,  equipped with the graph norm,  its spectrum by $\sigma(a)$ and its resolvent set by ${\rm res}(a)$. We will similarly denote by $\Dom Q$ the domain of a sesquilinear form $Q$. If $Q$ is closeable we denote by $Q^{\rm cl}$ its closure.

\subsubsection{Scale of Hilbert spaces associated to a selfadjoint operator}\label{ilatu}
  If $a$ is selfadjoint on $\cH$, we write $a>0$ if $a\geq 0$ and $\Ker\, a= \{0\}$.  If $a>0$ and $s\in \rr$ we denote by $a^{s}\cH$ the completion of $\Dom a^{-s}$ for the norm $\|a^{-s}u\|_{\cH}$. Equipped with the scalar product $(u|v)_{s}= (a^{-s}u| a^{-s}v)_{\cH}$, it  is a Hilbert space. The spaces $a^{s}\cH$ and $a^{-s}\cH$ form a dual pair for the duality pairing $\langle u| v\rangle= (a^{-s}u| a^{s}v)_{\cH}$.

We define similarly the spaces $\langle a\rangle^{s}\cH$ for any selfadjoint operator $a$ on $\cH$. We have $\langle a\rangle^{-s}\cH= \Dom |a|^{s}$ for $s>0$.  We have $\langle a\rangle^{-s}\cH\subset \cH\subset \langle a\rangle^{s}\cH$ for $s\geq 0$ and $\langle a\rangle^{s}\cH= |a|^{s}\cH$ if $0\not\in \sigma(a)$.

The notation $\langle a\rangle^{s}\cH$ or $a^{s}\cH$ is very convenient but somewhat ambiguous because usually $a\cH$ denotes the image 
of $\cH$ under the linear operator $a$. Let us explain how to reconcile these two meanings.

Let $\cH_{\rm c}$ be the space of $u\in \cH$ such that $u= \one_{I}(a)u$, for some compact $I\subset \rr\backslash \{0\}$. We equip $\cH_{\rm c}$ with its natural topology by saying  that $u_{n}\to u$ in $\cH_{\rm c}$ if there exists $I\subset \rr\backslash \{0\}$ compact such that $u_{n}= \one_{I}(a)u_{n}$ for all $n$ and  $u_{n}\to u$ in $\cH$. We denote by 
$\cH_{\rm loc}$ the topological anti-dual of $\cH_{\rm c}$. Then  $|a|^{s}$ and $\langle a\rangle^{s}$ preserve  $
\cH_{\rm c}$ and $\cH_{\rm loc}$, and  $\langle a\rangle^{s}\cH$, resp. $|a|^{s}\cH$ are the images in $\cH_{\rm loc}$ of $\cH$ under $\langle 
a\rangle^{s}$, resp. $|a|^{s}$. It follows that these
spaces are  subspaces (equipped with finer topologies) of $\cH_{\rm loc}$,  in particular they are pairwise compatible and one can consider their intersections inside $\cH_{\rm loc}$. It is easy to verify, using the spectral decomposition of $a$ that for example
\beq\label{zlat}
\cH\cap a^{s}\cH= \Dom a^{-s}, \ s\in\rr
\eeq
if we equip $\cH\cap a^{s}\cH$ with the norm $\| \cdot\|+ \|\cdot\|_{s}$.

\subsubsection{Operator inequalities}
If  $a_{1}, a_{2}$ are selfadjoint on $\cH$ with $a_{1}, a_{2}>0$ we write $a_{1}\lesssim a_{2}$ if $\Dom a_{1}^{\12}\supset\Dom a_{2}^{\12}$ and $a_{1}\leq c a_{2}$  on $\Dom a_{2}^{\12}$ for some $c>0$.  We write $a_{1}\sim a_{2}$ if $a_{1}\lesssim a_{2}$ and $a_{2}\lesssim a_{1}$.

 If $a_{1}\sim a_{2}$ the Kato-Heinz theorem implies  that $a_{2}^{-1}\sim a_{1}^{-1}$ and that $a_{1}^{s}\cH= a_{2}^{s}\cH$  as Banach spaces for $s\in [-\12, \12]$. 
 
 Similarly if $q_{1}, q_{2}$ are two  positive quadratic forms with $q_{i}(u, u)=0\Rightarrow u=0$,  we write $q_{1}\lesssim q_{2}$ if $\Dom q_{1}\supset \Dom q_{2}$ and $q_{1}\leq c q_{2}$  on $\Dom q_{2}$ and  we write $q_{1}\sim q_{2}$ if $q_{1}\lesssim q_{2}$ and $q_{2}\lesssim q_{1}$.

\subsubsection{Differential operators on manifolds}

  If $X$ is a smooth manifold and $a, b$ are differential operators on $X$ the composition $a\circ b$ is denoted by $ab$. If $a$ is a differential operator on $X$ and $u\in \cinf(X)$, then $au$ denotes the composition of $a$ with the operator of multiplication by $u$, while $(au)\in \cinf(X)$  denotes the image of $u$ under $a$.
\subsubsection{Spaces of distributions}\label{arlito}
Let $X$ be a smooth manifold. Fixing a smooth density we identify distributions  and distributional densities on $X$. If  $\Omega\subset X$ is an open set with smooth boundary and  $F(X)\subset \cD'(X)$ is a vector space, we denote by $\bar{F}(\Omega)\subset \cD'(\Omega)$ the space of {\em restrictions} of elements of $F(X)$ to $\Omega$.

We denote by $\delta_{a}\in\cD'(\rr)$ the Dirac distribution at $a\in \rr$.

\section{Spacetimes with a stationary bifurcate Killing horizon}\label{sec10}\init
In this section we recall the definition of spacetimes with  bifurcate Killing horizons, see \cite{B, KW}. We express various natural objects, like the lapse function, shift vector field and induced Riemannian metric in Gaussian coordinates near the bifurcation surface.  

We then  consider the {\em Wick rotated metric} $\rg^{\rm eucl}$, obtained by the Wick rotation $t\to \i s$ in the Killing time $t$, and show that if $s$ belongs to the circle $\bS_{(2\pi)\kappa^{-1}}$ of length $(2\pi)\kappa^{-1}$, for $\kappa$ the surface gravity of the horizon, $\rg^{\rm eucl}$ has a smooth extension up  to the bifurcation surface $\cB$. This  fundamental fact, already known for static horizons, see \cite[Sect. 2.2]{S1} lies at the basis of the construction of the HHI state in later sections.
\subsection{Bifurcate Killing horizons}\label{sec10.1}
\begin{definition}\label{def10.1}
  A {\em spacetime with a bifurcate Killing horizon} is a triple $(M, \rg, V)$ such that :
  \ben
  \item $(M, \rg)$ is a globally hyperbolic spacetime,
  \item $V$ is a smooth, complete Killing vector field on $(M, \rg)$,
  \item $\cB\defeq\{x\in M: V(x)=0\}$ is an orientable submanifold  of codimension $2$, called the {\em bifurcation surface},
  \item there exists a  smooth, space-like Cauchy hypersurface $\Sigma$ with $\cB\subset \Sigma$.
  \een
\end{definition}
If $n$ is  the future directed unit normal vector field to $\Sigma$, one defines the {\em lapse function} $N\in \cinf(\Sigma)$ and {\em shift vector field} $\rw$, which is a smooth vector field tangent to  $\Sigma$, by
\[
V= N n + \rw\hbox{ on }\Sigma,
\]
ie 
\[
N\defeq - V\dual \rg n, \ \rw\defeq V- Nn\hbox{ on }\Sigma.
\]
Let us denote by $y$ the elements of $\Sigma$.
The Cauchy surface $\Sigma$ is then decomposed as
\[
\Sigma= \Sigma^{-}\cup\cB\cup \Sigma^{+}, \ \Sigma^{\pm}\defeq\{y\in \Sigma: \pm N(y)>0\}, 
\]
ie $V$ is future/past directed over $\Sigma^{\pm}$.

 The spacetime $M$ splits as 
\[
M= {\mathcal M}^{+}\cup {\mathcal M}^{-}\cup \overline{\mathcal{F}}\cup\overline{\mathcal{P}},
\]
where  the {\em future cone} $\mathcal{F}\defeq I^{+}(\cB)$, {\em the past cone} $\mathcal{P}\defeq I^{-}(\cB)$, the {\em right/left wedges} ${\mathcal M}^{\pm}\defeq D(\Sigma^{\pm})$, are all globally hyperbolic when equipped  with  $\rg$.

\begin{figure}[H]
\centering\includegraphics[width=0.5\linewidth]{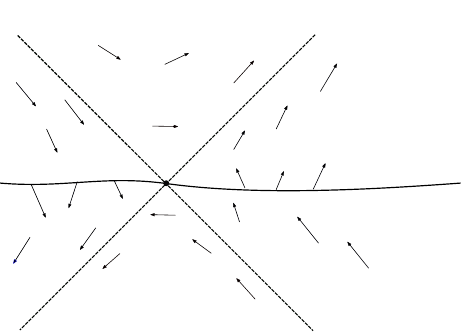}
\put(-10, 50){$\Sigma$ }
\put(-50, 75){${\mathcal M}^{+}$}
\put(-160, 70){${\mathcal M}^{-}$}
\put(-115, 85){$\mathcal{F}$}
\put(-115, 20){$\mathcal{P}$}
\put(-70, 95){$\mathcal{H}$}
\put(-150, 95){$\mathcal{H}$}
\put(-65, 10){$\mathcal{H}$}
\put(-150, 10){$\mathcal{H}$}
\put(-110, 55){$\mathcal{B}$}
\caption{\label{fig1}}
\end{figure}

The future cone $\mathcal{F}$ may be  a black hole.
The {\em bifurcate Killing horizon} is then
\[
\cH\defeq \p \mathcal{F}\cup \p \mathcal{P}.
\]
The Killing vector field  $V$ is tangent to $\cH$.  In Figure \ref{fig1} below  the vector field $V$ is represented by arrows.
The following definition is due to Sanders \cite{S1}.
\begin{definition}\label{def10.2}
  A triple $(M, \rg, V)$ as in Def. \ref{def10.1} is called a  spacetime with a {\em stationary}, resp. {\em static} bifurcate Killing horizon if one can find a Cauchy hypersurface $\Sigma$ as in Def. \ref{def10.1} such that in addition 
$V$ is {\em time-like} on $\Sigma\setminus \cB$, resp. $g${\em-orthogonal} to $\Sigma\setminus \cB$.
\end{definition}
\subsection{Wedge reflection}\label{sec10.2}
Additionally  one assumes  the existence of a {\em wedge reflection}, see \cite[Def. 2.6]{S1}.
\begin{definition}\label{def10.3}
 A {\em wedge reflection} $R$ for a spacetime $(M, \rg,V)$ with a stationary Killing horizon is a diffeomorphism $R:\cM^{-}\cup U\cup\cM^{+}\tosim \cM^{-}\cup U\cup\cM^{+}$, where  $U$ is a neighborhood of $\cB$ in $M$ such that:
 \ben
 \item $R$ is an isometry of $(\cM^{-}\cup U\cup\cM^{+}, \rg)$ which reverses the time orientation, 
 \item $R\circ R= Id$, $R= Id$ on $\cB$,
 \item $R^{*}V= V$.
 \een
\end{definition}
\subsubsection{Weak wedge reflection}\label{sec10.2.1}
It is known, see \cite[Prop. 2.7]{S1} that  if $R$ is a wedge reflection, one can find a Cauchy surface $\Sigma$ as in Def. \ref{def10.1} such that $R: \Sigma\tosim \Sigma$. The map $r\defeq R_{|\Sigma}$ is called a {\em weak wedge reflection}. If the Riemannian metric $\rh$ is the restriction of $\rg$ to $\Sigma$, one has:
\ben
\item $r$ is an isometry of $(\Sigma, \rh)$ with $r\circ r= Id$, 
\item $r= Id$ on $\cB$,
\item $r^{*}N= - N$, $r^{*}\rw= \rw$.
\een
By (3) above we have $r: \Sigma^{\pm}\tosim \Sigma^{\mp}$.
Henceforth, $\Sigma$ will denote a Cauchy surface with all these properties.

Let us  summarize the assumptions we will make on the spacetime $(M, \rg, V)$:

\medskip\noindent

\medskip\noindent

 (H1i) $(M, g , V)$ is a spacetime with a stationary bifurcate Killing horizon, admitting a wedge reflection $R$, 

(H1ii) the bifurcation surface $\cB$ is {\em compact and connected}.
\subsection{Klein-Gordon operators}\label{sec10.22}
We fix a real  function $m\in \cinf(M)$.  We will assume the following:

\medskip
\medskip
\[
\begin{array}{l}
{\rm (H2i)}\ V^{a}\nabla_{a}m(x)=0, \ m\circ R(x)= m(x), \ x\in {\mathcal M}^{+}\cup {\mathcal M}^{-}\cup U,\\[2mm]
{\rm (H2ii)}\ m(x)\geq m_{0}^{2}>0, \ x\in M,
\end{array}
\]ie as in \cite{S1} we assume that $m$ is stationary w.r.t. the Killing vector field $V$ and invariant under the wedge reflection, but we  consider only {\em massive} Klein-Gordon fields, where in \cite{S1} the weaker condition $m(x)>0$ as allowed.

The {\em Klein-Gordon operator} is
\begin{equation}
\label{e10.4bb}
P= - \Box_{\rg}+ m.
\end{equation}
\subsection{Conditions near infinity on $\Sigma$}\label{sec10.23}
It will be necessary, in order to control various energy spaces in Sect. \ref{sec12}, to impose conditions on the Killing vector field $V$ near infinity on $\Sigma$. 
\[
\begin{array}{rl}
&\exists \ U\hbox{ neighborhood of }\cB\hbox{ in }\Sigma\hbox{ such that:}\\[2mm]
({\rm H3i})&V+ \delta \rw\hbox{ is time-like on }\Sigma\setminus U\hbox{ for some }\delta>0, \\[2mm]
({\rm H3ii})& N^{-2}\rw^{i}\dual(\nabla_{i}^{\rh} N), \  N^{-1}\nabla_{i}^{\rh}\rw^{i}\hbox{ are bounded on }\Sigma\setminus U,
\end{array}
\]
where we recall that $\rh$ is the restriction of $\rg$ to $\Sigma$.

 (H3i)  has a clear geometrical meaning since it means  that  $V$ is {\em uniformly time-like} near infinity on $\Sigma$.   
(H3ii)  depends only on the asymptotic behavior of  $\rg$ on $\Sigma$ near infinity, so it is easy to check in practice. 

Finally in order to prove the purity of the HHI state constructed in Sect. \ref{sec14}, we will need the following condition
\[
({\rm H4})\ (\Sigma, \rh) \hbox{ is complete}.
\]
\subsection{The surface gravity}\label{sec10.3}
The surface gravity is defined by:
\[
\kappa^{2}= - \12
(\nabla^{(\rg)b}V^{a}\nabla_{b}^{(\rg)}V_{a})_{| \cB}, \ \kappa>0.
\]
It is a fundamental fact, see \cite[Sect. 2]{KW}, that $\kappa$ is  constant on $\cB$ and actually on the whole horizon $\cH$.

For $\omega\in \cB$ let $n_{\omega}\in T_{\omega}\Sigma$  be the unit normal to $\cB$ for $\rh$ pointing towards $\Sigma^{+}$. We introduce Gaussian normal coordinates to $\cB$ in $(\Sigma, \rh)$ by:
\[
\chi:\begin{array}{l}
]-\delta, \delta[\times \cB\to \Sigma\\[2mm]
(u, \omega)\mapsto \exp^{\rh}_{\omega}(un_{\omega})
\end{array}
\]
which is a smooth diffeomorphism from $]-\delta, \delta[\times \cB$ to a relatively compact neighborhood $U$ of $\cB$ in $\Sigma$.  In the next proposition we express $\rh$, $N$, $w$ and the wedge reflection $r$ in the local coordinates $(u, \omega)$ on $U$. We recall that  the elements of $\Sigma$ are denoted by $y$.
\begin{proposition}\label{prop10.0}
  On $U$ one has:
  \beq\label{e10.1}
r(u, \omega)= (-u, \omega),
\eeq
and
  \beq\label{e10.0}
\begin{array}{l}
\rh_{ij}(y)dy^{i}dy^{j}= du^{2}+ \rk_{\alpha\beta}(u, \omega)d\omega^{\alpha}d\omega^{\beta}, \\[2mm]
\rw^{i}(y)\p_{y^{i}}= \rw^{0}(u, \omega)\p_{u}+ \rw^{\alpha}(u, \omega)\p_{\omega^{\alpha}}, \\[2mm]
N(y)= N(u, \omega),\\[2mm]
m(y)= m(u, \omega),
\end{array}
\eeq
where $\rk_{\alpha\beta}(u, \omega)d\omega^{\alpha}d\omega^{\beta}$ is a smooth, $u-$dependent Riemannian metric on $\cB$ with:
\begin{equation}
\label{e10.4}
\begin{array}{l}
N(u, \omega)= u(\kappa+u^{2}d(u^{2}, \omega)),\\[2mm]
\rw^{0}(u, \omega)= u^{3}b(u^{2}, \omega), \ 
\rw^{\alpha}(u, \omega)= u^{2}c^{\alpha}(u^{2}, \omega),\\[2mm]
\rk_{\alpha\beta}(u, \omega)= {\rm d}_{\alpha\beta}(u^{2}, \omega),\\[2mm]
m(u, \omega)= n(u^{2}, \omega)
\end{array}
\end{equation}
for smooth maps  $ b, d,n,  c^{\alpha}, {\rm d}_{\alpha\beta}$ defined on $[-\epsilon, \epsilon]\times \cB$  for some $\epsilon>0$ with
\[
n(0, \omega)\geq c>0 \hbox{ for some } c>0. 
\]
\end{proposition}
The proof of Prop. \ref{prop10.0} is given in Appendix \ref{secapp1.0}. 
\subsection{The metric in $\cM^{+}$}\label{sec10.3b}
 Let us denote by $\Phi_{t}$ the flow of the Killing vector field $V$. We  identify  $\rr\times \Sigma^{+}$ with $\cM^{+}$
 by
 \[
\chi: \rr\times \Sigma^{+}\ni (t, y)\mapsto \Phi_{t}(y)\in \cM^{+}.
\]
We have $\chi^{*}V= \dfrac{\p}{\p t}$ and
\[
\begin{array}{rl}
\chi^{*}\rg=& - N^{2}(y)dt^{2}+ (dy^{i}+ \rw^{i}(y)dt)\rh_{ij}(y)(dy^{j}+ \rw^{j}(y)dt)\\[2mm]
=& -v^{2}(y)dt^{2}+ \rw_{i}(y)dy^{i}dt+ \rw_{j}(y)dt dy^{j}+ \rh_{ij}(y)dy^{i}dy^{j},
\end{array}
\]
for $v^{2}(y)=(N^{2}(y)- \rw^{i}(y)\rh_{ij}(y)\rw^{j}(y))$. Note that 
the fact that $V$ is time-like in $\cM^{+}$ is equivalent to
\begin{equation}
\label{e1.3}
N^{2}(y)>\rw^{i}(y)\rh_{ij}(y)\rw^{j}(y), y\in \Sigma^{+}.
\end{equation}
The unit normal vector field to the foliation $\Sigma_{t}= \{t\}\times \Sigma$ is
\beq\label{e0.01d}
 n= N^{-1}(\dfrac{\p}{\p t}- w),
\eeq
 Denoting $\chi^{*}\rg$ on $\rr\times \Sigma^{+}$ simply by $\rg$, we have 
 $|\rg|= N^{2}|\rh|$  and 
\begin{equation}
\label{e10.10}
\rg^{-1}=- N^{-2}\p_{t}^{2}+ N^{-2}(\rw^{i}\p_{y^{i}}\p_{t}+ \rw^{j}\p_{t}\p_{y^{j}})+ (\rh^{ij}- N^{-2}\rw^{i}\rw^{j})\p_{y^{i}}\p_{y^{j}}.
\end{equation}

Since  the potential $m$ is invariant under the Killing vector field, we have $m= m(y)$.
 
 \subsection{The Wick rotated metric}\label{sec10.4}
 \subsubsection{Complex metrics}\label{sec10.4.1}
If $X$ is a smooth manifold, we denote by ${\rm T}^{p}_{q}(X)$ the space of smooth, real  $(p,q)$ tensors on $X$ and by $\cc {\rm T}^{p}_{q}(X)$ its complexification. 
An element $\rk= \rk_{ab}(x)dx^{a}dx^{b}$ of $\cc {\rm T}^{0}_{2}(X)$ which is symmetric and non-degenerate will be called a {\em complex metric} on $X$.  
\subsubsection{The Wick rotated metric}
We denote by  $\bS_{\beta}= [-\b2, \b2[$ with endpoints identified   the circle of length $\beta$  and
 \[
\Me\defeq \bS_{\beta}\times \Sigma^{+},
\]
with variables $(s, y)$. Replacing   $t$  by $\i s$
we obtain the complex metric on $\Me$:
\beq\label{e0.6}
\begin{array}{rl}
\rg^{\rm eucl}=& N^{2}(y)ds^{2}+ (dy^{j}+ \i \rw^{j}(y)ds)\rh_{jk}(y)(dy^{k}+ \i \rw^{k}(y)ds)\\[2mm]
=&v^{2}(y)ds^{2}+  \i \rw_{j}(y)dy^{j}ds+ \i \rw_{j}(y)ds dy^{j}+ \rh_{jk}(y)dy^{j}dy^{k}.
\end{array}
\eeq
We embed $\Sigma\setminus \cB$ into $\Me= \bS_{\beta}\times \Sigma^{+}$  by the map
 \[
\hat{\psi}:y\mapsto \begin{array}{l}
 (0, y)\hbox{ for }y\in \Sigma^{+},\\
 (\b2, r(y))\hbox{ for }y\in \Sigma^{-},
\end{array}
\]
where $r: \Sigma\to \Sigma$ is the weak wedge reflection.

\subsection{The smooth extension}\label{sec10.5}
\begin{proposition}\label{prop10.1}
 Assume that $\beta= (2\pi)\kappa^{-1}$. Then there exists a smooth manifold $M^{\rm eucl}_{\rm ext}$ equipped with a smooth complex metric $\rg^{\rm eucl}_{\rm ext}$ and
 \ben
 \item a smooth  embedding $\psi:\Sigma\to M^{\rm eucl}_{\rm ext}$,
 \item a smooth isometric embedding $\chi: (\Me, \rg^{\rm eucl})\to (M^{\rm eucl}_{\rm ext}\setminus \cB_{\rm ext}, \rg^{\rm eucl}_{\ext})$, where $\cB_{\rm ext}= \psi(\cB)$,
 \item an open set $\Omega_{\rm ext}$ such that $\p \Omega_{\rm ext}= \psi(\Sigma)$ and  
 $\chi: ]0, \b2[\times \Sigma^{+}\tosim \Omega_{\rm ext}\setminus \cB_{\rm ext}$,
 \item a smooth function $m_{\rm ext}: M^{\rm eucl}_{\rm ext}\to \rr$ with $m_{\rm ext}\geq m_{0}^{2}>0$,
 \een
 such that:
 \[
\psi\traa{\Sigma\setminus\cB}= \chi\circ \hat{\psi}, \ 
\chi^{*}m_{\rm ext}= m\traa{\Me}.
\]
\end{proposition}
\begin{figure}[H]
\includegraphics[width=0.7\linewidth]{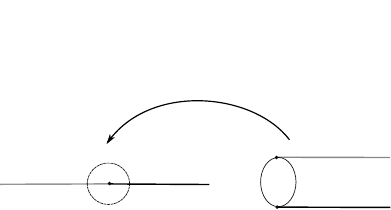}
\put(-135, 17){$\Sigma^{+}$}
\put(-220, 17){$\Sigma^{-}$}
\put(-180, 17){$\mathcal{B}$}
\put(-180, 35){$\Omega_{\rm ext}$}
\put(-100, 17){$\mathbb{S}_{\beta}$}
\put(-35, 34){$\Sigma^{+}\!\!\sim \!\!r(\Sigma^{-})$}
\put(-35, 7){$\Sigma^{+}$}
\put(-67,-15){$\mathbb{S}_{\beta}\times \Sigma^{+}$}
\put(-74, -8){$0$}
\put(-74, 21){$\b2$}
\put(-150, -15){$\rr^{2}\times \cB$}
\put(-120, 50){$\chi$}
\caption{The embedding $\chi$}
\end{figure}
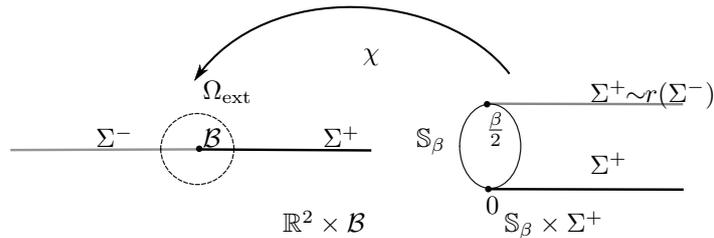
The proof of Prop. \ref{prop10.1} is given in Appendix \ref{secapp1.2}.

\section{Free Klein-Gordon fields}\label{secf}\init
In this section we briefly recall some well-known background material on free quantum Klein-Gordon fields on globally hyperbolic spacetimes.  We follow the presentation in \cite[Sect. 2]{GW1} based on {\em charged fields}. There is no loss of generality to restrict oneself to charged fields and gauge invariant states, see eg the discussion in \cite[Sect. 2]{GW1}. 

\subsection{{\rm CCR}$*$-algebras}\label{secf.1}
\subsubsection{Polynomial {\rm CCR} $*$-algebra}
Let $\cY$ a complex vector space and $q\in L_{\rm h}(\cY,\cY^{*})$ a Hermitian form on $\cY$,  possibly degenerate.  Setting $\cX= \cY_{\rr}, \sigma= {\rm Im}q$, $(\cX, \sigma)$ is a pre-symplectic space.

The {\em CCR} $*$-{\em algebra} $\CCR(\cY, q)$  is the complex $*$-algebra generated by symbols $\one, \psi(y), \psi^{*}(y)$, $y\in \cY$ and the relations
\[
\begin{array}{l}
\psi(y_{1}+ \lambda y_{2})= \psi(y_{1})+ \overline{\lambda}\psi(y_{2}),\\[2mm]
\psi^{*}(y_{1}+ \lambda y_{2})= \psi(y_{1})+ \lambda\psi^{*}(y_{2}), \quad y_{1}, y_{2}\in \cY, \lambda \in \cc, \\[2mm]
[\psi(y_{1}), \psi(y_{2})]= [\psi^{*}(y_{1}), \psi^{*}(y_{2})]=0, \\[2mm]
 [\psi(y_{1}), \psi^{*}(y_{2})]= \overline{y}_{1}\cdot q y_{2}\one, \quad y_{1}, y_{2}\in \cY,\\[2mm]
 \psi(y)^{*}= \psi^{*}(y), \quad y\in \cY.
\end{array}
\]

\subsubsection{Weyl $*$-algebra}
 The {\em Weyl} $*$-algebra ${\rm Weyl}(\cY, q)$ is  the $*$-algebra 
generated by the elements $\one, W(x)$, $x\in \cY_{\rr}$, with relations
\[
\begin{array}{rl}
&W(0)= \one, \quad W(x)^{*}= W(-x), \\[2mm]
&W(x_{1})W(x_{2})= \e^{-\frac{\i}{2} x_{1}{\cdot}\sigma
x_{2}}W(x_{1}+ x_{2}), 
\end{array}\quad x, x_{1}, x_{2}\in \cY_{\rr},
\]
where we recall that $\sigma= {\rm Im}q$.  

\subsubsection{Weyl $C^{*}$-algebra}
Setting
\[ 
 \| A\|= \sup_{\omega\in \cF}\omega(A^{*}A)^{\12}, \ A\in {\rm Weyl}(\cY, q),
\]
  where $\cF$ is the set of states on ${\rm Weyl}(\cY, q)$ one can show that $\|\cdot\|$ is a $C^{*}$-norm, see eg \cite{MSTV} and we denote by 
   ${\rm Weyl}^{C^{*}}(\cY, q)$ the completion of ${\rm Weyl}(\cY, q)$ for this norm.
\subsubsection{Quasi-free states}
 \begin{definition}
  A state $\omega$ on $\CCR(\cY, q)$ is {\em (gauge invariant) quasi-free} if 
 \[
\omega(\prod_{i=1}^{p}\psi(y_{i})\prod_{j=1}^{q}\psi^{*}(y'_{j}))=\left\{  \begin{array}{l}
0\hbox{ if }p\neq q,\\
\sum_{\sigma\in S_{p}}\prod_{i=1}^{p}\omega(\psi(y_{i})\psi^{*}(y'_{\sigma(i)}))\hbox{ if }p=q.
\end{array} \right.
\]
\end{definition}
The  {\em (complex) covariances} $\lambda^{\pm}\in L_{\rm h}(\cY, \cY^{*})$ of $\omega$ are defined as:
\[
\begin{array}{l}
\omega(\psi(y_{1})\psi^{*}(y_{2}))\eqdef \bar{y}_{1}\dito \lambda^{+} y_{2}, \\[2mm]
 \omega(\psi^{*}(y_{2})\psi(y_{1}))\eqdef \bar{y}_{1}\dito \lambda^{-} y_{2},
 \end{array} \ y_{1}, y_{2}\in \cY.
\]
The following result is  well-known, see eg \cite[Sect. 17.1]{DG}.
\begin{proposition}\label{propoto1}
 Two  Hermitian  forms $\lambda^{\pm}\in L_{\rm h}(\cY, \cY^{*})$ are the covariances of a quasi-free state $\omega$  on $\CCR(\cY, q)$ iff
 \begin{equation}
\label{ef.1}
\lambda^{\pm}\geq 0, \ \lambda^{+}- \lambda^{-}=q.
\end{equation}
\end{proposition}
If we set
\beq\label{ef.2}
\eta= \12\Re(\lambda^{+}+ \lambda^{-}), 
\eeq
then  we have
\[
\eta\geq 0, \ |x_{1}\dual x_{2}|\leq 2 (x_{1}\dual \eta x_{1})^{\12} (x_{2}\dual x_{2})^{\12}, \ x_{1}, x_{2}\in\cY_{\rr}.
\]

 It follows that $\omega$ induces a quasi-free state on ${\rm Weyl}(\cY, q)$, or equivalently on ${\rm Weyl}^{C^{*}}(\cY, q)$ defined by
 \[
 \omega(W(x))= \e^{- \12 x\dual \eta x}, \ x\in \cY_{\rr}.
 \]
 \subsubsection{Pure quasi-free states}
 \begin{definition}
 A  quasi-free state $\omega$ on $\CCR(\cY, q)$ is said {\em pure} if the induced state on ${\rm Weyl}^{C^{*}}(\cY, q)$ is pure.
\end{definition}
In the sequel we consider a quasi-free state $\omega$ on $\CCR(\cY, q)$ with covariances $\lambda^{\pm}$. We will assume that
\[
\Ker (\lambda^{+}+ \lambda^{-})= \{0\}.
\]
By Prop. \ref{propoto1}  this is true  if $q$ is non degenerate. 

Let  $\cY^{\rm cl}$ be the completion of $\cY$ for     the norm $\| y\|_{\omega}= (\bar{y}\dual\lambda^{+}y+ \bar{y}\dual\lambda^{-}y)^{\12}$.  Since $\lambda^{\pm}\geq 0$ and $q= \lambda^{+}- \lambda^{-}$, $q, \lambda^{\pm}$ uniquely extend as bounded Hermitian forms $q^{\rm cl}, \lambda^{\pm{\rm cl}}$ on $\cY^{\rm cl}$. Note that $q^{\rm cl}$ may be degenerate, even is $q$ is not. 
\begin{proposition}\label{propoto2}
The state $\omega$  on $\CCR(\cY, q)$ is pure  iff  there exist projections $c^{\pm}\in L(\cY^{\rm cl})$ such that 
\beq\label{e3.21}
c^{+}+ c^{-}= \one, \lambda^{\pm{\rm cl}}= \pm q^{\rm cl}\circ c^{\pm}.
\eeq
\end{proposition}
\begin{remark}\label{jetemmerde}
If $\omega$ is pure on $\CCR(\cY, q)$ then $q^{\rm cl}$ is non degenerate. In fact since $\lambda^{\pm{\rm cl}}$ and $q^{\rm cl}$ are Hermitian we obtain  $\lambda^{\pm{\rm cl}}= \pm c^{\pm*}q^{\rm cl}$ hence $\Ker q^{\rm cl}\subset \Ker (\lambda^{+{\rm cl}}+ \lambda^{-{\rm cl}})= \{0\}$. 
\end{remark}

{\bf Proof of Prop. \ref{propoto2}.}
Assume first that $\cY$ is complete for $\| \cdot\|_{\omega}$ ie $\cY^{\rm cl}= \cY$. This means that $\cY_{\rr}$ is complete for the euclidean scalar product $\eta$ in \eqref{ef.2}. In that situation we know from \cite[Thm. 17.13]{DG} that $\omega$ is pure iff $(2\eta, \sigma)$ is K\"{a}hler, ie if there exists an anti-involution $\ii$ on $\cY_{\rr}$ with $\ii^{\t}\sigma\ii = \sigma$ and $2 \eta= \sigma\ii$. This is equivalent to the existence of projections $c^{\pm}$ satisfying
\begin{equation}
\label{turlutututututu}
c^{+}+ c^{-}=\one, \ c^{+*}q c^{-}= 0, \ \lambda^{\pm}= \pm q \circ c^{\pm},
\end{equation}
as requested, see \cite[Prop. 2.7]{GW1}. 

Let us now consider the general case. We recall from  \cite[Thm. 2.3.19]{BR} that a state $\omega$ on a $C^{*}$-algebra $\mathfrak{A}$ is pure iff its GNS representation $(\cH, \pi)$ is irreducible, i.e. iff $\cH$ does not contain non-trivial closed subspaces invariant under $\pi(\mathfrak{A})$.

We set  
$\mathfrak{A}= {\rm Weyl}^{C^{*}}(\cY, q), \mathfrak{A}^{\rm cl}= {\rm Weyl}^{C^{*}}(\cY^{\rm cl}, q^{\rm cl})$ $\omega^{\rm cl}$ the quasi-free state on $ \mathfrak{A}^{\rm cl}$ with covariances $\lambda^{\pm{\rm cl}}$ and let  $(\cH^{({\rm cl})}, \pi^{({\rm cl})}, \Omega^{({\rm cl})})$ be the GNS triple for $(\mathfrak{A}^{({\rm cl})}, \omega^{({\rm cl})})$.  Using that $\cY$ is dense in $\cY^{\rm cl}$ for $\|\!\cdot\!\|_{\omega}$, we first obtain that $\cH= \cH^{\rm cl}$, $\Omega= \Omega^{\rm cl}$ and $\pi^{\rm cl}|_{\mathfrak{A}}= \pi$. 

We also easily obtain that $\pi(\mathfrak{A})$ is strongly dense in $\pi^{\rm cl}(\mathfrak{A}^{\rm cl})$. In fact, if $A= \sum_{1}^{N}\alpha_{i}\pi^{\rm cl}(W(y_{i}))\in \pi^{\rm cl}(\mathfrak{A}^{\rm cl})$ and $y_{i, n}\in \cY$ with $y_{i, n}\to y_{i}$ for $\|\!\cdot\!\|_{\omega}$, we obtain that $A_{n}= \sum_{1}^{N}\alpha_{i}\pi(W(y_{i,n}))$ is bounded in norm by $\sum_{1}^{N}|\alpha_{i}|$ and converges strongly to $A$ on the dense subspace $\pi(\mathfrak{A})\Omega$, hence on $\cH^{\rm cl}$.

From this fact we see that a closed subspace $\cK\subset \cH=\cH^{\rm cl}$ is invariant under $\pi(\mathfrak{A})$ iff it is invariant under $\pi^{\rm cl}(\mathfrak{A}^{\rm cl})$, hence $\omega$ is pure iff  its extension  $\omega^{\rm cl}$ to $\mathfrak{A}^{\rm cl}$ is pure.   We are hence reduced to the
already treated case when $\cY= \cY^{\rm cl}$. \qed

Let us now state another criterion for purity.

 Prop. \ref{purity} below is due to Kay and Wald \cite{KW} in the real case (see \cite[Prop. 2.1]{GW2} for the complex case).
\begin{proposition}\label{purity}
 The state $\omega$ with covariances $\Lambda^{\pm}$ is {\em pure} iff:
 \beq\label{arbo}
\bar{y}_{1}\dual (\lambda^{+}+ \lambda^{-})y_{1}= \sup_{y_{2}\in \cY, y_{2}\neq 0}\frac{|\bar{y}_{1}\dual q y_{2}|^{2}}{\bar{y}_{2}\dual (\lambda^{+}+ \lambda^{-})y_{2}}, \ \ \forall y_{1}\in \cY.
\eeq
\end{proposition}
\proof From \cite{KW} we know that $\omega$ is a pure state iff
\[
x_{1}\dual \eta x_{1}= \frac{1}{4}\sup_{x_{2}\in \cY_{\rr}, x_{2}\neq 0}\frac{|x_{1}\dual\Im q x_{2}|^{2}}{x_{2}\dual \eta x_{2}}.
\]
Using that $\eta=  \12 \Re(\lambda^{+}+ \lambda^{-})$ and that $q$ is sesquilinear, this is equivalent to \eqref{arbo}. \qed

\subsection{Free Klein-Gordon fields}\label{secf2}
Let $P= -\Box_{\rg}+ m$, $m\in \cinf(M, \rr)$  be a Klein-Gordon operator on a globally hyperbolic spacetime $(M, \rg)$.  Let  $G_{\rm ret/adv}$  be the retarded/advanced inverses of $P$ and  $G\defeq G_{\rm ret}- G_{\rm adv}$.   We apply the above framework to
\[
\cY= \frac{\coinf(M)}{P\coinf(M)}, \ \overline{[u]}\dito q[v]= \i (u| Gv)_{M},
\]
where $(u|v)_{M}= \int_{M}\bar{u}v dVol_{\rg}$.
Denoting by $\Sol(P)$ the space of smooth space-compact solutions of $P\phi=0$, it is well known that 
\[
[G]:\left(\frac{\coinf(M)}{P\coinf(M)}, \i(\cdot| G \cdot)_{M}\right)\ni [u]\mapsto Gu\in   (\Sol(P), q)
\]
is unitary for 
\beq\label{defdeq}
\overline{\phi}_{1}\dito q \phi_{2}\defeq\i\int_{\Sigma}(\nabla_{\mu}\bar{\phi}_{1}\phi_{2}- \bar{\phi}_{1}\nabla_{\mu}\phi_{2})n^{\mu}d\sigma_{\Sigma},
\eeq
where $\Sigma$ is any spacelike Cauchy hypersurface, $n^{\mu}$ is the future directed unit normal vector field to $\Sigma$ and $d\sigma_{\Sigma}$ the induced surface density.  Setting
\[
\varrho: \cinf_{\rm sc}(M)\ni \phi\mapsto \col{\phi\tra\Sigma}{\i^{-1}n^{\mu}\p_{\mu}\phi\tra\Sigma}= f\in \coinf(\Sigma; \cc^{2})
\]
 the map
\[
\left(\frac{\coinf(M)}{P\coinf(M)}, \i(\cdot| G \cdot)_{M}\right)\ni [u]\mapsto \varrho Gu\in (\coinf(\Sigma; \cc^{2}), q)
\]
is unitary  for
\begin{equation}
\label{ef.3b}
\bar{f}\dito q f\defeq\int_{\Sigma} (\bar{f}_{1} f_{0}+ \bar{f}_{0}f_{1})d \sigma_{\Sigma}, \ f= \col{f_{0}}{f_{1}}.
\end{equation}
In the sequel the $*$-algebra $\CCR(\cY, q)$ where $(\cY, q)$ is any of the above equivalent non degenerate Hermitian spaces will be denoted by $\CCR(P)$.
\subsection{Quasi-free states}\label{quasimodo}
One restricts attention to quasi-free states on $\CCR(P)$ whose covariances are given by distributions on $M\times M$, i.e.  such that there exists $\Lambda^{\pm}\in \cD'(M\times M)$ with
\beq\label{ef.2b}
\begin{array}{l}
\omega(\psi([u_{1}])\psi^{*}([u_{2}]))= (u_{1}| \Lambda^{+}u_{2})_{M}, \\[2mm]
 \omega(\psi^{*}([u_{2}])\psi([u_{1}]))= (u_{1}| \Lambda^{-}u_{2})_{M},
\end{array}
 \ u_{1}, u_{2}\in \coinf(M).
\eeq
In the sequel the distributions $\Lambda^{\pm}\in \cD'(M\times M)$ will be called the {\em spacetime covariances} of  the state $\omega$.

In \eqref{ef.2b} we identify distributions on $M$ with distributional densities using the density $dVol_{\rg}$ and use the notation $(u|\varphi)_{M}$, $u\in \coinf(M)$, $\varphi\in \cD'(M)$ for the duality bracket. We have then
\beq\label{tralala}
\begin{array}{l}
P(x, \p_{x})\Lambda^{\pm}(x, x')= P(x', \p_{x'})\Lambda^{\pm}(x, x')=0,\\[2mm]
 \Lambda^{+}(x, x')- \Lambda^{-}(x, x')= \i G(x, x').
\end{array}
\eeq
\subsection{Cauchy surface covariances}\label{csc}
Using $(\coinf(\Sigma; \cc^{2}), q)$ instead of $(\frac{\coinf(M)}{P\coinf(M)}, \i(\cdot| G \cdot)_{M})$ one can associate  to a quasi-free state its {\em Cauchy surface covariances} $\lambda^{\pm}$ defined by:
 \begin{equation}
\label{ef.4}
\lambda^{\pm}\defeq (\varrho_{\Sigma}^{*}q)^{*} \Lambda^{\pm} (\varrho_{\Sigma}^{*}q).
\end{equation}
 Using the canonical scalar product  $(f|f)_{\Sigma}\defeq\int_{\Sigma} (\bar{f}_{1}f_{1}+ \bar{f}_{0}f_{0})d\sigma_{\Sigma}$ we identify $\lambda^{\pm}$ with operators, still denoted by $\lambda^{\pm}: \coinf(\Sigma; \cc^{2})\to \cD'(\Sigma; \cc^{2})$.  Conversely we obtain $\Lambda^{\pm}$ from $\lambda^{\pm}$ by:
 \[
 \Lambda^{\pm}= (\varrho G)^{*} \lambda^{\pm}( \varrho G).
 \]

 \subsection{Hadamard states}
  A quasi-free  state is called a {\em Hadamard state}, (see \cite{R} for the neutral case and \cite{GW1} for the complex case) if 
 \def\WF{{\rm WF}}\def\cN{\mathcal{N}}
 \begin{equation}
\label{ef.3}
\WF(\Lambda^{\pm})'\subset \cN^{\pm}\times \cN^{\pm},
\end{equation}
where $\WF(\Lambda)'$ denotes the 'primed' wavefront set of $\Lambda$, i.e.  $S'\defeq\{((x, \xi), (x', -\xi')): ((x, \xi), (x', \xi'))\in S\}$ for $S\subset T^{*}M\times T^{*}M$, and $\cN^{\pm}$ are the two connected components (positive/negative energy shell) of  the characteristic manifold:
\beq\label{defdechar}
\cN\defeq \{(x, \xi)\in T^{*}M\setminus\zero:  \xi_{\mu}g^{\mu\nu}(x)\xi_{\nu}=0\}.
\eeq
We recall that  $T^{*}X\setminus\zero$ denotes the cotangent bundle of $X$ with the zero section removed.

Large classes of Hadamard states were constructed in terms of their Cauchy surface covariances in \cite{GW1, GOW} using pseudodifferential calculus on $\Sigma$, see below for a short summary.
\subsection{Pseudodifferential operators}\label{sec3.10}
We briefly recall the notion of (classical) pseudodifferential operators on a manifold, referring to \cite[Sect. 4.3]{Sh} for details.

For $m\in \rr$ we denote by $\Psi^{m}(\rr^{d})$ the space of classical pseudodifferential operators  on $\rr^{d}$, associated with poly-homogeneous symbols of order $m$, see eg \cite[Sect. 3.7]{Sh}. 

Let $X$ be a smooth, $d-$di\-mensional   manifold.  Let $U\subset X$  be a precompact   chart open set and $\psi: U\to \tilde{U}$ a chart diffeomorphism,  where $\tilde{U}\subset \rr^{d}$ is precompact, open.  We denote by $\psi^{*}: \coinf(\tilde{U})\to \coinf(U)$ the map $\psi^{*} u(x)\defeq u\circ \psi(x)$.\begin{definition}
 A linear continuous map $A: \coinf(X)\to \cinf(X)$ belongs to $\Psi^{m}(X)$ if  the following condition holds:
 
(C) Let    $U\subset X$    be precompact open, $\psi: U\to \tilde{U}$ a chart diffeomorphism, $\chi_{1}, \chi_{2}\in \coinf(U)$ and $\tilde{\chi}_{i}= \chi_{i}\circ \psi^{-1}$. Then there exists $\tilde{A}\in \Psi^{m}(\rr^{d})$ such that
 \beq\label{eapp.-4}
(\psi^{*})^{-1} \chi_{1}A \chi_{2}\psi^{*}= \tilde{\chi}_{1}\tilde{A}\tilde{\chi}_{2}.
\eeq
 Elements of $\Psi^{m}(X)$ are called {\em (classical) pseudodifferential operators} of order $m$ on $X$. 
 
 The subspace of $\Psi^{m}(X)$ of pseudodifferential operators with {\em properly supported kernels} is denoted by $\Psi^{m}_{\rm c}(X)$.
\end{definition}
Note that  if $\Psi^{\infty}_{(\rm c)}(X)\defeq \bigcup_{m\in \rr}\Psi^{m}_{(\rm c)}(X)$, then $\Psi^{\infty}_{\rm c}(X)$ is an algebra, but $\Psi^{\infty}(X)$ is not, since without the proper support condition, pseudodifferential operators cannot in general be composed.

To $A\in \Psi^{m}(X)$ one can associate its {\em principal symbol} $\sigma_{\rm pr}(A)\in \cinf(T^{*}X\setminus\zero)$, which is homogeneous of degree $m$ in the  fiber variable $\xi$ in $T^{*}X\setminus\zero$. $A$ is called {\em elliptic}  in $\Psi^{m}(X)$ at $(x_{0}, \xi_{0})\in T^{*}X\setminus\zero$ if $\sigma_{\rm pr}(A)(x_{0}, \xi_{0})\neq 0$.

If $A\in \Psi^{m}(X)$ there exists (many) $A_{\rm c}\in \Psi^{m}_{\rm c}(X)$ such that $A-A_{\rm c}$ has a smooth kernel.

\subsection{The Cauchy surface covariances of Hadamard states}
We now state a result which follows directly from a construction of Hadamard states in \cite[Subsect. 8.2]{GW1}.
\begin{theoreme}\label{allhad}
 Let $\omega$ be any Hadamard state for the free Klein-Gordon field on $(M, \rg)$ and $\Sigma$ a smooth space-like Cauchy surface. Then its Cauchy surface covariances $\lambda^{\pm}$ are $2\times 2$ matrices  with entries in $\Psi^{\infty}(\Sigma)$. 
\end{theoreme}
We refer the reader to \cite[Thm. 3.2]{G} for the proof. 

\section{Green operators and \Calderon projectors}\label{sec1}\init
 In this section we collect some formulas expressing the Green operators, i.e.  inverses for abstract operators of the form $\p_{s}+b$, where $s$ belongs either to $\rr$ or to the circle $\bS_{\beta}$. We also compute various \Calderon projectors. The formulas in this section will be used later in Sect. \ref{sec8} to express \Calderon projectors for second order elliptic operators obtained from abstract Klein-Gordon operators by Wick rotation.

\subsection{Green operators and \Calderon projectors}\label{sec1.1b}

Let $b$  be a selfadjoint operator on a Hilbert space $\ch$ with $\Ker\, b = \{0\}$.  We recall that   $\bS_{\beta}= [-\b2, \b2[$  is   the circle of length $\beta$. For $0<\beta\leq\infty$  we set    
\beq\label{e.4.not}
\ch_{\beta}=L^{2}(\bS_{\beta})\otimes \ch, \hbox{ for }\beta<\infty, \  \ch_{\infty}= L^{2}(\rr)\otimes \ch.
\eeq
The operator $\p_{ s}$ is anti-selfadjoint on $\ch_{\beta}$ with its natural domain. Denoting still by $b$ the extension of $b$ to $\ch_{\beta}$  we see that  $B_{\beta}= \p_{ s}+ b$ with domain $\Dom \p_{ s}\cap \Dom b$ is normal.  

To better understand the operator $B_{\beta}$ and its inverse $B_{\beta}^{-1}$ one can of course use the spectral decomposition of $b$, or the Fourier transform in $s$ (on $\rr$ or $\bS_{\beta}$), or both together.

%
\subsubsection{Green operators}\label{greenops}
If $0\in \sigma(b)$ then $0\in \sigma(B_{\beta})$ but we can still make sense out of $B_{\beta}^{-1}$ as
\[
B_{\beta}^{-1}: \ch_{\beta}\to (-\p_{s}^{2}+ b^{2})^{-\12}\ch_{\beta},
\]
or
\[
B_{\beta}^{-1}:  (-\p_{s}^{2}+ b^{2})^{\12}\ch_{\beta}\to \ch_{\beta}.
\]
Note that 
\[
\begin{array}{rl}
u= B_{\beta}^{-1}v, \ v\in \ch_{\beta}&\Leftrightarrow u\in |b|^{-1}\ch_{\beta}, \p_{s}u\in \ch_{\beta}, (\p_{s}+ b)u= v\in \ch_{\beta}\\[2mm]
&\Leftrightarrow u\in |b|^{-1}\ch_{\beta},  (\p_{s}+ b)u= v\in \ch_{\beta}.
\end{array}
\]
Let us set:
\beq\label{e8.5}
G_{\infty}( s)\defeq \e^{-  s b}\left(\one_{\rr^{+}}( s)\one_{\rr^{+}}(b)- \one_{\rr^{-}}( s)\one_{\rr^{-}}(b)\right).
\eeq
Note that
\beq\label{e8.5aa}
\int_{\rr}| G_{\infty}(s)| ds= |b|^{-1},
\eeq
where $|G_{\infty}(s)|\in B(\ch)$ is defined by functional calculus. 
A straightforward computation shows  then that:
\beq\label{e8.5a}
B_{\infty}^{-1}f( s)= \int_{\rr} G_{\infty}( s- s')f( s')d s', \ f\in  L^{2}(\rr;\ch)= \ch_{\infty}.
\eeq
The identity \eqref{e8.5aa} and the well-known fact that $L^{1}(\rr)\star L^{2}(\rr)\subset L^{2}(\rr)$, where $\star$ denotes the convolution, show that 
\[
B_{\infty}^{-1}: \ch_{\infty}=L^{2}(\rr; \ch)\to L^{2}(\rr; |b|^{-1}\ch)= |b|^{-1}\ch_{\infty}.
\]
Similarly  for $\beta<\infty$ let us define $G_{\beta}(s)$ as follows: 
we set
\[
G_{\beta}( s)\defeq \e^{- s b}\left(\one_{\rr^{+}}( s)(1- \e^{- \beta b})^{-1}- \one_{\rr^{-}}( s)(1- \e^{ \beta b})^{-1}\right), \  s\in[-\b2,\b2],
\]
(note that $G_{\beta}(\b2)= G_{\beta}(- \b2)$) and extend it to $s\in \rr$ by $\beta-$periodicity. In particular we have:
\beq\label{e8.6}
G_{\beta}( s)= \e^{-  s b}\left(\one_{\rr^{+}}( s)(1- \e^{- \beta b})^{-1}- \one_{\rr^{-}}( s)(1- \e^{ \beta b})^{-1}\right), \  s\in[-\beta, \beta].
\eeq
we have
\beq\label{e8.5ab}
\int_{\rr}|G_{\beta}(s)| ds= |b|^{-1},
\eeq
and
 \beq\label{e8.5c}
B_{\beta}^{-1}f( s)= \int_{\bS_{\beta}}G_{\beta}( s-  s')f( s')d s', \  \ f\in  L^{2}(\bS_{\beta};\ch)= \ch_{\beta}.
\eeq
Again  from \eqref{e8.5ab} we obtain that
\[
B_{\beta}^{-1}: \ch_{\beta}=L^{2}(\bS_{\beta}; \ch)\to L^{2}(\bS_{\beta}; |b|^{-1}\ch)= |b|^{-1}\ch_{\beta}.
\]
The r.h.s. of \eqref{e8.5a}, resp. \eqref{e8.5c} extends to $f\in \cE'(\rr; \ch)$, resp. $f\in \cD'(\bS_{\beta}; \ch)$ and give unique  extensions of  $B_{\infty}^{-1}$, resp. $B_{\beta}^{-1}$ with
\beq\label{dprimecont}
\begin{array}{l}
B_{\infty}^{-1}:  \cE'(\rr; \ch)\to  \cD'(\rr; |b|^{-1}\ch) \hbox{ continuously},\\[2mm]
B_{\beta}^{-1}: \cD'(\bS_{\beta}; \ch)\to \cD'(\bS_{\beta}; |b|^{-1}\ch)\hbox{ continuously}.
\end{array}
\eeq
\subsubsection{\Calderon projectors for $B_{\infty}$}\label{caldeinf}
Let us first motivate the definition of \Calderon projectors in Prop. \ref{prop4.1} below. We set $I_{\infty}^{\pm}=\pm]0, +\infty[$. For  $F\in \overline{C^{0}}(I_{\infty}^{\pm}; \ch)$ we set
\[
\Gamma^{\pm}_{\infty}F= F(0^{\pm})= \lim_{s\to 0^{\pm}}F( s).
\]
Assume that    $(\p_{ s}+b)F=0$ in $\cD'(I_{\infty}^{\pm}; \ch)$.
Denoting by $i_{\infty}^{\pm}F$ the extension of $F$ by $0$ in $\rr\setminus I_{\infty}^{\pm}$ we have
\[
(\p_{ s}+b)i_{\infty}^{\pm}F=\pm\delta_{0}( s)\otimes \Gamma_{\infty}^{\pm}F\hbox{ in }\cD'(\rr; \ch).
\]
This implies  formally that $i_{\infty}^{\pm}F= \pm B_{\infty}^{-1}(\delta_{0}( s)\otimes f)$ for $f= \Gamma_{\infty}^{\pm}F$ and hence:
\[
f= \pm\Gamma^{\pm}_{\infty}\circ B_{\infty}^{-1}(\delta_{0}( s)\otimes f)
\]
if $f= \Gamma_{\infty}^{\pm}F$ for $F$ solving $(\p_{ s}+ b)F=0$ in $I_{\infty}^{\pm}$.   

Note that  $B_{\infty}^{-1}(\delta_{0}(s)\otimes f)$ for $f\in \ch$ is well defined by the discussion at the end of \ref{greenops}.
\begin{proposition}\label{prop4.1}
\ben
\item  $B_{\infty}^{-1}(\delta_{0}(s)\otimes f)$ belongs to $\overline{C^{0}}(I_{\infty}^{\pm}; \ch)$  for $f\in \ch$. It follows that
 \beq\label{e4.1}
C_{\infty}^{\pm}f= \pm\Gamma^{\pm}_{\infty}\circ B_{\infty}^{-1}(\delta_{0}( s)\otimes f), \ f\in \ch
\eeq
are well defined. 
\item One has
\begin{equation}
\label{cppvac}
C^{\pm}_{\infty}= \one_{\rr^{\pm}}(b),
\end{equation}
\een
It follows that  $C_{\infty}^{\pm}\in B(\ch)$ are bounded complementary projections on $\ch$,  called {\em \Calderon projectors}.
\end{proposition}
\proof  Let $\chi_{n}(s)= n\chi(ns)$ for $\chi\in \coinf(\rr)$ with $\int \chi(s)ds=1$. Clearly $\chi_{n}(s)\otimes f\to \delta_{0}(s)\otimes f$ in $\cE'(\rr; \ch)$ and hence
\[
B_{\infty}^{-1}( \delta_{0}(s)\otimes f)= \lim_{n\to \infty} B_{\infty}^{-1}(\chi_{n}(s)\otimes f) \hbox{ in }\cD'(\rr; |b|^{-1}\ch).
\]
For $s>0$, we have
\[
B_{\infty}^{-1}(\chi_{n}\otimes f)(s)= \int\e^{-(s-s')b}\one_{\rr^{+}}(s-s')\one_{\rr^{+}}(b)\chi_{n}(s')\otimes fds'
\]
for $n$ large enough. Letting $n\to \infty$ we obtain that
\[
B_{\infty}^{-1}(\delta_{0}\otimes f)(s)= \e^{- s b}\one_{\rr^{+}}(b)f\hbox{ in }s>0,
\]
hence $B_{\infty}^{-1}(\delta_{0}\otimes f)\in\overline{C^{0}}(I_{\infty}^{+}; \ch)$ and $\Gamma^{+}_{\infty}\circ B_{\infty}^{-1}(\delta_{0}(s)\otimes f)= \one_{\rr^{+}}(b)f$. We use the same argument   for $C_{\infty}^{-}$. \qed

\subsubsection{\Calderon projectors for $B_{\beta}$}\label{caldebeta}
For $\beta<\infty$ we set $I_{\beta}^{\pm}=\pm]0, \b2[$.  The boundary $\p I_{\beta}^{\pm}$ has two connected components $\{s=0\}$ and $\{s= \beta/2\}$, which complicates a little the computation of the \Calderon projectors.  Moreover  the kernel  $G_{\beta}$ of $B_{\beta}^{-1}$ has a infrared singularity at $b=0$, coming from the factors $(1-\e^{\mp \beta b})^{-1}$
 in \eqref{e8.6}.
 
 As before we start with the heuristic motivation for  the definition of \Calderon projectors in Prop. \ref{prop4.2} below.
For   $F\in \overline{C^{0}}(I_{\beta}^{\pm}; \ch)$ we set:
\beq\label{caldecaldo}
\begin{array}{l}
\Gamma_{\beta}^{+}F\defeq F(0^{+})\oplus F(\b2^{-})\eqdef\Gamma_{\beta}^{(0)+}F\oplus\Gamma_{\beta}^{(\b2)+}F,\\[2mm]
\Gamma_{\beta}^{-}F\defeq F(0^{-})\oplus F(-\b2^{+})\eqdef\Gamma_{\beta}^{(0)-}F\oplus \Gamma_{\beta}^{(\b2)-}F.
\end{array}
\eeq
Assume that   $(\p_{ s}+b)F=0$ in $\cD'(I_{\beta}^{\pm}; \ch)$. Then  
denoting by $i_{\beta}^{\pm}F$ the extension of $F$ by $0$ in $\bS_{\beta}\setminus I_{\beta}^{\pm}$, we have
\[
(\p_{ s}+b)i_{\beta}^{\pm}F=\pm(\delta_{0}( s)\otimes \Gamma_{\beta}^{(0)\pm}F- \delta_{\b2}( s)\otimes \Gamma_{\beta}^{(\b2)\pm}F)\hbox{ in }\cD'(\bS_{\beta}; \ch), 
\]
which formally implies that
\[
i_{\beta}^{\pm}F= B_{\beta}^{-1} ( \delta_{0}( s)\otimes f^{(0)}- \delta_{\b2}( s)\otimes f^{(\b2)})
\]
for $f=f^{(0)}\oplus f^{(\b2)}= \Gamma_{\beta}^{\pm}F$. Again   the r.h.s. above is well defined by the discussion at the end of \ref{greenops}.

\begin{proposition}\label{prop4.2}
\ben
\item  $B_{\beta}^{-1} ( \delta_{0}( s)\otimes f^{(0)}- \delta_{\b2}( s)\otimes f^{(\b2)})$ belongs to $\overline{C^{0}}(I_{\beta}^{\pm}; (1+|b|^{-1})\ch)$ for $f= f^{(0)}\oplus f^{(\b2)}\in \ch\oplus \ch$.  It follows that
\beq
 \label{e4.2}
C_{\beta}^{\pm}f\defeq \pm\Gamma^{\pm}_{\beta}\circ B_{\beta}^{-1} ( \delta_{0}( s)\otimes f^{(0)}- \delta_{\b2}( s)\otimes f^{(\b2)}), \ f= f^{(0)}\oplus f^{(\b2)}\in \ch\oplus \ch
\eeq
are well defined and belong to $B(\ch\oplus\ch, (1+ |b|^{-1})\ch\oplus (1+ |b|^{-1})\ch)$.  
\item One has:
\beq\label{cpp}
\begin{array}{l}
C^{+}_{\beta}= \mat{(1- \e^{- \beta b})^{-1}}{(1- \e^{\beta b})^{-1}\e^{\b2 b}}{(1- \e^{-\beta b})^{-1}\e^{- \b2 b}}{(1- \e^{\beta b})^{-1}},\\[2mm]
C_{\beta}^{-}=\mat{(1-\e^{\beta b})^{-1}}{-\e^{\b2 b}(1- \e^{\beta b})^{-1}}
{-\e^{-\b2 b}(1-\e^{- \beta b})^{-1}}{(1-\e^{- \beta b})^{-1}}.
\end{array}
\eeq
\item On $\one_{I}(b)\ch\oplus \one_{I}(b)\ch$ for any interval $I\Subset \rr^{*}$ one has:
\[
C_{\beta}^{\pm}C_{\beta}^{\pm} =C_{\beta}^{\pm}, \ C_{\beta}^{+}+ C_{\beta}^{-}= \one,
\]
\een
It follows that  $C_{\beta}^{\pm}$ are complementary projections on $\one_{I}(b)\ch\oplus \one_{I}(b)\ch$ called {\em \Calderon projectors}.
\end{proposition}
\proof 
The proof of (1) is analogous to the proof of (1) in Prop. \ref{prop4.1}.  The fact that $\ch$ is replaced by $(1+ |b|^{-1})\ch$ comes from the extra 'infrared singularity' of $(1- \e^{\mp\beta b})^{-1}$, since $(1- \e^{- \beta \lambda})^{-1}$ behaves like $\lambda^{-1}$ near $\lambda=0$. 

(2) is a routine computation using \eqref{e8.6}.
We check (3) using  the identity $(1-a)^{-1}+ (1- a^{-1})^{-1}= 1$
 for $a= \e^{- \beta b}$.  \qed

\section{Vacua and KMS states for stable symplectic dynamics}\label{sec3}\init
In this section we recall well-known formulas for the covariances of the vacuum and KMS states associated to a symplectic flow  on a symplectic space. The symplectic flow  has to be {\em stable}, i.e.  generated by a positive classical energy. In concrete situations the symplectic flow  is generated by  a time-like Killing vector field.  We also recall the definition of the {\em double} KMS state,  due to Kay \cite{K1, K2}, which is related to  the Araki-Woods representation of a KMS state.

The new result of this section is that the covariances of the vacuum and double KMS states can be expressed by  the {\em \Calderon projectors} introduced in Sect. \ref{sec1}. Note that only the double KMS states will be important for the construction of the HHI state later on. Nevertheless the case of vacuum state is simpler and we include it for pedagogical reasons.
 \subsection{Weakly stable symplectic dynamics}\label{sec3.1}

 We describe now a framework for symplectic dynamics, which can be found in \cite[Sect. 18.2.1]{DG}, called there a {\em weakly stable symplectic dynamics}.
 
Let $(\cY, q)$  be a non-degenerate Hermitian  space and $E\in L_{\rm h}(\cY, \cY^{*})$ with $E>0$,  the function $\cY\ni y\mapsto \overline{y}\dual E y$  being the  {\em classical energy}. The {\em energy space} $\cY_{\rm en}$ is  the completion of $\cY$ for  the scalar product $(y_{1}| y_{2})_{\rm en}= \overline{y}_{1}\dual E y_{2}$ and  is a complex Hilbert space.

Let   $r_{t}= \e^{\i tb}$ be a strongly continuous unitary group   on $\cY_{\rm en}$ with selfadjoint generator $b$.   We assume that $r_{t}: \cY\to \cY$, $\cY\subset\Dom b$.  From Nelson's invariant domain theorem it follows that $b$ is essentially selfadjoint on $\cY$.
We assume also that
\begin{equation}
\label{ecorrect.1}
\Ker b= \{0\}.
\end{equation}
If one applies this abstract framework to Klein-Gordon equations on stationary spacetimes, the stronger condition $0\not\in \sigma(b)$ correspond to the massive case.

\subsubsection{Dynamical Hilbert space}\label{sec3.1.1}
It is convenient, in connection with the quantization of the symplectic flow $\{r_{t}\}_{t\in \rr}$, to introduce the {\em dynamical Hilbert space}
\[
\cY_{\rm dyn}\defeq |b|^{\12}\cY_{\rm en},
\]
see \cite[Subsect. 18.2.1]{DG}, equipped with the scalar product 
\[
(y_{1}| y_{2})_{\rm dyn}= (|b|^{-\12}y_{1}| |b|^{-\12}y_{2})_{\rm en}.
\]
 The group $\{r_{t}\}_{t\in \rr}$
extends obviously as a unitary group on $\cY_{\rm dyn}$.   If  we denote the generator of $r_{t}$ on $\cY_{\rm en/dyn}$ by $b_{\rm en/dyn}$ then $b_{\rm en}=b$ and $b_{\rm dyn}= |b|^{\12} b_{\rm en}|b|^{-\12}$ since $|b|^{-\12}: \cY_{\rm dyn}\to \cY_{\rm en}$ is unitary. Therefore we will often denote both generators by the same letter $b$, when there is no risk of confusion.

 We will assume that 
\begin{equation}
\label{ecorrect.2}
\cY\subset \cY_{\rm dyn}, \hbox{ i.e. } \cY\subset \Dom |b_{}|^{-\12},
\end{equation}
where we consider $b$ as acting on $\cY_{\rm en}$, 
 and that:
\beq\label{e7.-210}
\overline{y}_{1}\dual q y_{2}= (y_{1}| b^{-1}y_{2})_{\rm en}= (y_{1}| {\rm sgn}(b)y_{2})_{\rm dyn}, \quad y_{1}, y_{2}\in \cY.
\eeq
Note that if $b: \cY\to \cY$ this implies that
\[
\overline{y}_{1}\dual E y_{2}= \overline{y}_{1}\dual qb y_{2}, \quad y_{1}, y_{2}\in \cY,
\]
which means that 
 $\{r_{s}\}_{s\in \rr}$can be seen as the symplectic evolution group associated to the classical energy $\overline{y} \dual E y$ and the symplectic form $\sigma= \i^{-1}q$.
 
 For Klein-Gordon equations on stationary spacetimes, conditions \eqref{ecorrect.2}, \eqref{e7.-210} will  be checked in Lemma \ref{lemma12.1}. 

We can equip $\cY_{\rm dyn}$ with the bounded Hermitian form
\beq\label{ecorrect.3}
\overline{y}_{1}\dual q_{\rm dyn} y_{2}\defeq (y_{1}| {\rm sgn}(b)y_{2})_{\rm dyn},
\eeq
which is non-degenerate by \eqref{ecorrect.1}, and we have $q= q_{\rm dyn}$ on $\cY$. Note that one cannot extend $q$ as a bounded Hermitian form on  $\cY_{\rm en}$, unless $0\not\in \sigma(b)$.

\subsection{Vacuum state}\label{sec3.2}
We now recall the definition of the vacuum state $\omega_{\rm vac}$ associated to  the  dynamics $\{r_{t}\}_{t\in \rr}$.
\begin{definition}\label{def3.01}
 The {\em vacuum state} $\omega_{\rm vac}$  is  the quasi-free state on $\CCR(\cY_{\rm dyn}, q_{\rm dyn})$ defined by the covariances:
 \beq\label{e1.3:3}
\overline{y}_{1}\dual \lambda_{\rm vac}^{\pm}y_{2}= (y_{1}| \one_{\rr^{\pm}}(b)y_{2})_{\rm dyn}.
\eeq
\end{definition}

\begin{remark}\label{remark-correct.1}
 Since $\cY\subset \cY_{\rm dyn}$ and $q_{\rm dyn}=q$ on $\cY$,  we see that $\omega_{\rm vac}$  is also  a quasi-free state on $\CCR(\cY, q)$. By Prop. \ref{propoto2} $\omega_{\rm vac}$ is a pure state on $\CCR(\cY_{\rm dyn}, q_{\rm dyn})$. However this does not imply that it is  a pure state on $\CCR(\cY, q)$.
\end{remark}

From \eqref{ecorrect.3} we obtain that  if $c^{\pm}_{\rm vac}$ are defined as in Prop. \ref{propoto2} by
\[
\lambda^{\pm}_{\rm vac}\eqdef \pm q_{\rm dyn}\circ c^{\pm}_{\rm vac},
\]
one has:
\beq\label{e1.4}
c^{\pm}_{\rm vac}= \one_{\rr^{\pm}}(b).
\eeq
It follows from Prop. \ref{prop4.1} that:
\begin{equation}
\label{e4.5}
c^{\pm}_{\rm vac}= C^{\pm}_{\infty},
\end{equation}
where the \Calderon projectors $C_{\infty}^{\pm}$ are defined in Prop. \ref{prop4.1}, for $\ch= \cY_{\rm dyn}$. 
\subsection{KMS state}\label{sec3.3}
Let us now  define the $\beta-$KMS state $\omega_{\beta}$ associated to the  dynamics $\{r_{t}\}_{t\in \rr}$.  We introduce the {\em thermal Hilbert space}:
\beq\label{zlot}
\cY_{\rm th}\defeq \cY_{\rm dyn}\cap |b|^{\12}\cY_{\rm dyn}= \Dom |b|^{-\12},
\eeq
see \eqref{zlat}.
\begin{definition}\label{def3.02}
The $\beta${\em -KMS state} $\omega_{\beta}$ is  the quasi-free state on $\CCR(\cY_{\rm th}, q_{\rm dyn})$ defined by the covariances:
\begin{equation}
\label{e3.10}\begin{array}{l}
\bar{y}_{1}\dual\lambda_{\beta}^{+}y_{2}= \overline{y}_{1}\dual q_{\rm dyn}(1- \e^{-\beta b})^{-1}y_{2}, \\[2mm]
 \bar{y}_{1}\dual\lambda_{\beta}^{-}y_{2}= \overline{y}_{1}\dual q_{\rm dyn}(\e^{\beta b}-1)^{-1}y_{2}.
\end{array}
\end{equation}
\end{definition}
Again if $0\in \sigma(b)$, then using that $(1-\e^{\lambda})^{-1}$ behaves like $\lambda^{-1}$ near $\lambda= 0$, we see that 
$\lambda^{\pm}_{\beta}$ are defined only on $\cY_{\rm th}$.  
\begin{remark}\label{remark-correct.2}
In order for $\lambda^{\pm}_{\beta}$ to define a state on $\CCR(\cY, q)$ we need the stronger infrared condition
\begin{equation}
\label{ecorrect.4}
\cY\subset \cY_{\rm th}.
\end{equation}
\end{remark}

Again for Klein-Gordon equations on stationary spacetimes \eqref{ecorrect.4} will  be checked in Lemma \ref{lemma12.1}. 
\subsection{Double $\beta-$KMS states}\label{sec4.7}
The {\em double }$\beta-${\em KMS state} see \cite{K1, K2} can easily be related to the {\em Araki-Woods} representation of $\omega_{\beta}$, 
see eg \cite[Subsect. 17.1.5]{DG}, that we first briefly recall.
\subsubsection{Araki-Woods representation}
Let us denote by $\cZ$ the space $\cY_{\dyn}$ as a real vector space,  equipped  with the complex structure 
\[
\ii\defeq \i \circ \sgn(b)
\]
and the scalar product:
\[
(z_{1}| z_{2})_{\cZ}\defeq (y_{1+}| y_{2+})_{\dyn}+ (y_{2-}| y_{1-})_{\dyn},
\]
for $y_{\pm}\defeq \one_{\rr^{\pm}}(b)y$ and $z= y$ (considered as an element of $\cZ$).  $\cZ$  is a Hilbert space equal to $\cY_{\dyn +}\oplus \overline{\cY}_{\dyn -}$ for  $\cY_{\dyn \pm}= \one_{\rr^{\pm}}(b)\cY_{\dyn}$.  Note that since $[b, \ii]=0$, $b$ induces a selfadjoint operator on $\cZ$, still denoted by $b$.   We set
  \begin{equation}
\label{e4.7}
\rho\defeq (\e^{\beta |b|}-1)^{-1}, 
\end{equation}
which is a selfadjoint operator on $\cZ$. We have $\Dom \rho^{\12}= \cZ\cap |b|^{\12}\cZ$, which also equals $\cY_{\rm th}$ as  real vector spaces.

We also introduce  the Hilbert space $\cZ\oplus \bZ$ and the bosonic Fock space 
$\Gamma_{\rm s}(\cZ\oplus \bZ)$, see eg \cite[Subsect. 3.3.1]{DG}. For $(z_{1},\bz_{2})\in \cZ\oplus \bZ$ we denote by $a^{(*)}(z_{1}, \bar{z}_{2})$
 the Fock creation/annihilation operators acting on $\Gamma_{\rm s}(\cZ\oplus \bZ)$.

 The {\em left/right  Araki-Woods} creation/annihilation operators are defined for$z\in \Dom \rho^{\12}$ by:
\beq\label{e4.7b}
\begin{array}{l}
a_{\rm l}^{*}(z)= a^{*}((1+\rho)^{\12}z, 0)+ a(0, \bar{\rho}^{\12}\bar{z}), \\[2mm]
a_{\rm l}(z)= a((1+\rho)^{\12}z, 0)+ a^{*}(0, \bar{\rho}^{\12}\bar{z}), \\[2mm]
a_{\rm r}^{*}(\bar{z})=a(\rho^{\12}z, 0)+ a^{*}(0, (1+\bar{\rho})^{\12}\bar{z}), \\[2mm]
a_{\rm r}(\bar{z})=a^{*}(\rho^{\12}z, 0)+ a(0, (1+\bar{\rho})^{\12}\bar{z}).
\end{array}
\eeq
One has:
\[
[a_{\rm l}(z_{1}), a^{*}_{\rm l}(z_{2})]=(z_{1}|z_{2})_{\cZ}\one,\ 
[a_{\rm r}(\bar{z}_{1}), a_{\rm r}^{*}(\bar{z}_{2})]=(\bar{z}_{1}| \bar{z_{2}})_{\bar{\cZ}}\one, \ z_{1}, z_{2}\in \Dom \rho^{\12},
\]
all other commutators being equal to $0$. 
Setting  $z_{\pm}= y_{\pm}$ for $y\in \cY_{\dyn}$ we set
\beq\label{e4.9}
\begin{array}{l}
\psi_{\rm l}^{*}(y)\defeq a_{\rm l}^{*}(z_{+})+ a_{\rm l}(z_{-}),\ \psi_{\rm l}(y)\defeq a_{\rm l}(z_{+})+ a_{\rm l}^{*}(z_{-})\\[2mm]
\psi^{*}_{\rm r}(y)\defeq a_{\rm r}^{*}(\bar{z}_{-})+ a_{\rm r}(\bz_{+}), \ \psi_{\rm r}(y)\defeq a_{\rm r}(\bar{z}_{-})+ a_{\rm r}^{*}(\bz_{+}),
\end{array}\ y\in \cY_{\rm th}.
\eeq
An easy computation show that
\beq\label{e4.8}
\begin{array}{l}
[\psi_{\rm l}(y_{1}), \psi^{*}_{\rm l}(y_{2})]= \bar{y}_{1}\dual q_{\dyn} y_{2}, \ [\psi_{\rm r}(y_{1}), \psi^{*}_{\rm r}(y_{2})]= - \bar{y}_{1}\dual q_{\dyn} y_{2},
\end{array}
\eeq
all other commutators being equal to $0$. Moreover $\cY_{\rm th}\ni y\mapsto \psi^{*}_{\rm l/r}(y)$ is $\cc$-linear.
This means that $\cY_{\rm th}\ni y\mapsto \psi_{\rm l/r}^{(*)}(y)$  induces  two commuting representations of $\CCR(\cY_{\rm th}, \pm q_{\dyn})$.

From \eqref{e4.9} we obtain that:
\[
\begin{array}{l}
(\Omega| \psi_\l{}(y_{1})\psi_\l{}^{*}(y_{2})\Omega)_{\Gamma_{\rm s}(\cZ\oplus \bZ)}=\bar{y}_{1}\dual \lambda^{+}_{\beta}y_{2},\\[2mm]
 (\Omega| \psi^{*}_\l{}(y_{2})\psi_\l{}(y_{1})\Omega)_{\Gamma_{\rm s}(\cZ\oplus \bZ)}=\bar{y}_{1}\dual \lambda^{-}_{\beta}y_{2},,\\[2mm]
(\Omega| \psi_\l{}(y_{2})\psi_\l{}(y_{1})\Omega)_{\Gamma_{\rm s}(\cZ\oplus \bZ)}=
(\Omega| \psi^{*}_\l{}(y_{2})\psi^{*}_\l{}(y_{1})\Omega)_{\Gamma_{\rm s}(\cZ\oplus \bZ)}=0, 
\end{array}
\]
where $\Omega$ is the vacuum vector in $\Gamma_{\rm s}(\cZ\oplus \bZ)$. If $\pi_{\rm AW, l}$ is the representation of $\CCR(\cY_{\dyn}\cap |b|^{\12}\cY_{\dyn}, q_{\dyn})$
defined by $\pi_{\rm AW, l}(\psi^{(*)}(y))= \psi^{(*)}_\l{}(y)$, then $(\pi_{\rm AW, l}, \Gamma_{\rm s}(\cZ\oplus \bZ), \Omega)$ 
is the GNS representation associated to the $\beta-$KMS state $\omega_{\beta}$.

\subsubsection{ The double $\beta-$KMS state}
To define the double $\beta-$KMS state associated to $\omega_{\beta}$ we  set
\[
(\cX, Q)\defeq (\cY_{\rm th}\oplus \cY_{\rm th}, q_{\rm dyn}\oplus -q_{\rm dyn}).
\]
Recalling that $\sigma= \i^{-1}q_{\rm dyn}$, this corresponds to adding to  the real symplectic space $(\cY_{{\rm th}, \rr}, {\rm Re}\,\sigma)$  its anti-symplectic copy $(\cY_{{\rm th}, \rr}, -{\rm Re}\,\sigma)$.
From \eqref{e4.8} we see that $\cX\ni x\mapsto \Psi_{\rm AW}^{(*)}(x)$  for 
\beq\label{e1.1z}
\Psi_{\rm AW}^{(*)}(x)\defeq \psi^{(*)}_\l{}(y)+ \psi^{(*)}_{\r}(y'), \ x= (y, y')\in \cX
\eeq
induces   a representation of $\CCR(\cX, Q)$.
\begin{definition}\label{def4.1}
 The {\em double} $\beta-${\em KMS state} $\omega_{\rm d}$ is the quasi-free state on $\CCR(\cX, Q)$ defined by
 \[
\omega_{\rm d}(\Psi^{(*)}(x_{1})\Psi^{(*)}(x_{2}))\defeq (\Omega| \Psi^{(*)}_{\rm AW}(x_{1})\Psi^{(*)}_{\rm AW}(x_{2})\Omega)_{\Gamma_{\rm s}(\cZ\oplus \bZ)}, \ x_{1}, x_{2}\in \cX.
\] 
\end{definition}

\begin{proposition}\label{prop4.1:1}
 $\omega_{\rm d}$ is a pure,  gauge invariant quasi-free state on $\CCR(\cX, Q)$. If $\lambda^{\pm}_{\rm d}$  are the covariances of $\omega_{\rm d}$ we have
 \[
\bar{x}_{1}\dual\lambda^{\pm}_{\rm d}x_{2}= \pm \bar{x}_{1}\dual Q C^{\pm}_{\beta}x_{2}, \ x_{1}, x_{2}\in \cX.
\]
where $C^{\pm}_{\beta}$ are the \Calderon projectors for $B_{\beta}$ defined in Prop. \ref{prop4.2}.
 \end{proposition}
 \begin{remark}\label{remark4.1}
 Let  us  denote  $(\cY_{\rm th}, q_{\rm dyn})$ by $(\cY_{1}, q_{1})$ and let $(\cY_{2}, q_{2})$ be another Hermitian space with  $I:  (\cY_{2}, q_{2})\to (\cY_{1}, -q_{1})$  unitary.  Then $\omega_{\rm d}$ induces a quasi-free state on $\CCR(\cY_{1}\oplus \cY_{2}, q_{1}\oplus q_{2})$. Its covariances are
 \[
\mat{\one}{0}{0}{I^{*}} \lambda^{\pm}_{\rm d} \mat{\one}{0}{0}{I}= \pm \mat{q_{1}}{0}{0}{q_{2}} \mat{\one}{0}{0}{I^{-1}}C_{\beta}^{\pm} \mat{\one}{0}{0}{I}.
\]
 \end{remark}

\begin{remark}
If $\cY\subset \cY_{\rm th}$ then $\omega_{\rm d}$ is also a (non necessarily pure) state on $\CCR(\cY\oplus \cY, q\oplus - q)$. 
\end{remark}

{\bf Proof of Prop. \ref{prop4.1:1}.}
We  obtain from \eqref{e4.9}, \eqref{e4.7b}:
\beq\label{e4.10b}
\begin{array}{rl}
&\Psi_{\rm AW}^{*}(x)\Omega= a_{\rm l}^{*}(z_{+})\Omega+ a_{\rm l}(z_{-})\Omega+ a_{\rm r}^{*}(\bz'_{-})\Omega+ a_{\rm r}(\bz'_{+})\Omega\\[2mm]
=& ((\rho+1)^{\12}z_{+}+ \rho^{\12}z'_{+}, \bro^{\12}\bz_{-}+ (\bro+1)^{\12}\bz'_{-}),\\[2mm]
&\Psi_{\rm AW}(x)\Omega=a_{\rm l}(z_{+})\Omega+ a_{\rm l}^{*}(z_{-})\Omega+ a_{\rm r}(\bz'_{-})\Omega+ a_{\rm r}^{*}(\bz'_{+})\Omega\\[2mm]
=&((\rho+1)^{\12}z_{-}+ \rho^{\12}z'_{-}, \bro^{\12}\bz_{+}+ (\bro+1)^{\12}\bz'_{+}),
\end{array}
\eeq
as elements of $\cZ\oplus \bZ$.  From \eqref{e4.10b} we immediately obtain that
\[
\omega_{\rm d}(\Psi_{\rm AW}(x_{1})\Psi_{\rm AW}(x_{2}))= \omega_{\rm d}(\Psi_{\rm AW}^{*}(x_{1})\Psi_{\rm AW}^{*}(x_{2}))=0,
\]
ie $\omega_{\rm d}$ is gauge invariant for the complex structure $\i\oplus \i$ of $\cX$.
We have next
\[
\begin{array}{rl}
&\omega_{\rm d}(\Psi_{\rm AW}(x_{1})\Psi_{\rm AW}^{*}(x_{2}))= \left(\Psi_{\rm AW}^{*}(x_{1})\Omega| \Psi^{*}_{\rm AW}(x_{2})\Omega\right)_{\cZ\oplus \bZ}\\[2mm]
=& \left((\rho+1)^{\12}y_{1+}+ \rho^{\12}y'_{1+}|(\rho+1)^{\12}y_{2+}+ \rho^{\12}y'_{2+} \right)_{\cY}
\\[2mm]
&+\left( \rho^{\12}y_{1-}+ (\rho+1)^{\12}y'_{1-}| \rho^{\12}y_{2-}+ (\rho+1)^{\12}y'_{2-}\right)_{\cY}
\end{array}
\]
 If  $\lambda_{\rm d}^{+}\eqdef Q\circ C_{\rm d}^{+}$, where $Q= q\oplus - q$, we obtain  from \eqref{ecorrect.3} that:
 \[
C_{\rm d}^{+}= \mat{(\rho+1)\one_{+}- \rho\one_{-}}{- \rho^{\12}(\rho+1)^{\12}\one_{+}+ \rho^{\12}(\rho+1)^{\12}\one_{-}}{ \rho^{\12}(\rho+1)^{\12}\one_{+}- \rho^{\12}(\rho+1)^{\12}\one_{-}}{- \rho \one_{+}+ (\rho+1)\one_{-}},
\]
for $\one_{\pm}= \one_{\rr^{\pm}}(b)$. 
We compute:
\[
\begin{array}{rl}
&(1+\rho)\one_{+}- \rho\one_{-}\\[2mm]
=& (1- \e^{- \beta b})^{-1}\one_{+}- \e^{\beta b}(1-\e^{\beta b})^{-1}\one_{-}= (1- \e^{- \beta b})^{-1};\\[2mm]
&- \rho^{\12}(1+\rho)^{\12}\one_{+}+ \rho^{\12}(1+\rho)^{\12}\one_{-}\\[2mm]
=& - \e^{- \beta b/2}(1- \e^{- \beta b})^{-1}+ \e^{\beta b/2}(1- \e^{ \beta b})^{-1}= \e^{\beta b/2}(1- \e^{\beta b})^{-1};\\[2mm]
&\rho^{\12}(1+\rho)^{\12}\one_{+}- \rho^{\12}(1+\rho)^{\12}\one_{-}\\[2mm]
=&-\e^{\beta b/2}(1- \e^{\beta b})^{-1}= \e^{-\beta b/2}(1- \e^{- \beta b})^{-1};\\[2mm]
&- \rho \one_{+}+ (1+\rho)\one_{-}\\[2mm]
=&- \e^{- \beta b}(1- \e^{- \beta b})^{-1}\one_{+}+ (1- \e^{ \beta b})^{-1}\one_{-}= (1- \e^{ \beta b})^{-1}.
\end{array}
\]
Therefore $C_{\rm d}^{+}= C^{+}_{\beta}$.  Since $C_{\rm d}^{+}+ C_{\rm d}^{-}=\one$, we have also $C_{\rm d}^{-}= C_{\beta}^{-}$. 

To see that $\omega_{\rm d}$ is pure, we have to check that the representation of the Weyl algebra $\CCR^{\rm Weyl}(\cX, Q)$ associated to $\Psi^{(*)}_{\rm AW}(x)$, $x\in \cX$ is irreducible. This follows from the definition \eqref{e1.1z} of $\Psi^{(*)}_{\rm AW}$ and statements (5), (7) in \cite[Thm. 17.24]{DG}. \qed

\section{Abstract Klein-Gordon equations}\label{sec8}\init
In this section we collect some results about  abstract Klein-Gordon equations of the form
\beq
\label{e8.0}
(\p_{t}+ \neww^{*})(\p_{t}- \neww)\newphi+\newh_{0}\newphi=0,
\eeq
where $\newphi: \rr\to \newcH$, $\newcH$ is some Hilbert space and $\newh_{0}, \neww$ are linear operators on $\newcH$.   Such Klein-Gordon equations  arise from stationary metrics on a spacetime $M = \rr\times S$, with Killing vector field equal to $\frac{\p}{\p t}$, when $\neww$ represent the shift vector field and the lapse function is equal to $1$. The case of general stationary Klein-Gordon operators will be considered later in Sect. \ref{sec12}.

We will also consider the Wick rotated operator $\tilde{K}_{\beta}$ obtained by setting $t= \i s$, where $s$ belongs either to $\rr$ or to $\bS_{\beta}$. Using sesquilinear form techniques we give a rigorous meaning to  its inverse $\tilde{K}_{\beta}^{-1}$ and relate it to the Green operators in Sect. \ref{sec1}.
\subsection{Hypotheses}
We will assume the following hypotheses:
\begin{equation}
\label{e8.1}
\begin{array}{rl}
i)&\newh_{0}\hbox{ is selfadjoint on }\newcH\hbox{ and }\newh_{0}>0,   \\[2mm]
 ii)&\Dom \neww= \Dom \newh_{0}^{\12}, \ \Dom \newh_{0}^{\12}\subset \Dom \neww^{*}\hbox{ and }\neww\newh_{0}^{-\12}, \neww^{*}\newh_{0}^{-\12}\in B(\newcH),\\[2mm]
 iii)&\hbox{if }\newh\defeq \newh_{0}- \neww^{*}\neww\hbox{, considered as a quadratic form on }\Dom \newh_{0}^{\12}, \hbox{ then }\newh\sim \newh_{0}.
\end{array}
\end{equation}
We can rewrite \eqref{e8.0} as
\begin{equation}
\label{e8.2}
\p_{t}^{2}\newphi- 2 \i \newk \p_{t}\newphi+ \newh \newphi=0,
\end{equation}
where $\newk= (2\i)^{-1}(\neww- \neww^{*})$, which was considered in \cite{GGH} in a more general situation.   

\subsection{Quadratic pencils}\label{sec8.1.}
One associates to \eqref{e8.2} the  {\em quadratic pencil}
\[
p(z)=  z(2\newk-z)+\newh= (\i z+ \neww^{*})(\i z -\neww)+ \newh_{0}\in B(\langle \newh_{0}\rangle^{-\12}\newcH, \langle \newh_{0}\rangle^{\12}\newcH), z\in \cc,
\]
obtained by replacing $\p_{t}$ by $\i z$ and denotes by   $\rho(\newh, \newk)$ the set of $z\in \cc$ such that $p(z):\langle \newh_{0}\rangle^{-\12}\newcH\xrightarrow{\sim}  \langle \newh_{0}\rangle^{-\12}\newcH$.  Note that by applying the Fourier transform in $t$, the equation
\[
(\p_{t}+ \neww^{*})(\p_{t}- \neww)u+\newh_{0}u=v,
\]
is formally equivalent to
\[
p(\lambda)\hat{u}(\lambda)= \hat{v}(\lambda), \ \lambda\in \rr.
\]
 The inverse $p(z)^{-1}$ for $z\in \rho(\newh, \newk)$  appears in the expression  \eqref{e8.3b} of  the resolvent $(z-\newH)^{-1}$, where $\newH$ is defined in \eqref{e8.01}.  

 \begin{proposition}\label{stupido}
 \ben
\item   The operator $p(z)$ on $\newcH$ with domain $\langle \newh\rangle^{-1}\newcH$  is closed with $p^{*}(z)= p(\bar{z})$;
\item $p(z): \langle \newh\rangle^{-1}\newcH\tosim \cH$ iff $p(z): \langle \newh\rangle^{-\12}\newcH\tosim \langle \newh\rangle^{\12}\newcH$;
\item $p(\i \lambda): \langle \newh\rangle^{-1}\newcH\tosim \cH$ for $\lambda\in \rr^{*}$ and
\[
\sup_{\lambda\in \rr^{*}}\| (\newh_{0}+ \lambda^{2})^{\12}p(\i \lambda)^{-1}(\newh_{0}+ \lambda^{2})^{\12}\|_{B(\newcH)}<\infty.
\]
\een
\end{proposition}
\proof 
Statements (1) and (2) are shown in \cite[Lemma 8.1]{GGH2}, using that $\newk$ is symmetric on $\langle \newh_{0}\rangle^{-\12}\newcH= \langle \newh\rangle^{-\12}\newcH$ and $\newk\langle \newh\rangle^{-\12}\in B(\newcH)$. 
Let us now prove (3). Since $\Re p(\i \lambda)= \newh+ \lambda^{2}$ we have
\[
\lambda^{2}\|u\|^{2}\leq \Re(u| p(\i\lambda)u), \ u\in \langle\newh\rangle^{-1}\newcH,
\]
hence $\Ker p(\i \lambda)= \{0\}$ for $\lambda\in \rr^{*}$. Since $p(\i\lambda)^{*}= p(-\i\lambda)$ we have also $\Ker p(\i \lambda)^{*}= \{0\}$ hence $p(\i \lambda): \langle\newh\rangle^{-1}\newcH\tosim \newcH$ using that $p(\i\lambda): \langle\newh\rangle^{-1}\newcH\to \newcH$ is  closed. We have then 
\[
\begin{array}{l}
(\newh + \lambda^{2})^{-\12}p(\i \lambda)(\newh+ \lambda^{2})^{-\12}=\one - \i A(\lambda),\\[2mm]
A_{\lambda}\defeq \i (\newh + \lambda^{2})^{-\12}(\lambda \neww- \neww^{*}\lambda)(\newh + \lambda^{2})^{-\12}.
\end{array}
 \]
 From hypotheses \eqref{e8.1}, we obtain that $A_{\lambda}\in B(\newcH)$ and $A_{\lambda}= A_{\lambda}^{*}$.  Therefore $\| (\one - \i A_{\lambda})^{-1}\|_{B(\newcH)}\leq 1$  and since 
 \[
 (\newh+ \lambda^{2})^{\12}p(\i \lambda)^{-1}(\newh+ \lambda^{2})^{\12}
 \]
 we have 
 \[
 \sup_{\lambda\in \rr^{*}}\| (\newh+ \lambda^{2})^{\12}p(\i \lambda)^{-1}(\newh+ \lambda^{2})^{\12}\|_{B(\newcH)}\leq 1.
 \]
 This proves (3) using that $\newh \sim \newh_{0}$ . \qed
   
\subsection{First order system}\label{sec8.4}
Setting 
\beq\label{e8.00}
\newf(t)= \newrho_{t}\newphi\defeq \col{\newphi(t)}{\i^{-1}(\p_{t}- \neww)\newphi(t)}=\col{\newf_{0}(t)}{\newf_{1}(t)},
\eeq
 \eqref{e8.0} is formally rewritten as
\beq\label{e8.01}
\p_{t}\newf= \i \newH \newf, \ \newH= \mat{-\i \neww}{\one}{\newh_{0}}{\i \neww^{*}}.
\eeq
The conserved energy is
\beq\label{e8.02}
\bar{\newf}\dual \newE \newf= \|\newf_{1}- \i \neww\newf_{0}\|^{2}+ (\newf_{0}| \newh \newf_{0}),
\eeq
which is positive definite by \eqref{e8.1}.
The Hilbert space associated to $\newE$  will be denoted by $\newcE$. It equals $\newh_{0}^{-\12}\newcH\oplus \newcH$ as a topological vector space.
 We set also $\newcE^{*}\defeq \newcH\oplus \newh_{0}^{\12}\newcH$. 
 
We recall from \ref{zlat} that
\beq\label{zlut}
\newcH\cap \newh_{0}^{\pm \12}\newcH= \Dom \newh_{0}^{\mp\12}\hbox{ hence }\newcE\cap \newcE^{*}= \Dom \newh_{0}^{\12}\oplus \Dom \newh_{0}^{-\12}.
\eeq
The following proposition will be proved  in Subsect. \ref{sec8.proofs}.
\begin{proposition}\label{prop8.0}
The operator $\tilde{{\rm H}}=\mat{-\i \neww}{\one}{\newh_{0}}{\i \neww^{*}}$  is bounded from $\newcE$ to $\newcE^{*}$. 
It induces  on $\newcE$ the operator $\newH$ defined by
\[
\Dom \newH= \{\newf\in \newcE: \tilde{{\rm H}}\newf\in \newcE\cap \newcE^{*}\}.
\]
$\newH$ is a densely defined selfadjoint operator on $\newcE$ with ${\rm res}(\newH)= \rho(\newh, \newk)$.
\end{proposition}
Note that $(\newcE, \newcE^{*})$ form a non degenerate dual pair for  the charge
\beq\label{e8.02b}
\bar{\newf}\dual \newq \newf'= (\newf_{1}| \newf'_{0})_{\newcH} + (\newf_{0}| \newf'_{1})_{\newcH}, \ \newf\in \newcE, \ \newf'\in \newcE^{*},
\eeq
and one has
\[
\bar{\newf}\dual \newE \newf= \bar{\newf}\dual \newq \tilde{{\rm H}}\newf, \newf\in \newcE.
\]
\subsection{The Wick rotated operator}\label{sec8.2}
Setting formally $t= \i  s$  we obtain the formal expression
\beq\label{e8.10}
\newK= -(\p_{ s}+ \i \neww^{*})(\p_{ s}- \i \neww)+\newh_{0}.
\eeq
To give a meaning to \eqref{e8.10}, we will use sesquilinear forms techniques.  Let us set as in Sect. \ref{sec1.1b} for $0<\beta\leq \infty$:
\beq\label{e8.10a}
\newcH_{\beta}=L^{2}(\bS_{\beta})\otimes \newcH, \hbox{ for }\beta<\infty, \  \newcH_{\infty}= L^{2}(\rr)\otimes \newcH.
\eeq
We consider the sesquilinear form associated to $\newK$:
\[
\newQ_{\beta}(u, u)= \| \p_{ s}u\|^{2}_{\newcH_{\beta}}+ (u| \newh u)_{\newcH_{\beta}}- \i (\p_{ s}u| \neww u)_{\newcH_{\beta}}- \i (\neww u| \p_{ s}u)_{\newcH_{\beta}},
\]
with domain $\Dom \newQ_{\beta}= \langle  -\p_{ s}^{2}+ \newh_{0}\rangle^{-\12}\newcH_{\beta}$, where $\p_{ s}$ is equipped with its natural domain on $\newcH_{\beta}$. From hypotheses \eqref{e8.1} we obtain that
\[
\Re \newQ_{\beta}(u, u)\sim \| \p_{ s}u\|^{2}_{\newcH_{\beta}}+ (u| \newh_{0}u)_{\newcH_{\beta}}, \ |\Im \newQ_{\beta}(u, u)|\leq C\Re \newQ_{\beta}(u, u),
\]
hence $\newQ_{\beta}$ is a closed sectorial form. 

If we apply the  Lax-Milgram theorem as stated in \cite[Lemma 12.15]{Gr} to the Hilbert space $(-\p_{ s}^{2}+ \newh_{0})^{-\12}\newcH_{\beta}$ and note that its topological anti-dual is canonically identified with $(-\p_{ s}^{2}+ \newh_{0})^{\12}\newcH_{\beta}$, we obtain that 
$\newQ_{\beta}$ induces a boundedly invertible operator
\begin{equation}
\label{e8.11}
\newK_{\beta}:(-\p_{ s}^{2}+ \newh_{0})^{-\12}\newcH_{\beta}\tosim (-\p_{ s}^{2}+ \newh_{0})^{\12}\newcH_{\beta}.
\end{equation}

We can apply the  results of Subsect. \ref{sec1.1b} setting $\ch= \newcE$, $b= \newH$, see \eqref{e.4.not} for the notation used,  and obtain an operator
\beq\label{etututu}
\p_{ s}+ \newH: \newcE_{\beta}\tosim (- \p_{ s}^{2}+ \newH^{2})^{\12}\newcE_{\beta}.
\eeq
The relation between $\newK_{\beta}^{-1}$ and $\p_{ s}+ \newH$ is given by the following proposition. Below we denote by $\pi_{i}$ the maps
$\pi_{i}\newf= \newf_{i}$ for $\newf= \col{\newf_{0}}{\newf_{1}}$, so that
\[
\pi_{1}^{*}u= \col{0}{u}, \ \pi_{0}^{*}u = \col{u}{0}.
\]
\begin{proposition}\label{prop8.1}
 One has
 \beq\label{e8.8}
 \begin{array}{rl}
 i)& \pi_{1}^{*}: (-\p_{ s}^{2}+ \newh_{0})^{\12}\newcH_{\beta}\to (\p_{ s}^{2}+ \hat{H}^{2})^{\12}\newcE_{\beta}\hbox{ continuously},\\[2mm]
 ii)&\pi_{0}(\p_{ s}+ \newH)^{-1}\pi_{1}^{*} :(-\p_{ s}^{2}+ \newh_{0})^{\12}\newcH_{\beta}\to (-\p_{ s}^{2}+ \newh_{0})^{-\12}\newcH_{\beta}\hbox{ continuously},
 \\[2mm]
iii)&\newK_{\beta}^{-1}= \pi_{0}(\p_{ s}+ \newH)^{-1}\pi_{1}^{*}\hbox{ in  }L((-\p_{ s}^{2}+ \newh_{0})^{\12}\newcH_{\beta}, (-\p_{ s}^{2}+ \newh_{0})^{-\12}\newcH_{\beta}).
\end{array}
\eeq
\end{proposition}
\subsection{Proofs of Props. \ref{prop8.0} and \ref{prop8.1}}\label{sec8.proofs}
\subsubsection{Preparations}
 We will prove Props. \ref{prop8.0}, \ref{prop8.1} using results in \cite{GGH}. There the form \eqref{e8.2} of the Klein-Gordon equation is used and instead of \eqref{e8.00} one sets:
\beq\label{e8.2b}
g\defeq\col{\newphi}{\i^{-1}\p_{t}\newphi},
\eeq
\eqref{e8.2} is formally rewritten as 
\[
\p_{t}g= \i \hat{H} g, \ \hat{H}= \mat{0}{\one}{\newh}{2\newk}.
\]
The conserved energy is
\[
\bar{g}\dual \hat{E}g= \|g_{1}\|^{2}+ (g_{0}| \newh g_{0}).
\]
The  Hilbert space $\thE$ naturally associated to $\hat{E}$   equals again $\newh_{0}^{-\12}\newcH\oplus \newcH$.

If $\newf$ is given by \eqref{e8.00} and $g$ by \eqref{e8.2b}  one has 
\[
\newf= Ug\hbox{ for }U= \mat{\one}{0}{\i \neww }{\one},
\]
 and 
\beq\label{e8.3}
 U: \thE\xrightarrow{\sim} \newcE, \  (Ug| Ug)_{\newcE}= (g| g)_{\thE}.
\eeq
Formally one has $\newH= U\hat{H}U^{-1}$, and since $U: \thE\to \newcE$ is unitary, Prop. \ref{prop8.0} follows if we prove the analogous result for $\hat{H}$.  One sets then
\[
\thE^{*}\defeq \newcH\oplus \newh_{0}^{\12}\newcH,
\]
which forms again a dual pair  with $\thE$ for
\[
\bar{g}\dual \hat{q} g'= (g_{1}+\i \neww  g_{0}| g'_{0})_{\newcH}+ (g_{0}| g'_{1}+ \i \neww  g_{0}'), \ g\in \thE,  g'\in \thE^{*}.
\]
We have of course $\hat{q}= U^{*}q U$.

\subsubsection{Proof of Prop. \ref{prop8.0}}
The matrix $\hat{H}$ induces  a bounded operator  $\hat{\rm H}: \thE\to \thE^{*}$.  One denotes by $\hat{H}$ the linear operator induced by $\hat{\rm H}$ on $\thE$.  Its domain is 
\[
\Dom \hat{H}=\{g\in \thE: \hat{\rm H}g\in \thE\cap \thE^{*}\},
\]
and as in \eqref{zlut} we have $\thE\cap \thE^{*}=\Dom\newh_{0}^{\12}\oplus \Dom \newh_{0}^{-\12}$.

It follows then from \cite[Prop. 5.8, Thm. 5.9]{GGH}, and the fact that  there exists  $z\neq 0$ in $\rho(\newh, \newk)$,  that $\hat{H}$ with the domain above is a densely defined selfadjoint operator on $\thE$ with  ${\rm res}(\hat{H})= \rho(\newh, \newk)$. Setting
\[
\newH\defeq U\hat{H}U^{-1}
\]
completes the proof of Prop. \ref{prop8.0}.

\subsubsection{Proof of Prop. \ref{prop8.1}}
We have seen that  $\newH= U\hat{H}U^{-1}$ and $U: \hat{\cE}\tosim \newcE$ is unitary. Moreover $\pi_{0}U^{-1}= \pi_{0}$ and $U\pi_{1}^{*}= \pi_{1}^{*}$, so it suffices to prove the proposition with $\newH, \newcE$ replaced by $\hat{H}, \hat{\cE}$.

One can express the resolvent $(\hat{H}-z)^{-1}$ using $p(z)$ as follows: if $z\in \rho(\newh, \newk)$ then:
\beq\label{e8.3b}
(\hat{H}-z)^{-1}= p(z)^{-1}\mat{z-2\newk}{\one}{\newh}{z}\in B(\thE, \thE).
\eeq
Note that \eqref{e8.3b} is different from the formula found in \cite[Prop. 5.8]{GGH}, because weaker assumptions on $\newh,\newk$ were used there. In our case using that $\newk|\newh_{0}|^{-\12}\in B(\newcH)$ one  deduces from \cite[Lemma 2.2]{GGH} that
\[
p(z): \newcH+ |\newh_{0}|^{\12}\newcH\to \langle \newh_{0}\rangle^{-\12}\newcH= \newcH\cap |\newh_{0}|^{-\12}\newcH, \ z\in \rho(\newh, \newk).
\]
Using this fact it is straightforward to show that the rhs in \eqref{e8.3b} maps $\thE$ into itself. 

 In general we have $0\not\in \rho(\newh,\newk)$  hence $0\in \sigma(\hat{H})$ but $\hat{H}^{-1}$ is well defined as  
 \beq\label{e8.4}
\hat{H}^{-1}= \mat{-2\newh^{-1}\newk}{\newh^{-1}}{\one}{0}\in B(\thE, |\newh|^{-1}\newcH\oplus \newcH)
\eeq
which corresponds to \eqref{e8.3b} for $z=0$.

We have $\Ker\, \hat{H}= \{0\}$ since $\hat{H}g=0$ implies $g_{1}= 0, \newh g_{0}=0$ and $\newh$ is injective. Therefore we can apply the results of Subsect. \ref{sec1.1b} to construct $(\p_{ s}+ \hat{H})^{-1}$  for  $b= \hat{H}$, $\ch= \thE$.
As before we introduce the Hilbert spaces $\newcH_{\beta}$ and $\thE_{\beta}$ for $\beta\in ]0, \infty]$.
 
 Let us now prove the proposition. 
We denote by $\cF$ the Fourier transform  in $s$ on $\rr$ or $\bS_{\beta}$ and set $\hat{v}(\lambda)= \cF v(\lambda)$, where $\lambda\in \rr$ or $\bS_{\beta}$ is the associated Fourier variable. Let
 \[
M\defeq\cF\circ (\p_{s}+ \hat{H})^{-1}\pi_{1}^{*}(-\p_{s}^{2}+ \newh_{0})^{\12}\circ \cF^{-1}= \int^{\oplus}M(\lambda)d\lambda. 
\]
From \eqref{e8.4}, \eqref{e8.3b} we obtain that
\[
M(\lambda)= p(\i \lambda)^{-1}(\lambda^{2}+ \newh_{0})^{\12}\col{\one}{\i \lambda\one}, \ \lambda\neq 0, M(0)= \col{\newh^{-1}\newh_{0}^{\12}}{0}.
\]
From (3) in Prop. \ref{stupido} we obtain that 
\[
\sup_{\lambda\in \rr}\|M(\lambda)\|_{B(\newcH, \hat{\cE})}<\infty,
\]
which proves statement {\it i)} and also statement {\it ii)} of the proposition. 
Next if $v\in (-\p_{ s}^{2}+ \newh_{0})^{\12}\newcH_{\beta}$  and  $g=(\p_{ s}+ \hat{H})^{-1}\pi_{1}^{*}v$  we have $g\in \thE_{\beta}$ and
\[
\p_{ s}g_{0}+ g_{1}=0, \ \newK g_{0}= v,
\]
hence $\p_{ s}g_{0}\in \newcH_{\beta}$, $\newh_{0}^{\12}g_{0}\in \newcH_{\beta}$ and $\newK g_{0}=v$, which shows that $g_{0}\in (-\p_{ s}^{2}+ \newh_{0})^{-\12}\newcH_{\beta}$ and  $g_{0}= \newK_{\beta}^{-1}v$.  This completes the proof of statement {\it iii)}. \qed

\section{Vacua and KMS states for abstract Klein-Gordon equations}\label{sec9}\init
In this section we consider vacuum and KMS states for abstract, time-independent Klein-Gordon equations, which can be reduced to the framework of Sect. \ref{sec3}. We will show that the covariances of the vacuum and double $\beta$-KMS states can be expressed by the \Calderon projectors defined in Sect. \ref{sec3}. 
\subsection{Vacua and KMS states}\label{sec9.1}
Let us consider an abstract Klein-Gordon equation
\[
(\p_{t}+ \neww^{*} )(\p_{t}-\neww )\newphi+ \newh_{0}\newphi=0,
\]
as in Sect. \ref{sec8}, where $\newphi: \rr\to \newcH$ and $\newcH$ is a Hilbert space.      We denote by
\[
\newP= (\p_{t}+ \neww^{*} )(\p_{t}-\neww )+ \newh_{0}
\]
the corresponding Klein-Gordon operator. In the sequel we use the notation introduced in Subsect. \ref{sec3.1}.

The assumptions corresponding to those in Subsect. \ref{sec3.1} are as follows:

We assume that there exists a dense subspace $\newcD\subset \newcH$  and set 
\beq\label{e9.1}
\newcY\defeq \newcD\oplus \newcD, \ \bar{\newf}\dual  \newq\newf\defeq (\newf_{1}|\newf_{0})+ (\newf_{0}| \newf_{1}), \ \newf= \col{\newf_{0}}{\newf_{1}}\in \newcY.
\eeq
We fix linear operators $\newh_{0},\neww , \neww^{*} $ on $\newcH$ with domain $\newcD$ such that:
\begin{equation}
\label{e9.1b}
\begin{array}{l}
\neww, \ \neww^{*}, \ \newh_{0}: \newcD\to \newcD,\\[2mm]
(\neww^{*} u | v)= (u| \neww v), \ u, v\in \newcD, \ (u| \newh_{0}u)>0 \hbox{ for }u\in \newcD, u\neq 0,\\[2mm]
 \| \neww u\|^{2}\leq (1- \delta)(u| \newh_{0}u), \ \| \neww^{*} u\|^{2}\leq c (u| \newh_{0}u), \ u\in \newcD\hbox{ for }c>0, 0<\delta<1.
\end{array}
\end{equation}
Setting $\newq_{0}(u, u)= (u| \newh_{0}u)$ with $\Dom \newq_{0}= \newcD$, it follows that $\newq_{0}$ is closeable and we still denote by $\newh_{0}$ the operator associated to $\newq_{0}^{\rm cl}$, i.e.  the Friedrichs extension of $\newh_{0}$ on $\newcD$.  We assume that $\Ker\, \newh_{0}= \{0\}$ and deduce then from \eqref{e9.1b}  that hypotheses 
 \eqref{e8.1} are satisfied by $\newh_{0}, \neww , \neww^{*} $. By construction $\newcD$ is dense in $\newh_{0}^{-\12}\newcH$.
 
We set then
\beq\label{e9.2}
\bar{\newf}\dual \newE \newf= (\newf| \newf)_{\newcE}= \|\newf_{1}- \i \neww \newf_{0}\|^{2}+ (\newf_{0}| \newh\newf_{0}), \ \newf\in \newcY,
\eeq
and by the density of $\newcD$ in $\newh_{0}^{-\12}\newcH$ we obtain that $\newcY_{\rm en}= \newcE$.
We set also
\[
b=\newH=  \mat{-\i \neww }{\one}{\newh_{0}}{\i \neww^{*} },
\]
where $\newH$ is defined as a selfadjoint operator on $\newcE$ by Prop. \ref{prop8.0}.

\subsubsection{Infrared condition}\label{seccorrect.1}
 To be able to apply the results of Sect. \ref{sec3} we need to check the hypotheses \eqref{ecorrect.2}, \eqref{e7.-210}  and \eqref{ecorrect.4}, which are used to ensure that the   vacuum and KMS states are well defined as quasi-free states on $\CCR(\newcY, \newq)$, see Remarks \ref{remark-correct.1} and \ref{remark-correct.2}.  This is done in Prop. \ref{prop-correct.1} below. For Klein-Gordon equations on stationary spacetimes, the assumption \eqref{ecorrect.5}  will be checked in Lemma \ref{lemma12.1}.

 In Subsect. \ref{sec3.1} we introduced the dynamical  and thermal Hilbert spaces $\newcY_{\rm dyn}, \newcY_{\rm th}$ which in the present situation equal $|\newH|^{\12}\newcE$ and $|\newH|^{\12}\newcE\cap |\newH|\newcE$ respectively.
\begin{proposition}\label{prop-correct.1}
 Assume that  \begin{equation}
 \label{ecorrect.5}
 \newcD\subset \Dom \newh_{0}^{-\12}.
 \end{equation}
 Then  $\newcY= \newcD\oplus \newcD$ satisfies:
 \ben
 \item $\newcY\subset \Dom \newH^{-1}$, hence $\newcY\subset \newcY_{\rm th}\subset \newcY_{\rm dyn}$,
 \item $\bar{\newf}\dual \newE b^{-1}\newf= \bar{\newf}\dual \newq f$ for $f\in \newcY$
  \item $\newcY\subset \Dom \newH$.
 \een
\end{proposition}
It follows from Prop. \ref{prop-correct.1} that  hypotheses \eqref{ecorrect.2}, \eqref{e7.-210}  and \eqref{ecorrect.4} in Sect. \ref{sec3} are satisfied.

\proof  To check (1) it suffices  to show that
\beq\label{ecorrect.7}
(\newH^{-1}\newf|\newH^{-1}\newf)_{\newcE}<\infty, \ \newf\in \newcY,
\eeq
since $\newcY\subset \newcE$.

Let us now check \eqref{ecorrect.7}.  We recall that in Subsect. \ref{sec8.4} we considered also $\newH$ as a bounded operator from $\newcE$ to $\newcE^{*}$, where $\newcE^{*}= \newcH\oplus \newh_{0}^{\12}\newcH$, (see Prop. \ref{prop8.0}, where $\newH$ with the above meaning was denoted by $\tilde{{\rm H}}$).  The equation $\newH g= f$ for $g\in \newcE$, $f\in \newcE^{*}$ is equivalent to
\beq\label{ecorrect.8}
\left\{\begin{array}{l}
(\newh_{0}- \neww^{*}\neww)g_{0}= f_{1}- \i \neww^{*}f_{0}\in \newh_{0}^{\12}\newcH, \\[2mm]
 g_{1}- \i \neww g_{0}= f_{0}\in \newcH,
\end{array}\right.
\eeq
The operator $\newh= \newh_{0}- \neww^{*}\neww\in B(\newh_{0}^{-\12}\newcH, \newh_{0}^{\12}\newcH)$ is boundedly invertible, since from \eqref{e9.1b} the quadratic forms associated to $\newh$ and $\newh_{0}$ are equivalent. Therefore \eqref{ecorrect.8} is equivalent to
\beq\label{ecorrect.9}
\left\{\begin{array}{l}
g_{0}= \newh^{-1}(f_{1}- \i \neww^{*}f_{0})\in \newh_{0}^{-\12}\cH, \\[2mm]
 g_{1}= g_{0}+ \i \neww \newh^{-1}(f_{1}- \i \neww^{*}f_{0})\in \newcH.
\end{array}\right.
\eeq
It follows that $\newH: \newcE\to \newcE^{*}$ is boundedly invertible. Now by hypothesis \eqref{ecorrect.5}, we have $\newcD\subset \newh_{0}^{\12}\newcH$, hence $\newcY\subset \newcE^{*}$. Therefore $\newH^{-1}: \newcY\to \newcE$, which proves \eqref{ecorrect.7} and completes the proof of (1).  (2) follows then by using  \eqref{ecorrect.9} and \eqref{e9.2}. 

It remains to prove (3).  We have $\newcY\subset \newcE$ and since $\newcD\subset \Dom \newh_{0}^{-\12}$ we have also $\newcY\subset \newcE^{*}= \newcH\oplus \newh_{0}^{\12}\newcH$.  Since $\newH: \newcY\to \newcY$ this implies (3), by the definition of $\Dom \newcH$ in  Prop. \ref{prop8.0}. \qed

 We can then  apply Subsects. \ref{sec3.2}, \ref{sec3.3}, \ref{sec4.7}  and define the vacuum state $\newomega_{\rm vac}$, the $\beta$-KMS state $\newomega_{\beta}$ and the double $\beta$-KMS state $\newomega_{\rm d}$ associated to the symplectic dynamics $r_{t}= \e^{\i t \newH}$.

\subsection{The \Calderon projectors}\label{sec9.2}
 In Subsect.  \ref{sec8.2} we defined the Wick rotated operators
 \[
\newK_{\beta}= - (\p_{ s}+ \i \neww^{*} )(\p_{ s}- \i \neww )+\newh_{0}
\]
 and the Hilbert spaces $\newcH_{\beta}$ for  $0<\beta\leq \infty$ defined in \eqref{e8.10a}.  We now define \Calderon projectors for $\newK_{\beta}$, which are similar to the \Calderon projectors for the operators $B_{\beta}= \p_{ s}+ \newH$, acting on  the Hilbert spaces $\newcE_{\beta}$, defined in \ref{caldeinf} and \ref{caldebeta}.
 
 In all this subsection we will assume that \eqref{ecorrect.5} holds ie that $\newcY\subset \Dom \newh_{0}^{-\12}$.
\subsubsection{\Calderon projectors for $\newK_{\infty}$}\label{sec9.2.1}
We follow the  notation in \ref{caldeinf}, in particular $I_{\infty}^{\pm}= \pm ]0, +\infty[$. 
We fix $\chi\in \coinf(\rr)$ with $\int \chi(s)ds=1$ and set $\chi_{n}= n\chi(n^{-1}s)$ for $n\in \nn$.
\[
\newgamma^{*}_{\infty, n}\newg\defeq \chi_{n}'( s)\otimes \newg_{1}+ \chi_{n}( s)\otimes (\newg_{0}- \i \neww^{*} \newg_{1}),
\]
for $\newg\in \newcD\oplus\newcD$ so that
\[
\newgamma_{\infty,n}^{*}\newg\to \newgamma_{\infty}^{*}\newg\hbox{ in }\cE'(\rr; \newcH), \hbox{ when }n\to +\infty,
\]
for:
\[
\newgamma_{\infty}^{*}\newg\defeq \delta_{0}'( s)\otimes \newg_{1}+ \delta_{0}( s)\otimes (\newg_{0}- \i \neww^{*} \newg_{1}).
\]
If $\newu\in \overline{C^{1}}(I^{\pm}_{\infty}; \newcH)\cap \overline{C^{0}}(I^{\pm}_{\infty}; \newh_{0}^{-\12}\newcH)$ satisfies $\newK \newu= 0$ in $I_{\infty}^{\pm} $ we set
\[
\newgamma^{\pm}_{\infty}\newu= \col{\newu(0^{\pm})}{-(\p_{ s}- \i \neww )\newu(0^{\pm})},
\] 
and
\[
\newS= \mat{2\i \neww^{*} }{-\one}{\one}{0}.
\]

\begin{proposition}\label{prop9.1}
Assume that  $\newf\in \newcD\oplus \newcD$. Then:
\ben
 \item  the limit
 \[
 \lim_{n\to +\infty}\newK_{\infty}^{-1}\newgamma_{\infty,n}^{*}\newS\newf\eqdef \newK_{\infty}^{-1}\newgamma_{\infty}^{*}\newS\newf
  \]
exists in $\cD'(I_{\infty}^{\pm}; \newh_{0}^{-\12}\newcH)$
and is independent on the choice of the cutoff function $\chi$.  
\item We have 
\[
 \newK_{\infty}^{-1}\newgamma_{\infty}^{*}\newS\newf\in \overline{C^{1}}(I_{\infty}^{\pm};\newh_{0}^{-\12}\newcH),\]
\item 
 \[
 \mp\newgamma_{\infty}^{\pm}\newK_{\infty}^{-1}\newgamma_{\infty}^{*}\newS\newf= C_{\infty}^{\pm}\newf,
\]
\een
where $C_{\infty}^{\pm}$ are the \Calderon projectors for $B_{\infty}= \p_{ s}+ b$,  with $b= \newH$,  $\ch= \newcE$, defined in  Prop. \ref{prop4.1}.
\end{proposition}

\begin{definition}\label{def9.1}
 The operators
 \[
\newc^{\pm}_{\infty}\defeq\mp \newgamma_{\infty}^{\pm}\newK_{\infty}^{-1}\newgamma_{\infty}^{*}\in L(\newcD\oplus \newcD; \newcE)
 \]
 are called the  {\em \Calderon projectors} for $\newK_{\infty}$.
 \end{definition}

{\bf Proof of Prop. \ref{prop9.1}.} We prove only the $+$ case.  Note first that if $\chi\in \coinf(\rr)$ and $u\in \Dom\newh_{0}^{-\12}$ then $\chi(s)\otimes u\in (- \p_{s}^{2}+ \newh_{0})^{\12}\cH_{\infty}$. In fact after Fourier transform in $s$, this amounts to prove that
\[
\int_{\rr}|\hat{\chi}(\lambda)|^{2}(u| (\lambda^{2}+ \newh_{0})^{-1}u)_{\newcH}d\lambda<\infty,
\]
which follows from the fact that $(u| \newh_{0}^{-1}u)_{\newcH}<\infty$. If  $\newf\in \Dom\newcD\oplus \newcD$, then $ \newf_{0},\newf_{1}- \i \neww^{*}\newf_{0} \in \newcD\subset \Dom \newh_{0}^{-\12}$ hence
\[\newS \newgamma^{*}_{\infty, n} \newf= \chi_{n}'(s)\otimes \newf_{0}+ \chi_{n}(s)\otimes (\newf_{1}- \i \neww^{*}\newf_{0})
\]
belongs to $(- \p_{s}^{2}+ \newh_{0})^{\12}\cH_{\infty}$.
Therefore we can apply Prop. \ref{prop8.1} and we have
\beq\label{sloubi1}
\newK_{\infty}^{-1}\newgamma^{*}_{\infty, n} \newS\newf= \pi_{0}B_{\infty}^{-1}\pi_{1}^{*}\newgamma^{*}_{\infty, n} \newS\newf,
\eeq
where $B_{\infty}= (\p_{s}+ \newH)$, acting on $\newcE_{\infty}$.   
Arguing as in the proof of Prop. \ref{prop4.1}, we have for $\newg\in \newcE$:
\[
\begin{array}{l}
\lim_{n\to \infty}B_{\infty}^{-1}(\chi_{n}\otimes \newg)(s)= \one_{\rr^{+}}(\newH)\e^{- s\newH}\newg,\\[2mm]
\lim_{n\to \infty}B_{\infty}^{-1}(\chi_{n}'\otimes \newg)(s)= -\newH\one_{\rr^{+}}(\newH)\e^{- s\newH}\newg\hbox{ in }\cD'(I^{+}_{\infty}; \newcE).
\end{array}
\]
An easy computation show then that \[
\lim_{n\to +\infty}-(B_{\infty}^{-1}\newgamma^{*}_{\infty, n} \newS\newf)(s)=\one_{\rr^{+}}(\newH)\e^{-s \newH}\newf\eqdef F(s)\hbox{ in }\cD'(I_{\infty}^{+}; \newcE)
\]
which using \eqref{sloubi1} proves (1).  Next  since $\newf\in \newcD\oplus \newcD\subset \Dom \newH$, we have $F\in \overline{C^{1}}(I_{\infty}^{+}; \newcE)$, which implies that 
\[
\pi_{0}F(s)\in \overline{C^{1}}(I_{\infty}^{+};\newh_{0}^{-\12}\newcH)
\]
which proves (2). Finally we check that $\newgamma_{\infty}^{+}\pi_{0}F(s)=\Gamma^{+}_{\infty}\newf$, which using the definition of $C_{\infty}^{+}$ in Prop. \ref{prop4.1} proves (3). \qed

From Subsect. \ref{sec3.2} we obtain the following result, expressing the covariances of the vacuum state $\newomega_{\rm vac}$  for $P$ in terms of the \Calderon projectors $\newc_{\infty}^{\pm}$ for  the Wick rotated operator $\newK_{\infty}$. 

We recall that from Prop. \ref{prop-correct.1} we know that  the vacuum state $\omega_{\rm vac}$ is well defined  on $\CCR(\newcY, \newq)$.
\begin{proposition}\label{prop9.1b}
 The covariances of the vacuum state $\omega_{\rm vac}$, considered as a quasi-free state on $\CCR(\newcY, \newq)$ are equal to:
 \[
\newlambda_{\rm vac}^{\pm}= \pm \newq\circ \newc_{\infty}^{\pm}.
\]
\end{proposition}
\subsubsection{\Calderon projectors for $\newK_{\beta}$}\label{sec9.2.2}
We follows now the construction and notation in \ref{caldebeta}, in particular $I_{\beta}^{\pm}= \pm]0, \b2[$.  For $\chi_{n}$ as  in \ref{sec9.2.1}, we set:
\[
\begin{array}{l}
\newgamma_{\beta,n}^{(0)*}\tilde{g}= \chi_{n}'( s)\otimes \tilde{g}_{1}+ \chi_{n}( s)\otimes(\tilde{g}_{0}- \i \neww^{*} \tilde{g}_{1}), \\[2mm]
  \newgamma_{\beta, n}^{(\b2)*}\tilde{g}= -\chi_{n}'( s- \b2)\otimes \tilde{g}_{1}+ \chi_{n}( s-\b2)\otimes(\tilde{g}_{0}+ \i \neww^{*} \tilde{g}_{1}),
\end{array}
\]
for $\tilde{g}\in \newcD\oplus\newcD$, so that
\[
\newgamma_{\beta,n}^{(\sharp)*}\tilde{g}\to \newgamma_{\beta}^{(\sharp)*}\tilde{g}\hbox{ in }\cE'(\rr; \newcH)\hbox{ when }n\to +\infty, \ \sharp = 0, \b2
\]
for:
\[
\begin{array}{l}
\newgamma_{\beta}^{(0)*}\tilde{g}= \delta_{0}'( s)\otimes \tilde{g}_{1}+\delta_{0}( s)\otimes(\tilde{g}_{0}- \i \neww^{*} \tilde{g}_{1}), \\[2mm]
  \newgamma_{\beta}^{(\b2)*}\tilde{g}= -\delta_{\b2}'( s)\otimes \tilde{g}_{1}+ \delta_{\b2}(s)\otimes(\tilde{g}_{0}+ \i \neww^{*} \tilde{g}_{1}).\end{array}
\]
We set also
\[
\gamma_{\beta, (n)}^{*}(\tilde{g}^{(0)}\oplus \tilde{g}^{(\b2)})\defeq \newgamma_{\beta, (n)}^{(0)*}\tilde{g}^{(0)}+ \newgamma_{\beta, (n)}^{(\b2)*}\tilde{g}^{(\b2)}.
\]
If $\tilde{u}\in \overline{C^{1}}(I_{\beta}^{\pm}; \newcH)\cap \overline{C^{0}}(I_{\beta}^{\pm}; \newh_{0}^{-\12}\newcH)$ satisfies $\newK \tilde{u}= 0$ in $I_{\beta}^{\pm}$ we set
\[
\newgamma^{\pm}_{\beta}\newu= \newgamma_{\beta}^{(0)\pm}\newu\oplus \newgamma_{\beta}^{(\b2)\pm}\newu,
\]
for
\[
\newgamma^{(0)\pm}_{\beta}\newu= \col{\newu(0^{\pm})}{-(\p_{ s}- \i \neww )\newu(0^{\pm})}, \ \newgamma_{\beta}^{(\b2)\pm}\newu= \col{\newu(\mp\b2)}{(\p_{ s}- \i \neww )\newu(\mp\b2)}.
\]
Note the change of sign in the second component of $\newgamma^{(\b2)\pm}\newu$, which corresponds to choosing the {\em exterior} normal derivative to $I_{\beta}^{+}$.
We set also:
\beq\label{caldolo}
\newS^{(0)}=  \mat{2\i \neww^{*} }{-\one}{\one}{0}, \ \newS^{(\b2)}=  \mat{-2\i \neww^{*} }{-\one}{\one}{0}.
\eeq
\begin{proposition}\label{prop9.2}
Assume that  $\newf= \newf^{(0)}\oplus \newf^{(\b2)}$ with $\newf^{(0/ \b2)}\in \newcD\oplus \newcD$. Then:
\ben
 \item  the limit
 \[
 \lim_{n\to +\infty}\newK_{\beta}^{-1}\newgamma_{\beta,n}^{*}(\newS^{(0)}\newf^{(0)}\oplus \newS^{(\b2)}\newf^{(\b2)})\eqdef \newK_{\beta}^{-1}\newgamma_{\beta}^{*}(\newS^{(0)}\newf^{(0)}\oplus \newS^{(\b2)}\newf^{(\b2)})  \]
exists in $\cD'(I_{\beta}^{\pm}; \newh_{0}^{-\12}\newcH)$
and is independent on the choice of the cutoff function $\chi$.  
\item We have 
\[
\newK_{\beta}^{-1}\newgamma_{\beta}^{*}(\newS^{(0)}\newf^{(0)}\oplus \newS^{(\b2)}\newf^{(\b2)})\in \overline{C^{1}}(I_{\beta}^{\pm};\newh_{0}^{-\12}\newcH).
\]
\item 
 \[
 \mp\newgamma_{\beta}^{\pm}\newK_{\beta}^{-1}\newgamma_{\beta}^{*}(\newS^{(0)}\newf^{(0)}\oplus \newS^{(\b2)}\newf^{(\b2)})= (\one \oplus T)\circ C_{\beta}^{\pm}\circ (\one \oplus T)^{-1}\newf,
 \]
\een
where $C_{\beta}^{\pm}$ are the \Calderon projectors for $B_{\beta}= \p_{ s}+ b$,  with $b= \newH$,  $\ch= \newcE$, defined in  Prop. \ref{prop4.2}.
\end{proposition}

Again this leads to the following definition.
\begin{definition}\label{def9.2}
 The operators
\[
\newc^{\pm}_{\beta}\defeq \mp \newgamma_{\beta}^{\pm}\newK_{\beta}^{-1} \newgamma_{\beta}^{*}(\newS^{(0)}\pi^{(0)}\oplus \newS^{(\b2)}\pi^{(\b2)}) \in L((\newcD\oplus\newcD)^{2}; \newcE\oplus\newcE),
\]
where $\pi^{(0/\b2)}\newg= \newg^{(0/\b2)}$  are called  {\em \Calderon projectors} for $\newK_{\beta}$.
\end{definition}
{\bf Proof of Prop. \ref{prop9.2}.}  We prove only the $+$ case. Let $\newf\in \newcD\oplus \newcD$. Then:
\[
\begin{array}{l}
\newgamma^{(0)*}_{\beta,n}\newS^{(0)}\newf= \chi_{n}'(s)\otimes\newf_{0}+ \chi_{n}(s)\otimes (\i \neww^{*}\newf_{0}- \newf_{1}),\\[2mm]
\newgamma^{(\b2)*}_{\beta,n}\newS^{(\b2)}\newf= - \chi_{n}'(s-\b2)\otimes \newf_{0}- \chi_{n}(s- \b2)\otimes( \newf_{1}+ \i \neww^{*}\newf_{0}),
\end{array}
\]
and as in the proof of Prop. \ref{prop9.1} we see that  $\newgamma^{(0)*}_{\beta,n}\newS^{(0)}\newf$ and $\newgamma^{(\b2)*}_{\beta,n}\newS^{(\b2)}\newf$ belong to $(- \p_{s}^{2}+ \newh_{0})^{\12}\newcH_{\beta}$. Therefore we can apply Prop. \ref{prop8.1} and obtain
\beq\label{sloubi2}
\newK_{\beta}^{-1}\newgamma_{\beta, n}^{(0/\b2)*}\newS^{(0/\b2)}\newf= \pi_{0}B_{\beta}^{-1}\pi_{1}^{*}\newgamma_{\beta, n}^{(0/\b2)*}\newS^{(0/\b2)}\newf,
\eeq
where $B_{\beta}= (\p_{s}+ \newH)$, acting on $\newcE_{\beta}$.  If $u\in \newcD$ then $\pi_{1}^{*}u\in \Dom \newH\cap \Dom \newH^{-1}$ by Prop. \ref{prop-correct.1}.
Therefore as in the proof of Prop. \ref{prop4.1} we obtain for $\newg= \pi_{1}^{*}u$ and $u\in \newcD$:
\[
\begin{array}{l}
\lim_{n\to \infty}B_{\beta}^{-1}(\chi_{n}\otimes \newg)(s)= \e^{-s\newH}(\one - \e^{- \beta \newH})^{-1}\newg, \\[2mm]
\lim_{n\to \infty}B_{\beta}^{-1}(\chi_{n}'\otimes \newg)(s)= -\newH\e^{-s\newH}(\one - \e^{- \beta \newH})^{-1}\newg, \\[2mm]
\lim_{n\to \infty}B_{\beta}^{-1}(\chi_{n}(\cdot-\b2)\otimes \newg)(s)= -\e^{-(s-\b2)\newH}(\one - \e^{ \beta \newH})^{-1}\newg, \\[2mm]
\lim_{n\to \infty}B_{\beta}^{-1}(\chi_{n}'(\cdot-\b2)\otimes \newg)(s)= \newH\e^{-(s-\b2)\newH}(\one - \e^{ \beta \newH})^{-1}\newg, \hbox{ in }\cD'(I^{+}_{\beta}; \newcE).
\end{array}
\]
An easy computation shows that for $\newf\in \newcD\oplus\newcD$ we have:
\[
\begin{array}{l}
\lim_{n\to +\infty}- (B_{\beta}^{-1}\pi_{1}^{*}\newgamma^{(0)*}_{\beta,n}\newS^{(0)}\newf)(s)= \e^{-s\newH}(1- \e^{- \beta \newH})^{-1}\newf\eqdef F^{(0)}(s),\\[2mm]
\lim_{n\to +\infty}- (B_{\beta}^{-1}\pi_{1}^{*}\newgamma^{(\b2)*}_{\beta,n}\newS^{(\b2)}\newf)(s)= \e^{-(s-\b2)\newH}(1- \e^{\beta H})^{-1}T\newf\eqdef F^{(\b2)}(s),\hbox{ in }\cD'(I^{+}_{\beta}; \newcE),
\end{array}
\]
which using \eqref{sloubi2} proves (1). 

 Next since $\newcD\oplus \newcD\subset \Dom \newH^{-1}$, we have $(1- \e^{\pm \beta \newH})^{-1}\newf\in \newcE$, and since $\newH$ preserves $\newcD\oplus \newcD$, we have also $\newH(1- \e^{\pm \beta \newH})^{-1}\newf\in \newcE$. Therefore $F^{(0)}, F^{(\b2)}\in \overline{C^{1}}(I^{+}_{\beta}; \newcE)$ which shows that 
 \[
 \pi_{0}F^{(0/\b2)}(s)\in\overline{C^{1}}(I_{\beta}^{+}; \newh_{0}^{-\12}\newcH)
  \] and proves (2).   Finally we check that for $F(s)\in  \overline{C^{1}}(I^{+}_{\beta};\newcE)$ with $(\p_{s}+\newH) F(s)=0$ in $I_{\beta}^{+}$ we have:
 \[
 \newgamma_{\beta}^{(0)+}\pi_{0}F= \Gamma_{\beta}^{(0)+}F, \ \newgamma_{\beta}^{(\b2)+}\pi_{0}F= T \Gamma_{\beta}^{(\b2)+}F,
 \]
 which proves (3).   \qed

As in Prop. \ref{prop9.1b} we can using Subsect. \ref{sec4.7}  express the covariances of the double $\beta$-KMS state $\newomega_{\rm d}$ for $\newP$ in terms of the \Calderon projectors $\newc_{\beta}^{\pm}$ for the Wick rotated operator $\newK_{\beta}$.

We recall that from Prop. \ref{prop-correct.1} we know that  the double $\beta$-KMS state $\newomega_{\rm d}$ is  well defined on $\CCR(\newcY\oplus \newcY, \newq\oplus-\newq)$.
\begin{proposition}\label{prop9.2b}
 The covariances  of the double $\beta$-KMS state for $\newP$, considered as a state on $\CCR(\newcY\oplus \newcY, \newq\oplus-\newq)$ are equal to
 \[
\newlambda_{\rm d}^{\pm}= \pm \newQ\circ (\one\oplus T)^{-1}\newc_{\beta}^{\pm}(\one\oplus T), \hbox{ for }\newQ= \newq\oplus -\newq.
\]
\end{proposition}

\section{ Klein-Gordon equations on stationary spacetimes}\label{sec12}\init
In this section we consider Klein-Gordon equations on stationary spacetimes. If the lapse function $N$ associated to the Killing vector field $\rw$ is equal to $1$, one can directly reduce oneself to the situation of Sect. \ref{sec9}.   In general one has to replace the Klein-Gordon operator $P$ by $\tilde{P}= N PN$, which has the same purpose as a conformal transformation.

As an application  we  consider the Klein-Gordon operator $P$ in $\cM^{+}$ and express the covariances of the double $\beta$-KMS state in $\cM^{-}\cup \cM^{+}$ using the \Calderon projectors for the elliptic operator $K_{\beta}$ obtained from $P$ by Wick rotation in the Killing time coordinate $t$.
\subsection{Klein-Gordon equations on stationary spacetimes}\label{sec12.1}

\subsubsection{Stationary metrics}
Let $(S, \rh)$  be a Riemannian manifold, $N\in \cinf(S)$, $N>0$ and $\rw^{i}$ a vector field on $S$. Let us denote by $y$ the elements of $S$. We define the Lorentzian metric $\rg$ on $M = \rr\times S$:
\[
\begin{array}{rl}
\rg=& - N^{2}(y)dt^{2}+ \rh_{ij}(y)(dy^{i}+ \rw^{i}(y)dt)(dy^{j}+\rw^{j}(y)dt).
\end{array}
\]
We assume that $ \{0\}\times S$ is a Cauchy surface for $(M, \rg)$.  Such spacetimes are called {\em standard stationary spacetimes} in the terminology of \cite{S2}.

The vector field $\dfrac{\p}{\p t}$ is Killing for $\rg$ and is time-like iff
\begin{equation}
\label{e12.3}
N^{2}(y)>\rw^{i}(y)\rh_{ij}(y)\rw^{j}(y), \ y\in S.
\end{equation}
We will need later to impose the following stronger condition:
\begin{definition}\label{def12.1}
 The Killing vector field $\dfrac{\p}{\p t}$ is {\em uniformly time-like} if there exists $0<\delta<1$ such that:
 \[
(1- \delta)N^{2}(y)\geq \rw^{i}(y)\rh_{ij}(y)\rw^{j}(y), \ x\in S.
\]
 \end{definition}
 We have:
 \beq\label{e12.01d}
 |\rg|= N^{2}|\rh|, n= N^{-1}(\dfrac{\p}{\p t}- w),
\eeq
where $n$ is the  future directed unit normal to  the foliation $S_{t}= \{t\}\times S$.

\subsubsection{Stationary Klein-Gordon operators}
We consider a stationary Klein-Gordon operator on $(M, \rg)$:
\beq\label{e12.01}
P= -\Box_{\rg}+ m(y), \ m\in \cinf(S;\rr).
\eeq
 We will always assume that
\beq\label{e12.01a}
m(y)\geq m_{0}^{2}, \ m_{0}>0,
\eeq
ie that the Klein-Gordon equation is {\em massive}. Setting
\beq\label{e12.01c}
h_{0}\defeq \nabla^{*} \rh^{-1}\nabla + m, \ w\defeq \rw^{i}\dual \p_{y^{i}},
\eeq
we have
\begin{equation}
\label{e12.01b}
P= (\p_{t}+w^{*})N^{-2}(\p_{t}-w)+ h_{0},
\end{equation}
where in \eqref{e12.01c}, \eqref{e12.01b} the adjoints are computed with respect to the scalar product
\[
(u|v)_{M}= \int_{M}\bar{u}v N|\rh|^{\12}dtdy.
\]
\subsubsection{Hilbert spaces}
We denote by $L^{2}(M)$ the Hilbert space associated to the scalar product $(\cdot| \cdot)_{M}$ and by $\cH= L^{2}(S, |\rh|^{\12}dy)$ the Hilbert space associated to the scalar product
\[
(u| v)_{\cH}= \int_{S}\bar{u}v|\rh|^{\12}dy.
\]
We will also need the Hilbert space $\newcH=L^{2}(S, N|\rh|^{\12}dy)$ associated to the scalar product
\[
(u| v)_{\newcH}= \int_{S}\bar{u}v N |\rh|^{\12}dy,
\]
so that $L^{2}(M)= L^{2}(\rr, dt; \newcH)$.

\subsubsection{An  operator inequality}
 The inequality in Lemma \ref{prop1.1} below is  understood as an operator inequality  on $\newcH$.
 \begin{lemma}\label{prop1.1} 
 Assume that $\dfrac{\p}{\p t}$ is  uniformly time-like. Then 
 \[
(1- \delta)h_{0}\geq w^{*}N^{-2}w\hbox{ on }\coinf(S).
\]
\end{lemma}
\proof  Let   $\cX$   be a real vector space, $\rk\in L_{\rm s}(\cX, \cX^{\t})$ strictly positive and  $c\in \cX$. Then for $\gamma= \rk c\in \cX^{\t}$ and $\xi\in \cc \cX^{\t}$ we have
\[
\begin{array}{rl}
&(\bxi- \langle\bxi| c\rangle \gamma)\dito \rk^{-1}(\xi- \langle\xi| c\rangle \gamma)\\[2mm]
=& \bxi\dito \rk^{-1}\xi- 2 {\rm Re}\,(\langle\bxi| c\rangle \gamma\dito \rk^{-1}\xi)+ |\langle\xi| c\rangle|^{2}\gamma\dito \rk^{-1}\gamma\\[2mm]
=& \bxi\dito \rk^{-1}\xi- (2- c\dito \rk c) |\langle\xi| c\rangle|^{2},
\end{array}
\]
hence
\[
\rk^{-1}- | c\rangle\langle c|\geq (1- c\dito \rk c)|c\rangle \langle c|.
\]
Replacing $\rk$ by $(1- \delta)^{-1}\rk$ shows that if $(1- \delta)\geq c\dual \rk c$ we have
\begin{equation}
\label{e12.1}
(1-\delta)\rk^{-1}\geq |c\rangle\langle c|.
\end{equation}
For $u\in \coinf(S)$ we write
\[
\begin{array}{rl}
&(u| ((1- \delta)h_{0}- w^{*}N^{-2}w)u)_{\newcH}\\[2mm]
\geq& \int_{S}\p_{y^{i}}\bar{u}((1- \delta)\rh^{ij}(y)- \rw^{i}(y)N^{-2}\rw^{j}(y))\p_{y^{j}}u(y) N|\rh|^{\12}dy.
\end{array}
\]
Applying \eqref{e12.1} under the integral sign
 for  $\rk= \rh_{ij}(y)$,  $c= N^{-1}(y)\rw^{i}(y)$  we obtain the lemma. \qed
 
 \subsection{Selfadjoint operators}
In the rest of this section we will assume  that $\frac{\p}{\p t}$ is uniformly time-like.
 
Let $q_{0}(u, u)=(u |h_{0}u)_{\newcH}$  with $\Dom q_{0}= \coinf(S)$. The form $q_{0}$ is closeable and we denote  still denote by $h_{0}$ the selfadjoint operator on $\newcH$ associated to $q_{0}^{\rm cl}$, i.e.  the Friedrichs extension of $h_{0}$ on $\coinf(S)$. We have:
\[
h_{0}: h_{0}^{-\12}\newcH\tosim h_{0}^{\12}\newcH.
\]
Note that $h_{0}^{-\12}\newcH\subset \newcH$ since $h_{0}\geq m_{0}^{2}$.
 We set  also
 \beq\label{e12.40}
\tilde{q}_{0}(u, u)= q_{0}(Nu, Nu)_{\newcH}, \ \Dom \tilde{q}_{0}= \coinf(S),
\eeq
and denote by $\newh_{0}$ the selfadjoint operator on $\newcH$ associated to $\tilde{q}_{0}^{\rm cl}$, which  formally equals $N h_{0}N$.  From \eqref{e12.40} we obtain that
\beq\label{e12.40a}
N: \newh_{0}^{-\12}\newcH\tosim h_{0}^{-\12}\newcH, \ N: h_{0}^{\12}\newcH\tosim \newh_{0}^{\12}\newcH,
\eeq
and we have:
\begin{equation}
\label{e12.41}
\newh_{0}= N h_{0}N\hbox{ as an identity in }B(\newh_{0}^{-\12}\newcH, \newh_{0}^{\12}\newcH).
\end{equation}
 We also set
 \begin{equation}
\label{e12.42}
\begin{array}{l}
\neww= N^{-1}wN=  N^{-1}\rw^{i}\dual\p_{y^{i}} N, \\[2mm]
\neww^{*}= N w^{*}N^{-1}= - |\rh|^{-\12}\p_{y^{i}}\dual \rw^{i} |\rh|^{\12},
\end{array}
\end{equation}
with domain $\coinf(S)$.

Let us introduce the assumption
\begin{equation}
\label{e12.42:2}
 N^{-2}\rw^{i}\dual(\nabla_{i}^{\rh} N), \  N^{-1}\nabla_{i}^{\rh}\rw^{i}\hbox{ are bounded on }S.
\end{equation}
\begin{lemma}\label{lemma12.1}
 Assume  \eqref{e12.42:2}. Then $\newh_{0}, \neww, \neww^{*}$ satisfy the conditions 
 \eqref{e9.1b}  and \eqref{ecorrect.5} for $\newcD= \coinf(S)$.
\end{lemma}
\proof
We have seen in Lemma \ref{prop1.1} that $w^{*}N^{-2}w\leq (1- \delta)h_{0}$ on $\coinf(S)$, which implies $\neww^{*}\neww\leq (1- \delta)\newh_{0}$ on $\coinf(S)$. Let $(y^{1}, \dots, y^{d})$ be local coordinates on $S$. We have
\[
\begin{array}{l}
\neww= \rw^{i}\dual\p_{y^{i}}+ N^{-1}\rw^{i}\dual(\p_{y^{i}}N),\\[2mm]
\neww^{*}= - \rw^{i}\dual \p_{y^{i}} - |\rh|^{-\12}(\p_{y^{i}}\rw^{i} |\rh|^{\12})= - \rw^{i}\dual \p_{y^{i}}- \nabla_{i}^{\rh}\rw^{i}.
\end{array}
\]
Condition \eqref{e12.42:2} implies that $\neww^{*}= - \neww+ r$, where $r\in \cinf(S)$ and  $r N^{-1}$ is bounded on $\newcH$. The inequality $\neww\neww^{*}\leq C \newh_{0}$ follows from $\neww^{*}\neww\leq (1- \delta)\newh_{0}$ and $m_{0}^{2}N^{2}\leq  \newh_{0}$. 

Applying the Kato-Heinz theorem to this last inequality we obtain that $\newh_{0}^{-1}\leq m_{0}^{-2}N^{-2}$, i.e.  $\Dom N^{-1}\subset \Dom \newh_{0}^{-\12}$. Since  $N>0$, $\coinf(S)\subset \Dom N^{-1}$, which proves \eqref{ecorrect.5}.\qed

\subsection{Associated first order system}\label{sec1b.3}
We set:
\beq\label{e12.44}
\varrho_{t}\phi= \col{\phi(t)}{\i^{-1}N^{-1}(\p_{t}-w)\phi(t)}= f= \col{f_{0}}{f_{1}}, 
\eeq
and rewrite $P\phi=0$ as:
\beq\label{en1.1}
N^{-1}\p_{t}f= \i H f, H= \mat{-\i N^{-1}w}{\one}{h_{0}}{\i w^{*}N^{-1}}, \ f\in \coinf(S; \cc^{2}).
\eeq
The conserved energy is
\beq\label{en1.zloto}
\overline{f}\dual Ef= \|f_{1}- \i N^{-1}wf_{0}\|_{\newcH}^{2}+ (f_{0}| hf_{0})_{\newcH}, \ h= h_{0}- w^{*}N^{-2}w,
\eeq
and the conserved charge is
\[
\overline{f}\dual qf= (f_{1}|f_{0})_{\cH}+ (f_{0}| f_{1})_{\cH}.
\]
The energy space $\cE$ associated to $E$ equals $h_{0}^{-\12}\newcH\oplus \newcH$ as topological vector spaces.

\subsection{Reduction}\label{sec12.2}
We now introduce the Klein-Gordon operator
\[
\newP= NPN= (\p_{t}+ \neww^{*})(\p_{t}- \neww)+ \newh_{0},
\]
which is of the form considered in Sects. \ref{sec8}, \ref{sec9}. The operators $\newrho_{t}, \newH$, the energy $\newE$ and charge $\newq$ are defined as  in Subsect. \ref{sec8.4}:
\[
 \newrho_{t}\newphi=\col{\newphi(t)}{\i^{-1}(\p_{t}- \neww)\newphi(t)},
\ \newH= \mat{- \i \neww}{\one}{\newh_{0}}{\i \neww^{*}},
\]
\[
\begin{array}{l}
\overline{\tf}\dual\newE \tf= \|\tf_{1}- \i \neww \tf_{0}\|^{2}_{\newcH}+ (\tf_{0}| \newh\tf_{0})_{\newcH},\\[2mm]
\overline{\tf}\dual\newq \tf= (\tf_{1}| \tf_{0})_{\newcH}+ (\tf_{0}| \tf_{1})_{\newcH}, \ \tf\in \coinf(S; \cc^{2}).
\end{array}
\]
Setting
\beq\label{e12.3:3}
Z\eqdef \mat{N}{0}{0}{\one}, \ Z'\defeq\mat{\one}{0}{0}{N^{-1}},
\eeq
we have: 
\begin{equation}
\label{e12.4}
\varrho_{t}N= Z\newrho_{t}, \ N^{-1}\p_{t}- \i H= Z'(\p_{t}- \i \newH)Z^{-1},
\end{equation}
\begin{equation}
\label{e12.5}
Z^{*}EZ = \newE, \ Z^{*}qZ= \newq\hbox{ on }\coinf(S; \cc^{2}).
\end{equation}
We saw that the energy space $\newcE$ associated to $\newE$ equals $\newh_{0}^{-\12}\newcH\oplus \newcH$, and  from \eqref{e12.40a} we obtain that:
\beq\label{sloubi6}
Z: \newcE\tosim \cE.
\eeq
\subsection{Vacuum and KMS states}\label{sec12.4}
In Subsect. \ref{sec9.1} we defined  vacuum and $\beta-$KMS states for $\tilde{P}$. We obtain the corresponding 
 vacuum and $\beta-$KMS states  for $P$   by  conjugation by the map $Z$. 
 
In Def. \ref{def12.1:1} below, $\omega_{\rm vac}$ and  $\omega_{\beta}$ are  states on $\CCR(\coinf(S; \cc^{2}), \newq)$, while  $\omega_{\rm d}$ is a state on $\CCR(\coinf(S; \cc^{2})\oplus \coinf(S; \cc^{2})^{2}, \newq\oplus - \newq)$.
\begin{definition}\label{def12.1:1}
   We define the {\em vacuum state} $\omega_{\rm vac}$, the $\beta${\em -KMS  state} $\omega_{\beta}$ and the {\em double} $\beta${\em -KMS state} $\omega_{\rm d}$,  by their covariances:
\[
\begin{array}{l}
\lambda^{\pm}_{\rm vac}= (Z^{-1})^{*}\newlambda^{\pm}_{\rm vac}Z^{-1}, \ \lambda^{\pm}_{\beta}= (Z^{-1})^{*}\newlambda^{\pm}_{\beta}Z^{-1}, \\[2mm]
\lambda^{\pm}_{\rm d}= (Z^{-1}\oplus Z^{-1})^{*}\newlambda^{\pm}_{\rm d}(Z^{-1}\oplus Z^{-1}),
\end{array}
\] 
where the covariances $\newlambda^{\pm}_{\rm vac}$, $\newlambda^{\pm}_{\beta}$ and  $\newlambda^{\pm}_{\rm d}$ are defined in Defs. \ref{def3.01}, \ref{def3.02}  and Prop. \ref{prop4.1:1} for $b= \newH$.
\end{definition}
\subsection{The Wick rotated operator}\label{sec12.5}
\subsubsection{The Wick rotated metric}\label{sec12.5.1}
Let us denote by $\rg^{\rm eucl}$ the complex metric on $\rr\times S$ obtained  from $\rg$ by the substitution $t= \i  s$. We have:
\[
\rg^{\rm eucl}= N^{2}(y)d s^{2} +\rh_{ij}(y)(dy^{i}+ \i \rw^{i}(y)d s)(dy^{j}+ \i \rw^{j}(y)d s),
\]
Using that $\dfrac{\p}{\p t}$ is  uniformly time-like we obtain that  there exists $C>0$ such that
\begin{equation}
\label{e12.550}
|{\rm Im }(\bar{\eta}\dual \rg^{\rm eucl}(y)\eta)|\leq C{\rm Re}( \bar{\eta}\dual \rg^{\rm eucl}(y)\eta), \ y\in S, \eta\in \cc T_{(s,y)}(\rr\times S),
\end{equation}
 Moreover  we have
 \begin{equation}
\label{e12.551}
|\rg^{\rm eucl}|(y)\hbox{ is real valued, positive, and } |\rg^{\rm eucl}|^{\12}(y)= N(y)|\rh|^{\12}(y).
\end{equation}
With the terminology in Def. \ref{def13.0}  this means that the complex metric $\rg^{\rm eucl}$ is {\em uniformly sectorial}. 

 If $\Omega= ]0, +\infty[\times S$, the outer unit normal vector field  to $\Omega$ for $\rg^{\rm eucl}$, see \ref{oun},    is
\beq\label{e12.50}
\nu= - N^{-1}(\frac{\p}{\p s}- \i \rw),
\eeq
while if $\Omega= ]0, \b2[\times S$ it equals
\beq\label{e12.51}
\nu^{(0/  \b2)}=  \mp N^{-1}(\frac{\p}{\p s}- \i \rw)\hbox{ on }\{0/\b2\}\times S.
\eeq
The real vectors ${\rm Im \nu}$,  ${\rm Im}\,\nu^{(0/\b2)}$ are tangent to $S$, i.e.  condition \eqref{e13.2} below  is satisfied.
\subsubsection{The Wick rotated operator}\label{defilo}
We  consider now the Wick rotated operator $K$ obtained  from $P$ by the substitution $t= \i  s$. We have:
\begin{equation}
\label{e7.1}
K=- \Delta_{\rg^{\rm eucl}}+ m(y) =-(\p_{ s}+ \i w^{*})N^{-2}(\p_{ s}- \i w)+ h_{0},
\end{equation}
acting on the Hilbert spaces $\newcH_{\beta}$ for  $0<\beta\leq \infty$ defined in \eqref{e8.10a}.    We refer the reader to  \ref{sec14.1.2} for the Laplacian $\Delta_{\rg^{\rm eucl}}$ associated to $\rg^{\rm eucl}$. We recall that
\[
\begin{array}{l}
\newcH_{\beta}= L^{2}(\bS_{\beta}\times S, N(y)|\rh|^{\12}(y)dyds), \hbox{ for }0<\beta<\infty, \\[2mm]
\newcH_{\infty}=  L^{2}(\rr\times S, N(y)|\rh|^{\12}(y)dyds).
\end{array}
\]
It follows from Lemma \ref{prop1.1} that  if $h= h_{0}- w^{*}N^{-2}w$ we have:
\[
h\sim h_{0},  \ w^{*}N^{-2}w\lesssim h, \hbox{ on }\coinf(\rr\times S),
\]
where we use the scalar product of $\newcH_{\beta}$ in the operator inequalities.
We have:
\begin{equation}
\label{e7.1a}
\begin{array}{rl}
(u| Ku)_{\newcH_{\beta}}=& \|N^{-1}\p_{ s}u\|^{2}_{\newcH_{\beta}}+ (u| h u)_{\newcH_{\beta}}\\[2mm]
&- \i (N^{-1}\p_{ s}u| N^{-1}w u)_{\newcH_{\beta}}- \i (N^{-1}w u| N^{-1}\p_{ s}u)_{\newcH_{\beta}},
\end{array}
\end{equation}
for $u\in \coinf(\rr\times S)$ if $\beta= \infty$ and $u\in\coinf(\bS_{\beta}\times S)$ if $\beta<\infty$.

The sesquilinear form associated to  the realization $K_{\beta}$  of $K$ is
\[
\begin{array}{rl}
Q_{\beta}(u, u)=& \|N^{-1}\p_{ s}u\|^{2}_{\newcH_{\beta}}+ (u| h u)_{\newcH_{\beta}}\\[2mm]
&- \i (N^{-1}\p_{ s}u| N^{-1}w u)_{\newcH_{\beta}}- \i (N^{-1}w u| N^{-1}\p_{ s}u)_{\newcH_{\beta}},
\end{array}
\]
with domain $\Dom Q_{\beta}= \langle K_{0}\rangle^{- \12}\newcH_{\beta}$ and $K_{0}= - N^{-2}\p_{ s}^{2}+ h_{0}$ with its natural domain on $\newcH_{\beta}$.  From \eqref{e12.40a} we obtain that
\[
N^{-1}: \Dom Q_{\beta}\tosim \Dom  \newQ_{\beta}, \ Q_{\beta}(Nu, Nu)=  \newQ_{\beta}(u, u),
\]
where $ \newQ_{\beta}$ is defined in Subsect. \ref{sec8.2}.  It follows that $Q_{\beta}$ is  a closed sectorial form and we denote as before by $K_{\beta}$:
\[
K_{\beta}: K_{0}^{-\12}\newcH_{\beta}\tosim K_{0}^{\12}\newcH_{\beta}
\]
the induced boundedly invertible operator. We have
\beq\label{sloubi5}
\begin{array}{rl}
i)&N: \newK_{0}^{-\12}\newcH_{\beta}\tosim K_{0}^{-\12}\newcH_{\beta}, \ N: K_{0}^{\12}\newcH_{\beta}\tosim  \newK_{0}^{\12}\newcH_{\beta}\\[2mm]
 ii)&\newK_{\beta}= N K_{\beta}N, \hbox{ as elements of }B( \newK_{0}^{-\12}\newcH_{\beta},  \newK_{0}^{\12}\newcH_{\beta}),
\end{array}
\eeq
where $ \newK_{\beta}$ is the operator defined in Subsect. \ref{sec8.2}. 

\subsection{\Calderon projectors}\label{sec12.6}
We now define  the \Calderon projectors for $K_{\beta}$ and relate them to those for $\newK_{\beta}$ defined in Subsect. \ref{sec9.2}.  We use the notation $I^{\pm}_{\beta}$, $i^{\pm}_{\beta}$, $\newgamma_{\beta}$, $\newgamma^{\pm}_{\beta}$ introduced in Subsect. \ref{sec9.2}. We recall that  $\chi_{n}\in \coinf(\rr)$ is a sequence of cutoff functions converging to $\delta_{0}$.

\subsubsection{\Calderon projectors for $K_{\infty}$}\label{sec12.6.1}
We set
\[
\gamma^{*}_{\infty}g\defeq \delta_{0}'(s)\otimes g_{1} +\delta_{0}(s)\otimes (N^{-1}g_{0}- \i w^{*}N^{-2}g_{1})\ g\in \coinf(S; \cc^{2}),
\]
and
\[
S\defeq \mat{2\i N w^{*}N^{-2}}{-\one}{\one}{0},
\]
so that
\beq\label{sloubi3}
N^{-1}\newgamma_{\infty}^{*}\newS Z^{-1}= \gamma_{\infty}^{*}S.
\eeq
If $N^{-1}u\in  \overline{C^{1}}(I^{\pm}_{\infty}; \newcH)\cap \overline{C^{0}}(I^{\pm}_{\infty}; \newh_{0}^{-\12}\newcH)$ and $Ku=0$ in $I_{\infty}^{\pm}$ we set:
\[
\gamma_{\infty}^{\pm}u = \col{u(0^{\pm})}{-N^{-1}(\p_{ s}- \i w)u(0^{\pm})},
\]
so that
\beq\label{sloubi4}
 \gamma^{\pm}_{\infty}N= Z\newgamma^{\pm}_{\infty}.
\eeq
From Prop. \ref{prop9.1} and \eqref{sloubi6},  \eqref{sloubi5} {\it ii)},  \eqref{sloubi3} and  \eqref{sloubi4}, we see that  the operators
\beq\label{sloubi7}
\mp \gamma_{\infty}^{\pm} K_{\infty}^{-1}\gamma_{\infty}^{*} S= \mp Z\circ (\newgamma^{\pm}_{\infty}\newK_{\infty}^{-1}\newgamma_{\infty}^{*}\newS) \circ Z^{-1}
\eeq
are well defined as linear operators from $\coinf(S; \cc^{2})$ to $\cE$.  
This leads to the following definition.
\begin{definition}\label{def12.2}
 The {\em \Calderon projectors} $c_{\infty}^{\pm}$ for $K_{\infty}$ are:
 \[
c_{\infty}^{\pm}\defeq \mp \gamma_{\infty}^{\pm} K_{\infty}^{-1}\gamma_{\infty}^{*} S\in L(\coinf(S; \cc^{2}); \cE).
\]
\end{definition}
\begin{proposition}\label{prop12.1}
  The covariances of the vacuum state $\omega_{\rm vac}$  are equal to:
  \[
\lambda_{\rm vac}^{\pm}= \pm q\circ c_{\infty}^{\pm}.
\]
\end{proposition}
\proof This follows from the identities:
\[
\begin{array}{rl}
i)&\tilde{\lambda}_{\rm vac}^{\pm}= Z^{*}\lambda^{\pm}_{\rm vac}Z, \ \tilde{q}= Z^{*}qZ, \\[2mm]
ii)& \newlambda^{\pm}_{\rm vac}= \pm \newq \tilde{c}_{\infty}^{\pm}, \ c_{\infty}^{\pm}= Z \tilde{c}_{\infty}^{\pm}Z^{-1}.
\end{array}
\]
The identities in {\it i)} are obvious, the first identity in {\it ii)} is shown in Prop. \ref{prop9.1b}, the second follows from \eqref{sloubi7}. \qed

\subsubsection{\Calderon projectors for $K_{\beta}$}\label{sec12.6.1:1}
We set
\[
\begin{array}{l}
 \gamma_{\beta}^{(0)*}g=\delta_{0}'( s)\otimes N^{-2}g_{1}+ \delta_{0}( s)\otimes (N^{-1}g_{0}- \i w^{*}N^{-2}g_{1}),\\[2mm]
  \gamma_{\beta}^{(\b2)*}g= -\delta'_{\b2}( s)\otimes N^{-2}g_{1}+ \delta_{\b2}( s)\otimes (N^{-1}g_{0}+ \i w^{*}N^{-2}g_{1}), \ g\in \coinf(S; \cc^{2})
\end{array}
\]
\[
S^{(0)}\defeq \mat{2\i N w^{*}N^{-2}}{-\one}{\one}{0}, \ S^{(\b2)}\defeq \mat{-2\i N w^{*}N^{-2}}{-\one}{\one}{0}.
\]
\[
\gamma^{*}_{\beta}(g^{(0)}\oplus g^{(\b2)})\defeq \gamma_{\beta}^{(0)*}g^{(0)}+ \gamma_{\beta}^{(\b2)*}g^{(\b2)}.
\]
 As before we have:
\begin{equation}
\label{sloubi8}
N^{-1}\newgamma_{\beta}^{(0/\b2)*}\newS^{(0/\b2)}Z^{-1}= \gamma_{\beta}^{(0/\b2*)}S^{(0/\b2)}.
\end{equation}
If $N^{-1}u\in \overline{C^{0}}(I_{\beta}^{\pm}; \newh^{-1}\newcH)$ and $\p_{s}N^{-1}u\in \overline{C^{0}}(I_{\beta}^{\pm}; \newh_{0}^{-\12}\newcH)$ with $Ku= 0$ in $I_{\beta}^{+}$, we set
 \[
\gamma^{\pm}_{\beta}u= \gamma_{\beta}^{(0)\pm}u\oplus \gamma_{\beta}^{(\b2)\pm}u,
\]
for
\[
\gamma^{(0)\pm}_{\beta}u= \col{u(0^{\pm})}{-N^{-1}(\p_{ s}- \i w )u(0^{\pm})}, \ \gamma_{\beta}^{(\b2)\pm}u= \col{u(\mp\b2)}{N^{-1}(\p_{ s}- \i w )u(\mp\b2)},
\]
and we have:
\beq\label{sloubi9}
\gamma^{\pm}_{\beta}N= (Z\oplus Z)\newgamma_{\beta}^{\pm}.
\eeq
Again from Prop. \ref{prop9.2} and  \eqref{sloubi6},  \eqref{sloubi5} {\it ii)},  \eqref{sloubi8} and  \eqref{sloubi9}, we see that  the operators
\beq\label{sloubi10}
\begin{array}{rl}
&\pm \gamma_{\beta}^{\pm} K_{\beta}^{-1}\gamma_{\beta}^{*} (\gamma_{\beta}^{(0)*} S^{(0)}\pi^{(0)}+ \gamma_{\beta}^{(\b2)*}S^{(\b2)}\pi^{(\b2)})\\[2mm]
=& \pm (Z\oplus Z)\circ (\newgamma^{\pm}_{\beta}\newK_{\beta}^{-1}\newgamma_{\beta}^{(0)*} \newS^{(0)}\pi^{(0)}+ \newgamma_{\beta}^{(\b2)*}\newS^{(\b2)}\pi^{(\b2)}) \circ (Z\oplus Z)^{-1}
\end{array}
\eeq
are well defined as linear operators from $\coinf(S; \cc^{2})\oplus \coinf(S; \cc^{2})$ to $\cE\oplus \cE$.

Again this leads to the following definition.
\begin{definition}\label{def12.3}
 The {\em \Calderon projectors} $c_{\beta}^{\pm}$ for $K_{\beta}$ are:
 \[
c_{\beta}^{\pm}\defeq\mp\gamma_{\beta}^{\pm}K_{\beta}^{-1}(\gamma_{\beta}^{(0)*} S^{(0)}\pi^{(0)}+ \gamma_{\beta}^{(\b2)*}S^{(\b2)}\pi^{(\b2)}),
\]
where $\pi^{(0/\b2)}g= g^{(0/\b2)}$.
\end{definition}
Using now Prop. \ref{prop9.2b}, the same argument as in \ref{sec12.6.1} gives the following proposition. 
\begin{proposition}\label{prop12.2}
  The covariances of the double $\beta$-KMS state $\omega_{\rm d}$ are equal to:
  \[
\lambda_{\rm d}^{\pm}= \pm Q\circ(\one\oplus T)^{-1} c_{\beta}^{\pm}(\one\oplus T), \ Q= q\oplus q,
\]
where $T= \mat{\one}{0}{0}{-\one}$.
\end{proposition}

\subsection{The double $\beta$-KMS state in $\cM^{+}\cup \cM^{-}$}\init\label{sec12.7}
We now apply the computations of the previous subsections to $S= \Sigma^{+}$.   In fact if $\phi_{t}$ is the flow of the Killing vector field $V$, the map
\[
\chi: \rr\times \Sigma^{+}\ni (t, y)\mapsto \phi_{t}(y)\in \cM^{+}
\] 
is a diffeomorphism such that $\chi^{*}\rg$ is as in Subsect. \ref{sec12.1}.

We first claim  that  $\frac{\p}{\p t}$ is uniformly time-like
 and \eqref{e12.42:2} holds on $\Sigma^{+}$.  
 
\begin{proposition}\label{propoto}
 Assume that hypothesis {\rm (H3)} holds. Then $\frac{\p}{\p t}$ is uniformly time-like
 and \eqref{e12.42:2} holds on $\Sigma^{+}$.  
\end{proposition}
\proof By {\rm (H3)} $\frac{\p}{\p t}$ is uniformly time-like
 and   \eqref{e12.42:2} holds on $\Sigma^{+}\setminus U$, where $U$ is any small neighborhood of 
$\cB$ in $\Sigma^{+}$, by hypotheses (H).  
 To check the conditions on $U$, we use Prop. \ref{prop10.0}. Recalling that $(u, \omega)$ are Gaussian normal coordinates to $\cB$ in $(\Sigma, \rh)$, we obtain
 \[
\rw\dual \rh \rw\in O(u^{4}), \rw\dual (\nabla N)\in O(u^{3}), (\nabla\dual \rw)\in O(u^{2}), \rw\dual \nabla(|\rh|^{\12})\in O(u^{2}),
\] 
from which our claim follows, since $N(y)=\kappa u+ O(u^{3})$. \qed
 
 \subsubsection{The double $\beta$-KMS state in $\cM^{+}\cup \cM^{-}$}
  Let us now define the double $\beta$-KMS state in $\cM^{+}\cup \cM^{-}$.

  The wedge reflection $R$ is an isometric involution from $(\cM^{-}, \rg)$ to $(\cM^{+}, \rg)$. It induces  on $\Sigma$ the weak wedge reflection $r$, which equals the identity on $\cB$ and maps $\Sigma^{-}$ bijectively on $\Sigma^{+}$. 
  
$R$ reverses the time orientation, hence  induces a unitary  involution:
  \[
\cR: (\frac{\coinf(\cM^{-})}{P \coinf(\cM^{-})}, \i G)\in [u]\mapsto [u\circ R]\in   (\frac{\coinf(\cM^{+})}{P \coinf(\cM^{+})}, -\i G).
\]
In a more familiar language, $\cR$ is anti-symplectic. Since 
\[
\varrho_{\Sigma^{\pm}}\circ G: \left(\frac{\coinf(\cM^{\pm})}{P \coinf(\cM^{\pm})}, \i (\cdot | G\cdot)\right)\tosim (\coinf(\Sigma^{\pm}; \cc^{2}), q)
\]
is unitary, $\cR$  induces the unitary involution
\[
\cR_{\Sigma}: (\coinf(\Sigma^{-}; \cc^{2}), q)\tosim (\coinf(\Sigma^{+}; \cc^{2}), - q).
\]
 The following expression for $\cR_{\Sigma}$ follows from the fact that $R$  reverses the time orientation.
 
\begin{lemma}\label{lemma12.2}
 One has
 \[
\cR_{\Sigma}f=  T r^{*}f,
\]
where $T= \mat{\one}{0}{0}{-\one}$  and $r^{*}f(y)= f(r(y))$.
\end{lemma}
  We have defined in Subsect. \ref{sec12.4} the double $\beta$-KMS state $\omega_{\rm d}$ through its Cauchy surface covariances $\lambda^{\pm}_{\rm d}$. The associated Hermitian  space is 
  \[(\coinf(\Sigma^{+}; \cc^{2}), q)\oplus (\coinf(\Sigma^{+}; \cc^{2}), -q).
  \]  
  
   $(\cM^{+}\cup \cM^{-}, \rg)$ is a (disconnected) globally hyperbolic spacetime with Cauchy surface $\Sigma^{+}\cup \Sigma^{-}$ and we denote a Cauchy data on $\Sigma^{+}\cup \Sigma^{-}$ as
   \[
f= f^{+}\oplus f^{-}, \ f^{\pm}\in \coinf(\Sigma^{\pm}; \cc^{2}).
\]
  Using Remark \ref{remark4.1}, we obtain from $\omega_{\rm d}$ a pure, quasi-free state $\omega_{\rm D}$ in $\cM^{+}\cup \cM^{-}$ as follows:

\begin{definition}
The  double $\beta$-KMS state $\omega_{\rm D}$ in $\cM^{+}\cup \cM^{-}$  is defined by the Cauchy surface covariances:
\[
\overline{f}\dual \lambda^{\pm}_{\rm D}f\defeq \overline{(\one \oplus \cR_{\Sigma})f}\dual \lambda^{\pm}_{\rm d}(\one \oplus \cR_{\Sigma})f, \ f= f^{+}\oplus f^{-}\in \coinf(\Sigma^{+}\cup \Sigma^{-}; \cc^{2}).
\]
\end{definition}
 From Prop. \ref{prop12.2} and Lemma \ref{lemma12.2} we obtain the following expresssion for $\lambda^{\pm}_{\rm D}$.

\begin{proposition}\label{prop12.3}
 One has:
 \[
\lambda^{\pm}_{\rm D}= \pm Q\circ (\one \oplus r^{*})^{-1}c_{\beta}^{\pm}(\one \oplus r^{*}), 
\]
where $c_{\beta}^{\pm}$ are the \Calderon projectors for $K_{\beta}$ defined in Def. \ref{def12.3} and $Q= q\oplus q$.
\end{proposition}

\section{The HHI state}\label{sec14}\init
In this section we construct the HHI state $\omega_{\rm HHI}$ in $M$ and prove that it is a pure Hadamard state, extending the double $\beta$-KMS state $\omega_{\rm D}$ in $\cM^{-}\cup \cM^{+}$ for $\beta= (2\pi)\kappa^{-1}$.  We use the expression of $\omega_{\rm D}$ by \Calderon projectors for the Wick rotated operator $K_{\beta}$, see Subsect. \ref{sec12.7}.   

Since $K_{\beta}$ is a Laplace operator for the complex metric $\rg^{\rm eucl}$ on $M^{\rm eucl}=\bS_{\beta}\times \Sigma^{+}$, one can if $\beta= (2\pi)\kappa^{-1}$ extend it to a Laplace operator $K_{\rm ext}$ on the smooth extension $(M^{\rm eucl}_{\rm ext}, \rg^{\rm eucl}_{\rm ext})$.

The boundary of the open set $\Omega_{\rm ext}$ extending $\Omega_{\beta}= ]0, \b2[\times \Sigma^{+}$ is diffeomorphic to the full Cauchy surface $\Sigma$, and we can use the  \Calderon projectors for $K_{\rm ext}, \Omega_{\rm ext}$  to define a pair of covariances $\lambda^{\pm}_{\rm HHI}$.
The fact that they define a  state is actually quite easy, using some standard continuity properties of the \Calderon projectors and density results in Sobolev spaces. The proof of the Hadamard property of $\omega_{\rm HHI}$ relies also on an easy argument using pseudodifferential  calculus, taken from \cite{G}.  

\subsection{Laplacians for complex metrics}\label{sec14.1}
 We recall that complex metrics on a manifold $X$  are defined in  \ref{sec10.4.1}.
 \begin{definition}\label{def13.0}
A complex metric $\rk$ on  a manifold $X$ is called {\em uniformly sectorial} if  
\ben
\item  there exists  $C>0$ such that 
\beq\label{e13.0}
|{\rm Im}\,(\bar{v}^{a}\rk_{ab}(x)v^{b})|\leq C {\rm Re}\,(\bar{v}^{a}\rk_{ab}(x)v^{b}), \ \forall\, x\in X, \ v\in \cc T_{x}X;
\eeq
\item $|\rk(x)| = {\rm det}(\rk_{ab}(x))>0$ $\forall x\in X$.
\een
\end{definition}
Note that if $\rk$ is uniformly sectorial, then 
 \beq\label{e13.1}
 |{\rm Im}\,(\bar{\xi}_{a}k^{ab}(x)\xi_{b})|\leq C {\rm Re}\,(\bar{\xi}_{a}k^{ab}(x)\xi_{b})\ \forall\, x\in X, \ \xi\in \cc T^{*}_{x}X,
 \eeq
 i.e.  $\rk^{-1}$ is also uniformly sectorial. In fact 
 if $\xi= \rk v$ we have $\bar{\xi}\dual \rk^{-1}\xi= \overline{\rk v}\dual v= \overline{\overline{v}\dual \rk v}$ and \eqref{e13.1} follows from \eqref{e13.0}.
\subsubsection{Laplacians for complex metrics}\label{sec14.1.2}
If $\rk$ is a complex metric on $X$, one  defines the Christoffel symbols:
\[
\Gamma^{c}_{ab}\defeq \12 \rk^{cd}(\p_{a}\rk_{cd}+ \p_{b}\rk_{ad}- \p_{d}\rk_{ab}),
\]
 the covariant derivative:
\[
\nabla^{(\rk)}_{a}T^{b}= \p_{a}T^{b}+ \Gamma_{ac}^{b}T^{c},
\]
 and  the Laplacian associated to $\rk$, acting on $\coinf(X)$:
 \[
\Delta_{\rk}\defeq  \nabla^{(\rk)}_{a}\rk^{ab}\nabla^{(\rk)}_{b}
\]
as for real metrics.  For $m\in \cinf(X, \rr)$, we set:
\[
K\defeq - \Delta_{\rk}+ m,
\]
and equip $\coinf(X)$ with the scalar product:
\[
(u|v)_{X}\defeq \int_{X}\bar{u}v|\rk|^{\12}dx.
\]
 We have
 \beq\label{e13.1b}
\nabla_{a}^{(\rk)}T^{a}= |\rk|^{-\12} \p_{a}(|\rk|^{\12}T^{a}),
\eeq
see Subsect.  \ref{secapp1.1} and hence 
\[
K= - |\rk|^{-\12}\p_{a}\rk^{ab}|\rk|^{\12}\p_{a}+ m.
\]
The formal adjoint of $K$ for the scalar product $(\cdot| \cdot)_{X}$ equals
\[
K^{*}= -|\rk|^{-\12}\p_{a}\overline{\rk}^{ba}|\rk|^{\12}\p_{b}+ m= - \Delta_{\rk^{*}}+ m.
\]
\def\ovk{\overline{\rk}}
\begin{proposition}\label{prop13.1}
 Assume that $\rk$ is uniformly sectorial and that $m_{0}^{2}\leq m(x)$ for $m_{0}>0$. Let
 \[
 \begin{array}{l}
Q(u, u)= (u|Ku)_{X}= \int_{X}(\p_{a}\overline{u}\rk^{ab}\p_{b}u+ m \bar{u}u)|\rk|^{\12}dx, \\[2mm]
Q^{*}(u, u)= (u|K^{*}u)_{X}= \int_{X}(\p_{a}\overline{u}\ovk^{ba}\p_{b}u+ m \bar{u}u)|\rk|^{\12}dx,
\end{array}
\]
with $\Dom Q= \Dom Q^{*}= \coinf(X)$. 
Then:
\ben
\item  $Q, Q^{*}$ are closeable with  $\Dom Q^{\rm cl}= \Dom Q^{*{\rm cl}}=H^{1}_{\rk}(X)$, equal to the completion of $\coinf(X)$ for the norm
\[
\|u\|^{2}_{1}=  \int_{X}(\overline{\p_{a}u}{\rm Re}\,\rk^{ab}\p_{b}u +m\bar{u}u) |\rk|^{\12}dx.
\]
\item  $Q^{\rm cl}, Q^{*{\rm cl}}$ are sectorial and induce  isomorphisms: 
\[
K^{\rm cl}, K^{*\cl}: H^{1}_{\rk}(X)\tosim H^{1}_{\rk}(X)^{*}, 
\]
with $K^{\rm cl}= K, K^{*{\rm cl}}= K^{*}$ on $\coinf(X)$. 
\item Let us denote also by $K^{\cl}, K^{*\cl}$ the associated linear operators on $L^{2}(X, |\rk|^{\12}dx)$ with domains
\[
\begin{array}{l}
\Dom K^{\cl}= H^{1}_{\rk}(X)\cap (K^{\cl})^{-1}L^{2}(X, |\rk|^{\12}dx), \\[2mm]
 \Dom K^{*\cl}= H^{1}_{\rk}(X)\cap (K^{*\cl})^{-1}L^{2}(X, |\rk|^{\12}dx).
\end{array}
\]
Then $0\not\in \sigma(K^{\cl})$, $0\not\in \sigma(K^{*\cl})$ and $(K^{*\cl})^{-1}= ((K^{\cl})^{-1})^{*}$.
\een
\end{proposition}
\proof
Using \eqref{e13.1} under the integral sign defining $Q$ or $Q^{*}$, we obtain  $|{\rm Im }Q^{*}(u, u)|\leq C {\rm Re}\,Q(u, u)$ and that $Q, Q^{*}$ are closeable.  The domain  of   their closures  equals  $H^{1}_{\rk}(X)$ by definition and their closures are sectorial. By  the Lax-Milgram theorem we obtain isomorphisms $K^{\cl}, K^{*\cl}$, which completes the proof of (1) and (2). The first properties of the linear operators $K^{\cl}$, $K^{*\cl}$ acting on $L^{2}(X, |\rk|^{\12}dx)$ are immediate. From  \cite[Thm. VI.2.5]{K} we have $K^{*\cl}= (K^{\cl})^{*}$ hence $(K^{*\cl})^{-1}= ((K^{\cl})^{-1})^{*}$. \qed

\subsubsection{Outer unit normal}\label{oun}
Let  $\Omega\subset X$  be an open set with a smooth boundary $\p \Omega$. We set
\[
\Omega^{+}\defeq \Omega, \ \Omega^{-}\defeq X\backslash \Omega^{\rm cl}.
\]
We can define the outer unit normal vector field to $\p\Omega$, denoted by $n\in \cc TX$ by the following conditions:
\[
\begin{array}{rl}
i)& n(x)\dual\rk(x)v= 0, \ \forall v\in T_{x}\p\Omega, \\[2mm]
ii)& n(x)\dual \rk(x)n(x)=1,\\[2mm]
iii)& {\rm Re}\,n(x) \hbox{ is outwards pointing}.
\end{array}
\]
If $\Omega$ is locally equal to  $\{f>0\}$ for $f\in \cinf(X, \rr)$ with $df\neq 0$ on $\{f=0\}$, we have:
\[
n^{a}= \frac{-\rk^{ab}\nabla_{b}f}{(\nabla_{a}f\rk^{ab}\nabla_{b}f)^{\12}},
\]
where in the denominator we take the usual determination of $z^{\12}$.

We also assume the following condition:
\beq\label{e13.2}
{\rm Im}\,n(x)\in T_{x}\p\Omega, \ x\in \p\Omega,
\eeq
which  is equivalent to $\nabla_{a}f\rk^{ab}\nabla_{b}f\in \rr$ on $\p\Omega$, if $\Omega= \{f>0\}$.
%
%
%

 The volume form $d{\rm Vol}_{\rk}= |\rk|^{\12}dx^{1}\wedge \cdots \wedge dx^{n}$ associated to $\rk$ is real, as is the associated density $d\mu_{\rk}= | d{\rm Vol}_{\rk}|= |\rk|^{\12}dx$. It is easy to see from \eqref{e13.2} that the induced density $d\sigma_{\rh}= |d{\rm Vol}_{\rh}|$ associated to the induced metric $\rh$ on $\p\Omega$ is also real valued.

\subsubsection{Gauss-Green formula}
One obtains by the same arguments as in the real case the Gauss-Green formula:
\beq\label{ecorrect.11}
\int_{\Omega}\nabla_{a}^{(\rk)}w^{a}d\mu_{\rk}= \int_{\p\Omega}w^{a}n_{a}d\sigma_{\rh},
\eeq
for $w$ a smooth vector field  on $X$.
\begin{lemma} We have:
\begin{equation}
\label{e1.1zob}
\int_{\Omega}(\bar{v}Ku- \overline{K^{*}v}u) d\mu_{\rk}= \int_{\p\Omega}(n^{a}\nabla_{a}^{(\rk)}\bar{v}u - \bar{v}n^{a}\nabla_{a}^{(\rk)}u)d\sigma_{\rh},\ u\in \overline{\coinf}(\Omega), \ v\in \overline{\cinf}(\Omega).
\end{equation}
\end{lemma}
\proof For $X^{a}= \bar{v}\rk^{ab}\nabla^{(\rk)}_{b}u$ we obtain $\nabla^{(\rk)}_{a}X^{a}= \nabla^{(\rk)}_{a}\bar{v}\rk^{ab}\nabla^{(\rk)}_{b}u- \bar{v}Ku$, hence by \eqref{ecorrect.11}:
\beq\label{e1.2z}
\int_{\Omega}\bar{v}Ku d\mu_{\rk}= \int_{\Omega}(\nabla^{(\rk)}_{a}\bar{v}\rk^{ab}\nabla^{(\rk)}_{b}u+ m\bar{v}u)d\mu_{\rk}- \int_{\p\Omega} \bar{v}n^{a}\nabla_{a}^{(\rk)}ud\sigma_{\rh}.
\eeq
The same identity replacing $\rk$ by $\bar{\rk}$ gives:
\[
\int_{\Omega}\bar{u}K^{*}v d\mu_{\bar{\rk}}= \int_{\Omega}(\nabla^{(\bar{\rk})}_{a}\bar{u}\bar{\rk}^{ab}\nabla^{(\bar{\rk})}_{b}v+ m \bar{u}v)d\mu_{\bar{\rk}}- \int_{\p\Omega} \bar{u}\bar{n}^{a}\nabla_{a}^{(\bar{\rk})}vd\sigma_{\overline{\rh}},
\]
which by taking complex conjugates gives:
\beq\label{e1.3z}
\int_{\Omega}u\overline{K^{*}v} d\mu_{\rk}= \int_{\Omega}(\nabla^{(\rk)}_{a}u\rk^{ab}\nabla^{(\rk)}_{b}\bar{v}+ mu\bar{v})d\mu_{\rk}- \int_{\p\Omega} un^{a}\nabla_{a}^{(\rk)}\bar{v}d\sigma_{\rh},
\eeq
since $\overline{\nabla_{a}^{(\rk)}u}= \nabla_{a}^{(\bar{\rk})}\bar{u}=\nabla_{a}^{(\rk)}\bar{u}$. Subtracting \eqref{e1.3z} from \eqref{e1.2z} gives \eqref{e1.1zob}. \qed

\subsubsection{Trace operators}
For $u\in \cinf(X)$ we set:
\[
\gamma u\defeq\col{u\tra{\p\Omega}}{\p_{n}u\tra{\p\Omega}}\in \cinf(\p\Omega; \cc^{2}).
\]
The formal adjoint $\gamma^{*}$ of $\gamma$ is given by:
\beq\label{e12.52b}
\gamma^{*}f=  (d\mu_{\rk})^{-1}f_{0}d\p\Omega+ (d\mu_{\rk})^{-1} (n^{\mu}\p_{\mu})^{*}f_{1}d\p\Omega,
\eeq
where  if $g\in \cinf(\p\Omega)$, $gd\p\Omega$ is the distributional density defined as
\[
\langle u|g d\p\Omega\rangle= \int_{\p\Omega}ug d\sigma_{\rh},\ u\in \coinf(X), 
\] 
and  
\[
\langle u|(n^{\mu}\p_{\mu})^{*}g d\p\Omega\rangle= \langle n^{\mu}\p_{\mu}u|g d\p\Omega\rangle.
\]
Similarly  for $u\in \overline{C^{\infty}}(\Omega^{\pm})$ we set:
\[
\gamma^{\pm}u\defeq\col{u\tra{\p\Omega}}{\p_{n}u\tra{\p\Omega}},
\]
where the trace is taken from $\Omega^{\pm}$.

 In the rest of this subsection, we assume that $\rk$ is uniformly sectorial and that \eqref{e13.2} holds.
\subsubsection{\Calderon projectors}

\begin{definition}\label{def13.1}
The {\em \Calderon projectors} $c^{\pm}$ associated to $(K, \Omega)$ are defined as
 \[
c^{\pm}= \mp \gamma^{\pm}\circ (K^{\rm cl})^{-1}\circ \gamma^{*}\circ S,
\]
where
\[
S= \mat{2\i b^{*}}{-\one}{\one}{0}, 
\]
 $b= {\rm Im}\,n^{a}\nabla_{a}$ and $b^{*}$ is the formal adjoint of $b$ in $L^{2}(\p\Omega, d\sigma_{\rh})$.
\end{definition}
Note that  the operator $S$ is well defined on $\coinf(\p\Omega; \cc^{2})$, since ${\rm Im }n$ is tangent to $\p\Omega$.

It is not a priori clear that $c^{\pm}$ are well defined, since even for $f\in \coinf(\p\Omega; \cc^{2})$, $\gamma^{*} Sf$ does not belong to $H^{1}_{\rk}(X)^{*}$. 

 To show that $c^{\pm}$ make sense, one  can apply the following proposition.  We denote by $H^{s}_{\rm c}(\p\Omega)$ resp. $H^{s}_{\rm loc}(\p\Omega)$ for $s\in \rr$, the compactly supported, resp. local Sobolev spaces on $\p\Omega$ and set:
\beq\label{e.9.not}
\cH^{s}_{\rm c/loc}(\p\Omega)= H^{s-\12}_{\rm c/loc}(\p\Omega)\oplus H^{s-1-\12}_{\rm c/loc}(\p\Omega), \ s\in \rr.
\eeq
 \begin{proposition}\label{prop14.1}
 \ben
 \item $c^{\pm}: \cH^{s}_{\rm c}(\p\Omega)\to \cH^{s}_{\rm loc}(\p\Omega)$ continuously for any $s\in \rr$,
 \item $c^{\pm}$ are $2\times 2$ matrices with entries in $\Psi^{\infty}(\p\Omega)$.
 \een
\end{proposition}
\proof
The differential operator $K$ is elliptic, since its principal symbol equals $\xi\dual \rk^{-1}(x)\xi$.  $K$ admits hence a properly supported parametrix $Q\in \Psi^{-2}_{\rm c}(X)$.
We claim that  $(K^{\rm cl})^{-1}- Q$ is a smoothing operator, i.e.  has a smooth distributional kernel.  In fact by the usual argument of commuting derivatives with $K$, we first obtain that $(K^{\rm cl})^{-1}: H^{s}_{\rm c}(\p\Omega)\to H^{s+2}_{\rm loc}(\p\Omega)$ is continuous for all $s\geq -1$. Next since $K Q= \one +R_{-\infty}$, where $R$ is smoothing and properly supported, we have 
\[
K^{\rm cl} ((K^{\rm cl})^{-1}-Q)u= Ru, \ u\in \coinf(\p\Omega),
\]
hence $(K^{\rm cl})^{-1}= Q + (K^{\rm cl})^{-1}R_{-\infty}$ on $\coinf(\p\Omega)$. This implies that $(K^{\rm cl})^{-1}$ extends  as a continuous map from 
 $H^{s}_{\rm c}(\p\Omega)$ to $H^{s+2}_{\rm loc}(\p\Omega)$ for all $s\in \rr$ and that $(K^{\rm cl})^{-1}- Q$ is smoothing.

It suffices hence to prove the proposition with $(K^{\rm cl})^{-1}$ replaced by $Q$ in the definition of $c^{\pm}$. Note that $Q$ is properly supported, and the support of its distributional kernel can be assumed to lie in an arbitrary small neighborhood of the diagonal in $X\times X$.

We claim that to prove the proposition, we can reduce ourselves to the case when $X$ and $\p\Omega$ are compact, which is the situation considered in \cite[Sect. 11.1]{Gr}. Let us now explain this reduction.

Let $(U_{i})_{i\in \nn}$  be an open cover of a neighborhood of $\p\Omega$  in $X$ and $\chi_{i}: U_{i}\to B_{n-1}(0,1)\times ]-1, 1[$,  an associated chart diffeomorphism with $\chi_{i}(U_{i}\cap \p\Omega)= B_{n-1}(0, 1)\times \{0\}$, where $n= {\rm dim}X$ and $B_{d}(0, 1)$ is the open unit ball in $\rr^{d}$. 
If $ V_{i}= U_{i}\cap \p\Omega$ and $\psi_{i}= \chi_{i | U_{i}\cap \p\Omega}$, then $(V_{i}, \psi_{i})_{i\in \nn}$ is an atlas of $\p\Omega$. Refining if necessary the cover $(U_{i})_{i\in \nn}$, we can assume that if $V_{i}\cap V_{j}\neq \emptyset$, then  there exists an chart open set $U\subset X$ with $U_{i}, U_{j}\Subset U $ and  a  chart diffeomorphism $\chi: U\to B_{n-1}(0, 1)\times ]-1, 1[$ with $\chi(U\cap \p\Omega)= B_{n-1}(0, 1)\times \{0\}$.

  Let  $1= \sum_{i\in \nn}\varphi_{i}$ with $\varphi_{i}\in\coinf(V_{i})$ a subordinate partition of unity of $\p\Omega$.  From the topologies $H^{s}_{\rm c}(\p\Omega)$ and $H^{s}_{\rm loc}(\p\Omega)$, we  see that  the continuity of a  map $A: H^{s}_{\rm c}(\p\Omega)\to H^{s'}_{\rm loc}(\p\Omega)$  is equivalent to  the continuity of $\varphi_{i}A\varphi_{j}: H^{s}_{\rm c}(\p\Omega)\to H^{s'}_{\rm loc}(\p\Omega)$   
for all $i, j\in \nn$.  

If $V_{i}\cap V_{j}= \emptyset$ then $\varphi_{i}c^{\pm}\varphi_{j}$ is smoothing, since  the kernel of $Q$ is smooth outside of the diagonal.  If $V_{i}\cap V_{j}\neq \emptyset$,   we use the open set $U$ and  diffeomorphism $\chi$ introduced above to reduce ourselves to the situation when $X= B_{n-1}(0, 1)\times ]-1, 1[$, $\p\Omega= B_{n-1}(0, 1)\times \{0\}$ and $V_{i}, V_{j}\Subset B_{n-1}(0, 1)$.  

We can then embed $B_{n-1}(0, 1)\times ]-1, 1[$ in a compact manifold $\tilde{X}$, $\p\Omega$ in a smooth, compact submanifold $\tilde{\p\Omega}$ of $\tilde{X}$, and extend $Q$ as a properly supported pseudodifferential operator $\tilde{Q}\in \Psi^{-2}_{\rm c}(\tilde{X})$ such  that $\varphi_{i}(Q- \tilde{Q})\varphi_{j}$ is a smoothing operator. Therefore we can replace $X, \p\Omega, Q$ by $\tilde{X}, \tilde{\p\Omega}, \tilde{Q}$ as claimed.

 A neighborhood $V$  of $\p\Omega$ in $X$ is  then diffeomorphic to $]-\delta, \delta[\times \p\Omega$, and  one can use coordinates $(s, y)$ on $]-\delta, \delta[\times \p\Omega$.  
In \cite[Sect. 11.1]{Gr} the trace operator  is  defined as
\[
\tilde{\gamma}u= \col{u(0, y)}{\i^{-1}\p_{s}u(0, y)}.
\]
Clearly we have $\gamma = L\circ \tilde{\gamma}$, where $L=\mat{\one}{ 0}{r}{\one}$ and $r$ is a first order differential operator on $\p\Omega$. This implies that $L: \cH^{s}_{\rm c/loc}(\p\Omega)\tosim \cH^{s}_{\rm c/loc}(\p\Omega)$.  The \Calderon projectors  $\tilde{C}^{\pm}$ in \cite[Sect. 11.1]{Gr}  are equal to $L^{-1}\circ c^{\pm}\circ L$, and \cite[Prop. 11.7]{Gr} implies that $\tilde{C}^{\pm}: \cH^{s}_{\rm c}(\p\Omega)\to \cH^{s}_{\rm loc}(\p\Omega)$ for all $s\in \rr$, which implies (1).

Property (2) is a standard fact, see \cite[Sect. 11.1]{Gr}. \qed

 \subsection{Construction of the HHI state}\label{sec14.2}

\subsubsection{ The Laplacian on $M^{\rm eucl}_{\rm ext}$}\label{sec14.2.1}
We now apply  the above framework to $(X, \rk)=(M^{\rm eucl}_{\rm ext}, \rg^{\rm eucl}_{\rm ext})$,  the smooth extension of $(M^{\rm eucl}, \rg^{\rm eucl})$  constructed in Prop. \ref{prop10.1}, for $\beta= (2\pi)\kappa^{-1}$.   We assume that hypothesis {\rm (H3)} in Subsect. \ref{sec10.23} holds.
By Prop. \ref{propoto}   the Wick rotated metric $\rg^{\rm eucl}$ satisfies the conditions in \ref{sec12.5.1}, i.e.  is uniformly sectorial. By continuity the same is true of its extension $\rg^{\rm eucl}_{\rm ext}$.
We denote by
\[
K_{\rm ext}= \Delta_{\rg^{\rm eucl}_{\rm ext}}+ m_{\rm ext},
\]
the associated Laplacian.  We choose the open set $\Omega_{\rm ext}\subset M_{\rm ext}^{\rm eucl}$, whose boundary $\p \Omega_{\rm ext}$ is diffeomorphic to $\Sigma$, see Prop. \ref{prop10.1}.   We saw in \ref{sec12.5.1} that if $\nu$ is the unit outer normal to $]0, \b2[\times \Sigma^{+}$, then ${\rm Im}\,\nu$ is tangent to $\p( ]0, \b2[\times \Sigma^{+})$. Again by continuity, the same is true of the unit outer  normal to $\Omega_{\rm ext}$, i.e.  condition \eqref{e13.2} is satisfied. 
Therefore we can apply the results of Subsect. \ref{sec14.1} to $K_{\rm ext}$ and $\Omega_{\rm ext}$.

We need one more result, which states that $K_{\rm ext}$ is the unique extension of $K_{(2\pi)\kappa^{-1}}$ to $L^{2}(M_{\rm ext}^{\rm eucl})$.
\begin{proposition}\label{prop14.2}
Let $U: \coinf(M^{\rm eucl})\to \coinf(M^{\rm eucl}_{\rm ext}\setminus \cB_{\rm ext})$ defined by:
\[
Uu = u\circ \chi^{-1},
\] 
where $\chi: M^{\rm eucl}\tosim M^{\rm eucl}_{\rm ext}\setminus \cB_{\rm ext}$ is the diffeomorphism constructed in Prop. \ref{prop10.1}.
 Then $U$ extends as a unitary operator
 \[
U: L^{2}(M^{\rm eucl}, N(y)|\rh|^{\12}(y)dy ds)\to L^{2}(M^{\rm eucl}_{\rm ext}, |g^{\rm eucl}_{\rm ext}|^{\12}dx),
\]
with $K_{\rm ext}= U K_{(2\pi)\kappa^{-1}} U^{*}$.
\end{proposition}
\proof  $U$ clearly extends as a unitary operator. Let us check the second statement. 

As a differential operator, $K_{(2\pi)\kappa^{-1}}$ equals   $-\Delta_{\rg^{\rm eucl}}+ m$.  As an unbounded operator,  $K_{(2\pi)\kappa^{-1}}$ is defined in  \ref{defilo} using the sesquilinear form $Q_{(2\pi)\kappa^{-1}}$, while $K_{\rm ext}$ is defined with the sesquilinear form $Q_{\rm ext}$ for $\rk= \rg^{\rm eucl}_{\rm ext}$ and $m = m_{\rm ext}$,  see Prop. \ref{prop13.1}.  

$Q_{(2\pi)\kappa^{-1}}$ is the closure of its restriction to $\coinf (M^{\rm eucl})$, while $Q_{\rm ext}$ is the closure of its restriction to $\coinf(M^{\rm eucl}_{\rm ext})$.  

Taking into account the isometry $\chi: M^{\rm eucl}\tosim M^{\rm eucl}_{\rm ext}\setminus \cB_{\rm ext}$, it suffices to check that 
$\coinf(M^{\rm eucl}_{\rm ext}\setminus \cB_{\rm ext})$ is a form core for $Q_{\rm ext}$, 
 i.e.  that  this space  is dense in  the space  $H^{1}_{\rk}(X)$ for  $(X, \rk)= (M^{\rm eucl}_{\rm ext}, \rg^{\rm eucl}_{\rm ext})$, see Prop. \ref{prop13.1}.
 
Using the coordinates $(X, Y, \omega)$ near $\cB_{\rm ext}\sim \{0\}\times \cB$, this follows from the fact that $\coinf(\rr^{2}\backslash \{0\})$  is dense in $H^{1}(\rr^{2})$, see eg \cite[Thm. 3.23]{A}. \qed
\subsubsection{The HHI state}\label{sec14.2.2}
Let us denote by $c^{\pm}_{\rm ext}$ the \Calderon projectors for $(K_{\rm ext}, \Omega_{\rm ext})$, defined as in Def. \ref{def13.1}.
The following theorem is a slightly more precise version of Thm. \ref{thm1.1}.
\begin{theoreme}\label{thm14.1}
Assume conditions {\rm (H1)},  {\rm (H2)},  {\rm (H3)}. Then: 
 \ben
 \item $\lambda^{\pm}_{\rm HHI}= \pm q\circ c^{\pm}_{\rm ext}$ are the Cauchy surface covariances of a  quasi-free state $\omega_{\rm HHI} $ for $P$ in $M$, called the {\em HHI state}.
 \item  The restriction of $\omega_{\rm HHI}$ to $\cM^{+}\cup \cM^{-}$ is the double $\beta-$KMS state $\omega_{\rm D}$ for $\beta= (2\pi)\kappa^{-1}$.
\item $\omega_{\rm HHI}$ is a {\em Hadamard state} in $M$.
\item Let $\omega$  be a quasi-free state for $P$ in $M$ whose restriction to $\cM^{+}\cup \cM^{-}$ equals $\omega_{\rm D}$ and  such that  its space-time covariances map $\coinf(M)$ into $\cinf(M)$ continuously. Then $\omega= \omega_{\rm HHI}$.
\item Assume moreover that  {\rm (H4)} holds. Then $\omega_{\rm HHI}$ is a pure state. 
\een
\end{theoreme}
Note that it follows from (4) above that $\omega_{\rm HHI}$ is the unique Hadamard state in $M$ whose restriction to $\cM^{+}\cup \cM^{-}$ equals $\omega_{\rm D}$.

\begin{remark}\label{remark-purity}
If we know that $\omega_{\rm D}$ is a pure state on $\CCR(\coinf(\Sigma\setminus \cB; \cc^{2}), q)$ then $\omega_{\rm HHI}$ is also a pure state on $\CCR(\coinf(\Sigma; \cc^{2}), q)$. In fact let $\cY^{\rm cl}$ be the completion of $\coinf(\Sigma\setminus \cB; \cc^{2})$ for the scalar product $\lambda^{+}_{\rm D}+ \lambda^{-}_{D}$.  In the proof of Thm. \ref{thm14.1} we show  that $\cY^{\rm cl}$ is also the completion of $\coinf(\Sigma; \cc^{2})$ for $\lambda^{+}_{\rm D}+ \lambda^{-}_{\rm D}= \lambda^{+}_{\rm HHI}+ \lambda^{-}_{\rm HHI}$.  Applying then Prop. \ref{propoto2} we obtain that $\omega_{\rm HHI}$ is also a pure state on $\CCR(\coinf(\Sigma; \cc^{2}), q)$.

However it seems difficult to prove directly that $\omega_{\rm D}$ is pure on $\CCR(\coinf(\Sigma\setminus \cB; \cc^{2}), q)$, because of the infrared problem caused by the fact that  the lapse function $N$ vanishes at $\cB$. Therefore we will prove directly that $\omega_{\rm HHI}$ is a pure state on $\CCR(\coinf(\Sigma; \cc^{2}), q)$, using arguments from \cite{GW2}.
\end{remark}
{\bf Proof of Thm. \ref{thm14.1}.}   We start by checking some useful claims.

{\em Claim 1}.
We claim that
\begin{equation}
\label{ecorrect.70}
\lambda^{\pm}_{\rm HHI}= \lambda_{\rm D}^{\pm} \hbox{ on }\coinf(\Sigma\setminus\cB; \cc^{2}).
\end{equation}
   We note that the map $(\one\oplus r^{*})$  in Prop. \ref{prop12.3} corresponds to the embedding of $\coinf (\Sigma^{+}\cup \Sigma^{-}; \cc^{2})$ into $\coinf(\Sigma\setminus \cB)$ obtained from  $\psi: \Sigma\to M^{\rm eucl}_{\rm ext}$ in Prop. \ref{prop10.1}.  The exterior normal  to $\Omega_{\rm ext}$   is   a smooth extension of the image under $\chi$ of the exterior normal to $]0, \pi\kappa^{-1}[\times \Sigma^{+}$   defined in \eqref{e12.51}. Therefore using also  Prop. \ref{prop14.2} we obtain that
 \beq\label{e14.1}
(\one\oplus r^{*})^{-1}c_{(2\pi/\kappa)}^{\pm}(\one\oplus r^{*})= c^{\pm}_{\rm ext}, 
\eeq
on $\coinf(\Sigma^{+}\cup \Sigma^{-}; \cc^{2})$. This implies \eqref{ecorrect.70}.

{\em Claim 2}. We claim that
\begin{equation}
\label{ecorrect.80}
\lambda_{\rm HHI}^{\pm}, q\hbox{ are continuous sesquilinear forms on }\cH^{1}_{\rm c}(\Sigma),
\end{equation}
where the spaces $\cH^{s}_{\rm c/  loc}(\Sigma)$ are defined in \eqref{e.9.not}. 
 Let us denote  by $\rh_{\rm ext}$  the metric induced by $\rg^{\rm eucl}_{\rm ext}$ on $\Sigma$ and  use the scalar product of $L^{2}(\Sigma, |\rh_{\rm ext}|^{\12}dy)\otimes \cc^{2}$ to identify sesquilinear forms with operators,  so that $q= \mat{0}{1}{1}{0}$.   Let us set
\[
\tilde{\cH}^{s}_{\rm c/loc}(\Sigma)= H^{s+\12}_{\rm c/loc}(\Sigma)\oplus H^{s+1+\12}_{\rm c/loc}(\Sigma), \ s\in \rr.
\]
Using that $H^{s}_{\rm c}(\Sigma)$ and $H^{-s}_{\rm loc}(\Sigma)$ form a dual pair, we see that  $\cH^{s}_{\rm c}(\Sigma)$ and $\tilde{\cH}^{-s}_{\rm loc}(\Sigma)$ form a dual pair. Moreover 
 $q: \cH^{s}_{\rm loc}(\Sigma)\to \tilde{\cH}^{s-2}_{\rm loc}(\Sigma)$ continuously, and by  Prop. \ref{prop14.1} (1) the same is true of $\lambda^{\pm}_{\rm HHI}= \pm q\circ c^{\pm}_{\rm ext}$. For $s=1$ we have $s-2= -s$, which proves \eqref{ecorrect.80}. 
 
 {\em Proof of (1) and (2)}. 
 Since $\omega_{\rm D}$ is a state we deduce from \eqref{ecorrect.70} that
 \begin{equation}
\label{ecorrect.90}
\lambda_{\rm HHI}^{+}- \lambda_{\rm HHI}^{-}= q, \ \lambda_{\rm HHI}^{\pm}\geq 0 \hbox{ on }\coinf(\Sigma\setminus \cB; \cc^{2}).
 \end{equation}
 It is a well-known fact that  since $\cB\subset \Sigma$ is of codimension 1,  $\coinf(\Sigma\setminus \cB)$ is dense in $H^{\pm\12}_{\rm c}(\Sigma)$.  Applying then \eqref{ecorrect.80} we deduce from \eqref{ecorrect.90} that:
 \[
\lambda_{\rm HHI}^{+}- \lambda_{\rm HHI}^{-}= q, \ \lambda_{\rm HHI}^{\pm}\geq 0 \hbox{ on }\cH_{\rm c}^{1}(\Sigma)= H^{\12}_{\rm c}(\Sigma)\oplus H^{-\12}_{\rm c}(\Sigma),
\]
 hence
 \[
\lambda_{\rm HHI}^{+}- \lambda_{\rm HHI}^{-}= q, \ \lambda_{\rm HHI}^{\pm}\geq 0 \hbox{ on }\coinf(\Sigma; \cc^{2}),
\]
which proves (1).  Statement (2) follows from \eqref{ecorrect.70}.

{\em Proof of (3)}.
 By Thm. \ref{allhad} there exists a reference Hadamard state $\omega_{\rm ref}$ for $P$ in $M$ whose Cauchy surface covariances on $\Sigma$, denoted by  $\lambda^{\pm}_{\rm ref}$ are $2\times 2$ matrices with entries in $\Psi^{\infty}(\Sigma)$. By Prop. \ref{prop14.1} the same is true for $\lambda^{\pm}_{\rm HHI}$.

The restriction of $\omega_{\rm HHI}$ to $\cM^{+}$ is a  Hadamard state for $P$, since   it  is a $(2\pi)\kappa^{-1}$-KMS state for a time-like, complete Killing vector field.  The restriction of $\omega_{\rm HHI}$ to $\cM^{-}$ is also a Hadamard state for $P$. 

In fact by Prop. \ref{prop12.3}, its Cauchy surface covariances on $\Sigma^{-}$ are the images of those of $\omega_{\rm D}$ on $\Sigma^{+}$ by  the weak wedge reflection 
$r$. Since $r^{*}\rh= \rh$, $r^{*}N= - N$ and $r^{*}w= w$, see \ref{sec10.2.1}, the expression \eqref{e12.01b} of $P$ in $\rr\times \Sigma^{-}$ shows that the restriction of $\omega_{\rm D}$ to $\cM^{-}$ is also a Hadamard state.

This implies that the restriction of $\omega_{\rm HHI}$ to $\cM^{+}\cup \cM^{-}$ is a Hadamard state. The same is true of the restriction of the reference Hadamard state $\omega_{\rm ref}$ to $\cM^{+}\cup \cM^{-}$.   
Denoting by $\Lambda^{\pm}_{\rm HHI / ref}$ the spacetime covariances of $\omega_{\rm HHI/ref}$, this implies that $\Lambda^{\pm}_{\rm HHI}- \Lambda^{\pm}_{\rm ref}\in \cinf((\cM^{+}\cup \cM^{-})^{2})$, since the spacetime covariances of two Hadamard states differ by smooth kernels. 
Passing to Cauchy surface covariances on $\Sigma^{+}\cup \Sigma^{-}$, this implies using \eqref{ef.4}  that if $\chi\in \coinf(\Sigma\setminus \cB)$, then
\[
\chi\circ( \lambda^{\pm}_{\rm HHI}- \lambda^{\pm}_{\rm ref})\circ \chi\hbox{ is a smoothing operator on }\Sigma.
\]
We claim that this implies that $\lambda^{\pm}_{\rm HHI}- \lambda^{\pm}_{\rm ref}$ is smoothing, which will imply that $\omega_{\rm HHI}$ is a Hadamard state. 

In fact let  $a$ be one of the entries of $\lambda^{\pm}_{\rm HHI}- \lambda^{\pm}_{\rm ref}$, which is a scalar pseudodifferential operator belonging to $\Psi^{m}(\Sigma)$ for some $m\in \rr$. We know that  $\chi\circ a\circ \chi$ is smoothing for any $\chi\in \coinf(\Sigma\backslash \cB)$.  Then its principal symbol $\sigma_{\rm pr}(a)$ vanishes on $T^{*}(\Sigma\backslash \cB)$ hence on $T^{*}\Sigma$ by continuity, so $a\in\Psi^{m-1}(\Sigma)$. Iterating this argument we obtain that $a$ is smoothing, which completes the proof of (3).

{\em Proof of (4)}. The proof of (4) is identical to \cite[Prop. 7.4]{G}. 

{\em Proof of (5)}. The proof of (5) will be given in Sect. \ref{secoto}.
\qed

\section{Purity of the Hartle-Hawking-Israel state}\init\label{secoto}
To prove that $\omega_{\rm HHI}$ is a pure state, we will follow the method in \cite{GW2}, by using the criterion for purity recalled in Prop. \ref{purity}. We start by some preparations.
\subsection{Preparations}
In all this section the manifold $M_{\rm eucl}^{\rm ext}$ will be denoted by $X$,  the complex metric $\rg_{\rm ext}^{\rm eucl}$ by $\rk$, the elliptic operator $K_{\rm eucl}$ by $K$,   the open set $\Omega_{\rm ext}$ defined in Prop. \ref{prop10.1} by $\Omega$,   and the \Calderon projectors $c^{\pm}_{\rm ext}$ by $c^{\pm}$, in accordance with the notation used in  Subsect. \ref{sec14.1}.    We recall that
\[
\| u\|^{2}_{H^{1}_{\rk}(X)}= \int_{X}\p_{a}\bar{u}\Re\rk^{ab}\p_{b}u+ m\bar{u}u|\rk|^{\12}dx,
\]
and  will set 
\beq\label{ecorrect.12}
(u|v)_{\Omega}= \int_{\Omega}\bar{u}v|\rk|^{\12}dx, \ \| u\|^{2}_{H^{1}_{\rk}(\Omega)}= \int_{\Omega}\p_{a}\bar{u}\Re\rk^{ab}\p_{b}u+ m\bar{u}u|\rk|^{\12}dx.
\eeq
Note that the closure of $\overline{\coinf}(\Omega)$ for the norm $\| \cdot\|_{H^{1}_{\rk}(\Omega)}$ coincide with the space $\overline{H^{1}_{\rk}}(\Omega)$ of restrictions to $\Omega$ of elements of $H^{1}_{\rk}(\Omega)$.

From the construction of  the manifold $M_{\rm eucl}^{\rm ext}$ in Subsect. \ref{secapp1.2}, we see that  there exist  $U\Subset X$ and $V\subset \Sigma$ which are compact neigbhorhoods of $\{0\}\times \cB$, such that $X\setminus U$ equals $\bS_{\beta}\times(\Sigma^{+}\setminus V)$.  Recall also that the coordinates on $\bS_{\beta}\times\Sigma^{+}$ are denoted by $(s, y)$.
\begin{lemma}
 Assume hypothesis ({\rm H4)}. Then there exists a family $\chi_{n}\in \coinf(M^{\rm ext}_{\rm eucl})$ such that:
 \[
\begin{array}{rl}
i)&\chi_{n}= 1\hbox{ on }U, \\[2mm]
ii)&\chi_{n}= \chi_{n}(y)\hbox{ on }X\setminus U,\\[2mm]
iii)&\chi_{n}\to 1\hbox{ locally uniformly on }X, \\[2mm]
iv)&\|d\chi\|_{\infty}=\sup_{ X}|d\chi_{n}|\in O(n^{-1}), 
\end{array}
\]
where we set $|d\chi_{n}|= (d\chi_{n}\dual \rh^{-1}d\chi_{n})^{\12}$.
\end{lemma}
\proof 
If $\Sigma$ is compact it suffices to take $\chi_{n}=1$. If $\Sigma$ is not compact, then $(\Sigma, \rh)$ is complete by hypothesis (H4).
It follows that if we fix a point $y^{0}\in \Sigma$, (for example $y^{0}\in \cB$), the geodesic balls $B_{\rh}(y^{0}, n)$ are compact and $\bigcup_{n\in \nn}B_{\rh}(y^{0}, n)= \Sigma$.  By  \cite[Thm. 1]{AFLR},  there exists $r\in \cinf(\Sigma)$ such that:
 \[
 \12 d(y^{0}, y)\leq r(y)\leq 2 d(y^{0}, y)\  \|\nabla r\|_{\infty}\leq 2.
\]
We choose $F\in \coinf(\rr)$ with $F(s)= 1$ in $s\leq 1$, $F(s)= 0$ in $s\geq 2$, and $\psi\in \coinf(X)$ equal to $1$ near the neighborhood $U$ of $\{0\}\times\cB$. We choose
\[
\chi_{n}(x)= \psi(x)+ (1- \psi)(x)F(n^{-1}r(y)), 
\]
where we recall that outside $U$, $X=\bS_{\beta}\times\Sigma^{+}$ with coordinates  $(s, y)$. It is easy to check that $\chi_{n}$ has the required properties. \qed
\begin{lemma}\label{jenaimarre}
Let $\chi\in \coinf(X)$ with $\chi= 1$  near $\{0\}\times\cB$ and $\chi(x)= \chi(y)$ outside a neighborhood of $\{0\}\times\cB$.  Then 
\[
|(v| [K, \chi]u)_{\Omega}|\leq C\| d\chi\|_{\infty}\| v\|_{H^{1}_{\rk}(\Omega)}\| u\|_{H^{1}_{\rk}(\Omega)},
\]
where $\| d\chi\|_{\infty}= \sup_{\Sigma}(d \chi\dual \rh^{-1}d\chi)^{\12}$.
 \end{lemma}
\proof Let $U$ a neighbhorhood of $\{0\}\times \cB$ such that  $\chi=1$ on $U$.   We have  $K= (-\p_{s}+ \i \w^{*})N^{-2}(\p_{s}-\i \w)$ on $\supp\nabla \chi\subset X\setminus U$, see \eqref{e7.1}, and
\begin{equation}
\label{ecorrect.10}
\|u\|^{2}_{H^{1}_{\rk}(X)}\sim \int_{\bS_{\beta\times \Sigma^{+}}}(|N^{-1}\p_{s}u|^{2}+ \nabla\bar{u}\rh^{-1}\nabla u + m\bar{u}u)N|\rh|^{\12}dsdy, \ u\in \coinf(X\setminus U).
\end{equation}
Therefore
\beq\label{ecorrect.11b}
[K, \chi]= 2\i N^{-1}w\dual d \chi N^{-1}\p_{s}+ \nabla^{*}\dual \rh^{-1} d\chi- d\chi\dual \rh^{-1}\nabla.
\eeq
We have 
\[
|w\dual d \chi|\leq (w\dual \rh w)^{\12}( d\chi \dual \rh^{-1}d\chi)^{\12},
\]
hence
\[
N^{-1}| w\dual d \chi|\leq (N^{-2}w\dual \rh w)^{\12}(d \chi\dual \rh^{-1}d\chi)^{\12}\leq (d \chi\dual \rh^{-1}d\chi)^{\12}\leq \| d\chi\|_{\infty}, 
\]
since $w\dual\rh w\leq N^{2}$ by \eqref{e1.3}.  Combining \eqref{ecorrect.10} and \eqref{ecorrect.11b} we obtain the lemma. \qed

\begin{proposition}\label{propototo}
 Let $f\in \coinf(\Sigma)^{2}$. Then:
 \begin{equation}
\label{e.200}
\begin{array}{rl}
i)&\chi_{n}c^{+}f- c^{+}\chi_{n}c^{+}f\to 0,\\[2mm]
ii)&\chi_{n}c^{+}\chi_{n}c^{+}f- c^{+}\chi_{n}c^{+}\chi_{n}c^{+}f\to 0,
\end{array}
\end{equation}
in $\cD'(\Sigma)^{2}$ when $n\to \infty$.
\end{proposition}
\proof 
Let us note that the identity \eqref{ecorrect.11} is of course still valid if $u\in \overline{\coinf}(\Omega)$ and $v\in \overline{\cinf}(\Omega)$.  It is also valid if $u\in \overline{\coinf}(\Omega)$ and $v\in \overline{H^{2}_{\rm loc}}(\Omega)$, since all the terms in the identity are still well defined.

For $f\in \coinf(\Sigma)^{2}$ we set
\[
V^{+}f\defeq -r_{\Omega}(K^{\rm cl})^{-1}\gamma^{*}Sf,
\]
where $r_{\Omega}$ is the operator of restriction to $\Omega$ and we recall that $S$ is defined in Def. \ref{def13.1}.
Let $w\in \overline{\coinf}(\Omega)$ and
\[
v= (K^{*{\rm cl}})^{-1}e_{\Omega}w= (K^{\rm cl})^{-1*}e_{\Omega}w,
\]
where $e_{\Omega}$ is the operator of extension by $0$ in $X\setminus \Omega$.

We know that $v\in H^{1}_{\rk}(X)$ by Prop. \ref{prop13.1}, and $v\in H^{2}_{\rm loc}(X)$ using that $e_{\Omega}w\in L^{2}_{\rm loc}(X)$ and elliptic regularity.  For $u\in\overline{\coinf}(\Omega^{+})$ we obtain from \eqref{e1.1zob}:
\begin{equation}
\label{e.20}
(v|Ku)_{\Omega}- (K^{*}v|u)_{\Omega}= (\gamma^{+}v| S\gamma^{+}u)_{\Sigma}= (\gamma v| S\gamma^{+}u)_{\Sigma},
\end{equation}
where in the last equality we use that   $\gamma^{\pm}v= \gamma v$. In fact  $\gamma v$ is well defined as an element of $H^{3/2}_{\rm loc}(\Sigma)\oplus H^{1/2}_{\rm loc}(\Sigma)$ since $v\in H^{2}_{\rm loc}(X)$. Next, we obtain:
\begin{equation}
\label{e.21}
\begin{array}{rl}
(\gamma v| S\gamma^{+}u)_{\Sigma}&= (v| \gamma^{*}S\gamma^{+}u)_{X}= ((K^{\rm cl})^{-1*}e_{\Omega}w| \gamma^{*}S\gamma^{+}u)_{X}\\
&= (e_{\Omega}w|(K^{\rm cl})^{-1}\gamma^{*}S\gamma^{+}u)_{X}= - (w| V^{+}\gamma^{+}u)_{\Omega}.
\end{array}
\end{equation}
From \eqref{e.20}, \eqref{e.21} we obtain:
\begin{equation}
\label{e.22}
(w| u - V^{+}\gamma^{+}u)_{\Omega}
=(v|Ku)_{\Omega}, \ \ u,w\in \overline{\coinf}(\Omega).
\end{equation}
We now fix $f\in \coinf(\Sigma)^{2}$ and $u = V^{+}f$. By  the same argument as in \cite[Lemma A.1]{GW2} we know that $u\in \overline{H^{1}_{\rk}}(\Omega)\cap \overline{\cinf}(\Omega)$.  We now apply \eqref{e.22} replacing $u$ by $u_{n}= \chi_{n}u$, which belongs to $\overline{\coinf}(\Omega)$.

Since $Ku= 0$ in $\Omega$, we have $K\chi_{n}u= [K, \chi_{n}]u$ and by Lemma \ref{jenaimarre} we obtain:
\beq\label{ecorrect.13}
|(v| [K, \chi_{n}]u)_{\Omega}|\leq C n^{-1}\|v\|_{H^{1}_{\rk}(\Omega)}\|u\|_{H^{1}_{\rk}(\Omega)}.
\eeq
Using \eqref{e.22} this yields:
\begin{equation}
\label{e.24}
|(w| \chi_{n}u- V^{+}\gamma^{+}\chi_{n}u)_{\Omega}|\leq 
Cn^{-1}\| v\|_{H^{1}_{\rk}(\Omega)}\|u\|_{H^{1}_{\rk}(\Omega)}\leq Cn^{-1}\|e_{\Omega}w\|_{H^{1}_{\rk}(X)^{*}}\|u\|_{H^{1}_{\rk}(\Omega)},
\end{equation}
since by Prop. \ref{prop13.1} $K^{*{\rm cl}}: H^{1}_{\rk}(X)\tosim H^{1}_{\rk}(X)^{*}$. 

We recall that the space $\overline{H^{1}_{\rk}}(\Omega)$ is the space of restrictions to $\Omega$ of elements in $H^{1}_{\rk}(X)$. 
Now for $g\in \coinf(X)$ we have
\[
|(e_{\Omega}w|g)_{\Omega}|= (w| r_{\Omega}g)_{\Omega}\leq C\| w\|_{\bar{H^{1}_{\rk}}(\Omega)^{*}}\| r_{\Omega}g\|_{\bar{H^{1}_{\rk}}(\Omega)}\leq  C\| w\|_{\bar{H^{1}_{\rk}}(\Omega)^{*}}\| g\|_{H^{1}_{\rk}(X)},
\]
which implies that
\begin{equation}
\label{arsou}
\| e_{\Omega}w\|_{H^{1}_{\rk}(X)^{*}}\leq C \| w\|_{\bar{H^{1}_{\rk}}(\Omega)^{*}}.
\end{equation}
Therefore we deduce from \eqref{e.24} by duality that
if
\[
r_{1,n}\defeq \chi_{n}u- V^{+}\gamma^{+}\chi_{n}u,
\]
we have
\begin{equation}
\label{e.25}
\|r_{1, n}\|_{\bar{H^{1}_{\rk}}(\Omega)}\leq C n^{-1}\| u\|_{\bar{H^{1}_{\rk}(}\Omega)}.
\end{equation}
Hence, 
\begin{equation}
\label{e.25b}
r_{1, n} \to 0\hbox{ in }\bar{H^{1}_{\rk}}(\Omega)\hbox{ as }n\to\infty. 
\end{equation}
We now apply \eqref{e.22} once again replacing $u$ by $v_{n}= \chi_{n}V^{+}\gamma^{+}\chi_{n}u$. Note that $V^{+}\gamma^{+}\chi_{n}u\in \bar{\cinf}(\Omega)$, so $v_{n}\in \bar{\coinf}(\Omega)$. We obtain since $KV^{+}\gamma^{+}\chi_{n}u=0$ in $\Omega$:
\[
(w| v_{n}- V^{+}\gamma^{+}v_{n})_{\Omega}= (v| [K, \chi_{n}]V^{+}\gamma^{+}\chi_{n}u)_{\Omega},
\]
hence by \eqref{ecorrect.13} and \eqref{arsou}
\[
|(w| v_{n}- V^{+}\gamma^{+}v_{n})_{\Omega}|\leq Cn^{-1}\| w\|_{\bar{H^{1}_{\rk}}(\Omega)^{*}}\| V^{+}\gamma^{+}\chi_{n}u\|_{H^{1}_{\rk}(\Omega)}\leq Cn^{-1}\| \chi_{n}u\|_{\bar{H^{1}_{\rk}(}\Omega)},
\]
where we use \eqref{e.25} in the last inequality.   Since 
\[
\| \chi_{n} u\|_{\bar{H^{1}_{\rk}(}\Omega)}\leq C (\| \chi\|_{\infty}+ \| d\chi\|_{\infty})\| u\|_{\bar{H^{1}_{\rk}(}\Omega)}\leq c \| u\|_{\bar{H^{1}_{\rk}(}\Omega)},
\]
we obtain finally that if
\[
r_{2, n}\defeq \chi_{n}V^{+}_{0}\gamma^{+}\chi_{n}u- V^{+}_{0}\gamma^{+}\chi_{n}V^{+}_{0}\gamma^{+}\chi_{n}u
\]
we have:
\begin{equation}
\label{e.26}
r_{2, n} \to 0\hbox{ in }\bar{H^{1}_{\rk}}(\Omega)\hbox{ as }n\to\infty. 
\end{equation}
We note then that if $V\Subset X$ and $V^{+}= V\cap\Omega$, we have  $K\chi_{n}u=0$ in $V^{+}$ for $n\gg 1$, hence $Kr_{1, n}= Kr_{2, n}=0$ in $V^{+}$ for $n\gg 1$.

We can introduce local coordinates on $\Sigma$ near $y^{0}\in \Sigma$, and map $V$ to a neighborhood $\tilde{V}$ of $(0, 0)$ in $\rr^{1+d}$ for $d= \dim\Sigma$. Denoting by $H^{m, \ell}(\rr^{1+d})$ the space of $u\in \cS'(\rr^{1+d})$ such that $\langle D_{x}\rangle^{m}\langle D_{y}\rangle^{\ell}u\in L^{2}(\rr^{1+d})$, and then using the coordinates we define the spaces  ${H^{1+k, -k}_{\rm loc}}(X)$ and $\overline{H^{1+k, -k}_{\rm loc}}(\Omega)$ (the definition depends on the choice of coordinates, but this is not important here). Then one deduces, using that $Kr_{i ,n}=0$ in $V^{+}$ that $r_{i, n}\to 0$ also in  $\overline{H^{1+k, -k}_{\rm loc}}(\Omega)$ for any $k\in \nn$, see eg \cite[page 311]{Gr}.

We can then  safely apply  the boundary value operator $\gamma^{+}$ and deduce from \eqref{e.25b}, \eqref{e.26} that:
\[
\gamma^{+}r_{i, n}\to 0\hbox{ in }\cD'(\Sigma; \cc^{2}).
\]
Since  $c^{+}f= \gamma^{+}V^{+}f= \gamma^{+}u$ this yields:
\[
\begin{array}{l}
\chi_{n}c^{+}f- c^{+}\chi_{n}c^{+}f\to 0,\\[2mm]
\chi_{n}c^{+}\chi_{n}c^{+}f- c^{+}\chi_{n}c^{+}\chi_{n}c^{+}f\to 0,
\end{array}
\]
in $\cD'(\Sigma; \cc^{2})^{2}$ when $n\to \infty$, which completes the proof. \qed

\subsection{Purity of the Hartle-Hawking-Israel state}
We will follow the arguments in \cite[Subsect. 4.4]{GW2}.

By Prop. \ref{purity}  it suffices to find, for each $f\in \coinf(\Sigma; \cc^{2})$, a sequence $f_{n}\in\coinf(\Sigma; \cc^{2})$ such that:
\beq\label{e.27}
\lim_{n\to +\infty}\frac{| \bar{f}\dual q f_{n}|^{2}}{ \bar{f}_{n}\dual(\lambda^{+}+ \lambda^{-})f_{n}}= \bar{f}\dual(\lambda^{+}+ \lambda^{-})f.
\eeq
We take
\[
f_{n}= \chi_{n}(c^{+}- c^{-})f,
\]
which belongs to $\coinf(\Sigma; \cc^{2})$ since $c^{\pm}: \coinf(\Sigma; \cc^{2})\to \cinf(\Sigma; \cc^{2})$.  Since $\chi_{n}f=f$ for $n\gg 1$  we have
\beq\label{e.28}
\bar{f}\dual q f_{n}= \overline{f}\dual (\lambda^{+}+ \lambda^{-})f \hbox{ for }n\gg 1.
\eeq
Let us now compute the limit of the denominator in \eqref{e.27}.  We have $f_{n}= \chi_{n}(2c^{+}-1)f= (2\chi_{n}c^{+}-1)f$ for $n\gg 1$ and using that $qc^{+}$ is Hermitian by Thm. \ref{thm14.1}, we obtain that $ \bar{f}_{n}\dual(\lambda^{+}+ \lambda^{-})f_{n}= \bar{f}qg_{n}$ for $n\gg 1$ for
\[
\begin{array}{rl}
&g_{n}=(2c^{+}\chi_{n}-1)(2c^{+}-1)(2\chi_{n}c^{+}-1)f\\[2mm]
=&(8 c^{+}\chi_{n}c^{+}\chi_{n}c^{+}- 8 c^{+}\chi_{n}c^{+}- 4 c\chi_{n}^{2}c^{+}+6c-1)f\hbox{ for }n\gg 1.
\end{array}
\]
We apply Prop. \ref{propototo} {\it ii)} and {\it i)} (if necessary replacing $\chi_{n}$ by $\chi_{n}^{2}$) and obtain that
\[
\begin{array}{rl}
\bar{f}\dual q g_{n}=& \bar{f}\dual q (8\chi_{n}c^{+}\chi_{n}- 8 \chi_{n}c^{+}- 4 \chi_{n}^{2}c^{+}+6c^{+}-1)f+ o(1)\\[2mm]
=& \bar{f}\dual q (2c^{+}-1)f+ o(1)= \bar{f}\dual (\lambda^{+}+ \lambda^{-})f+o(1)
\end{array}
\]
when $n\to \infty$. Using \eqref{e.28} this  completes the proof of \eqref{e.27}. \qed

\appendix
\section{}\init\label{secapp1}
\subsection{Proof of Prop. \ref{prop10.0}}\label{secapp1.0}
Since $r$ is an isometry of $(\Sigma, \rh)$, $r_{|\cB}= Id$ and $r: \Sigma^{+}\to \Sigma^{-}$ we obtain \eqref{e10.1}.
 The first identity in \eqref{e10.0} follows from the fact that $(u, \omega)$ are normal Gaussian coordinates to $\cB$ for $\rh$, the other are
 tautologies.

We obtain  from \eqref{e10.1} and \ref{sec10.2.1} that $N$, $\rw^{0}$ are  odd in $u$,
 $\rw^{\alpha}$, $\rk_{\alpha\beta}$ are even in $u$ with $\rw^{\alpha}(0, \omega)=0$.  The function $m$ is even in $u$ by invariance under $r$. We now use  Killing's equation 
 \beq\label{e-1.0}\nabla_{a}V_{b}+ \nabla_{b}V_{a}=0,
 \eeq
 noting  that since $V=0$ on $\cB$ we have
\beq\label{e10.3b}
\nabla_{a}V_{b}= \p_{a}V_{b}\hbox{ on }\cB.
\eeq
If we work in Gaussian normal coordinates to $\Sigma$ for $\rg$, so that
\[
\rg= - dt^{2}+ \rh_{ij}(t,y)dy^{i}dy^{j}, \ V= - N(t, y)\p_{t}+ \rw^{0}(t, y)\p_{u}+ \rw^{\alpha}(t, y)\p_{\omega^{\alpha}}
\]
 and  $y= (u, \omega)$, we obtain from \eqref{e-1.0}, \eqref{e10.3b} that: 
\[
\p_{u}V_{0}(0, \omega)=0\Rightarrow \p_{u}\rw^{0}(0, \omega)=0.
\]
Summarizing we have:
\begin{equation}
\label{e10.44}
\begin{array}{l}
N(u, \omega)= ua(u^{2}, \omega),\\[2mm]
\rw^{0}(u, \omega)= u^{3}b(u^{2}, \omega),\ \rw^{\alpha}(u, \omega)= u^{2}c^{\alpha}(u^{2}, \omega),\\[2mm]
\rk_{\alpha\beta}(u, \omega)= {\rm d}_{\alpha\beta}(u^{2}, \omega),\ m(u, \omega)= n(u^{2}, \omega),
\end{array}
\end{equation}
for smooth functions $a, b, c^{\alpha}, {\rm d}_{\alpha\beta}:]-\epsilon, \epsilon[\times \cB\to \rr$ with 
\[
n(0, \omega)\geq c>0, c^{-1}\one \leq [{\rm d}_{\alpha\beta}(0, \omega)]\leq c\one, \hbox{ for some }c>0. 
\]
To complete the proof of the proposition it remains to show that $\kappa= a(0, \omega)$. 

To do this we reexpress the surface gravity $\kappa$. By \cite[Lemma 2.5]{S1} we have:
\[
\kappa^{2}= (\rh^{ij}\p_{i}N \p_{j}N)_{|\cB}- \12 (\rh^{ij}\rh^{kl}\nabla_{i}^{(\rh)}\rw_{l}\nabla_{j}^{(\rh)}\rw_{k})_{|\cB},
\]

which using \eqref{e10.44}  gives $\kappa= a(0, \omega)$. \qed
\subsection{Proof of Prop. \ref{prop10.1}}\label{secapp1.2}
 We recall that  we defined the coordinates $(u, \omega)\in ]-\delta, \delta[\times \cB$ on  a small neighborhood $U$ of $\cB$ in $\Sigma$.  $U\cap \Sigma^{+}$ is diffeomorphic to $]0, \delta[\times \cB$ using the coordinates $(u, \omega)$.
If
\[
X= u\cos(\kappa s), Y= u \sin(\kappa  s),
\]
we have:
\[
 du= u^{-1}(X dX+ Y dY), \ d s=\kappa^{-1}u^{-2}( XdY- YdX).
\]
By Prop. \ref{prop10.0} we obtain:
\[
\begin{array}{l}
\rk_{\alpha\beta}(u, \omega)d\omega^{\alpha}d\omega^{\beta}= d_{\alpha\beta}(X^{2}+ Y^{2}, \omega)d\omega^{\alpha}d\omega^{\beta},\\[2mm]
\i \rw_{\alpha}(u, \omega)d\omega^{\alpha}d s= \i \kappa^{-1}b_{\alpha}(X^{2}+ Y^{2}, \omega)(XdY- YdX)d\omega^{\alpha},\\[2mm]
\i \rw_{0}(u, \omega)du d s= \i \kappa^{-1}b_{0}(X^{2}+ Y^{2}, \omega)(X dY- YdX)(XdX+ YdY),\\[2mm]
v^{2}(u, \omega)d s^{2}+ du^{2}\\[2mm]
= u^{2}\kappa^{2}(1+ u^{2}d(u^{2}, \omega))\kappa^{-2}u^{-4}(X dY- YdX)^{2}+ u^{-2}(XdX+ YdY)^{2}\\[2mm]
= dX^{2}+ dY^{2}+ d(X^{2}+ Y^{2}, \omega)(XdY- YdX)^{2}.
\end{array}
\]
Let us  denote by $B_{2}(0, \delta)= \{(X, Y)\in \rr^{2}: X^{2}+ Y^{2}\leq \delta^{2}\}$ the open disk of center $0$ and radius $\delta$ in $\rr^{2}$.  
If $\beta = (2\pi)\kappa^{-1}$, then $(u, \kappa  s)\in ]0, \delta[\times \bS_{2\pi}$ are polar coordinates on $B_{2}(0, \delta)\setminus \{0\}$. The expression \eqref{e0.6} for $\rg^{\rm eucl}$ and  the estimates above show that $\rg^{\rm eucl}$ extends as a smooth complex metric  on $B_{2}(0, \delta)\times \cB$.

We then construct  $M^{\rm eucl}_{\rm ext}$ by gluing $B_{2}(0, \delta)\times \cB$ with $M^{\rm eucl}=\bS_{\beta}\times \Sigma^{+}$ over $ \{(X, Y)\in \rr^{2}: \12 \delta^{2}<X^{2}+ Y^{2}<\delta^{2}\}\times \cB$ using the map:
\beq\label{titi}
 \begin{array}{l}
\bS_{\beta}\times ]0, \delta[\times \cB\to B_{2}(0, \delta)\times \cB\\
( s, u, \omega)\mapsto (u\cos(\kappa s),  u \sin(\kappa  s), \omega).
\end{array}
\eeq
  The complex metric $\rg^{\rm eucl}$ defined on $\bS_{\beta}\times \Sigma^{+}$ extends to a smooth complex metric $\rg^{\rm eucl}_{\rm ext}$ on $M^{\rm eucl}_{\rm ext}$. By Prop. \ref{prop10.0} we have $m= n(X^{2}+ Y^{2}, \omega)$, hence $m$ extends as a smooth function on $M^{\rm eucl}_{\rm ext}$.
  
 Let us now  embed $\Sigma$ isometrically into $M^{\rm eucl}_{\rm ext}$.  In  the coordinates $(u, \omega)$ on $\Sigma$ near $\cB$ the embedding $\hat{\psi}$ becomes
\[
(u,\omega)\mapsto  \left\{\begin{array}{l}
 (0, u, \omega)\hbox{ for }0<u<\delta,\\
 (\b2,-u, \omega)\hbox{ for }-\delta<u<0,
\end{array}\right.
\]
which smoothly extends to $u=0$, the image of $\Sigma$ under this extension being locally equal to $\{Y=0\}$.  

The open set $\Omega_{\rm ext}$ is obtained by gluing $\{Y>0\}$ with $]0, \b2[\times \Sigma^{+}$ using the map \eqref{titi}.
This completes the proof.
\subsection{Proof of \eqref{e13.1b}}\label{secapp1.1}
 A mechanical computation gives:
 \[
 \begin{array}{rl}
\sum_{i}\nabla_{i}T^{i}=&\sum_{i} \p_{i}T^{i}+ \12 \sum_{i,k,l}\rk^{il}(\p_{i}\rk_{kl}+ \p_{k}\rk_{il}- \p_{l}\rk_{ik})T^{k}\\[2mm]
=&\sum_{i} \p_{i}T^{i} +\12 \sum_{i,k,l}\rk^{il} \p_{k}\rk_{il}T^{k}\eqdef I,
\end{array}
\]
using that $\rk^{li}= \rk^{il}$, $\rk_{kl}= \rk_{lk}$.
Next
\[
\sum_{i} |\rk|^{-\12} \p_{i}(|\rk|^{\12}T^{i})= \sum_{i}\p_{i}T^{i} + \12 \sum_{i}|\rk|^{-1}\p_{i}|\rk|T^{i}\eqdef II.
\]
Since
\[
{\rm det}A(t)^{-1}\frac{d}{dt}{\rm det}A(t)= {\rm Tr}(A(t)^{-1}\frac{d}{dt}A(t)), 
\]
 we get that $\p_{i}|\rk|= |\rk|{\rm Tr}(\rk^{-1}\p_{i}\rk)$. Next  we compute:
\[
(\rk^{-1}\p_{i}\rk)^{j}_{k}= \sum_{l}\rk^{jl}\p_{i}\rk_{lk},\ {\rm Tr}(\rk^{-1}\p_{i}\rk)= \sum_{k,l}\rk^{kl}\p_{i}\rk_{lk},
\]
which  shows that $I= II$. \qed

\end{document}